\documentclass[12pt]{article}

\usepackage[
  top=1in, bottom=1in, left=1in, right=1in,
  textheight=600pt
]{geometry}
\usepackage{setspace}
\usepackage{xr}
\usepackage{amsmath}
\usepackage{amssymb}
\usepackage{amsfonts}
\usepackage{amsthm}
\usepackage{mathtools}
\usepackage{fixmath}
\usepackage{bm}

\usepackage{booktabs}
\usepackage{array}
\usepackage{multirow}
\usepackage{threeparttable}
\usepackage[table]{xcolor}

\usepackage{graphicx}
\usepackage{subcaption}
\usepackage{float}
\usepackage{rotating}
\usepackage{tikz}

\usepackage{natbib}
\usepackage{hyperref}
\hypersetup{
    colorlinks=true,
    citecolor=blue,
    linkcolor=blue,
    urlcolor=blue
}
\bibpunct{(}{)}{;}{a}{,}{,}

\usepackage{enumitem}
\usepackage{xcolor}
\usepackage{hyperref}
\usepackage{authblk}

\newtheorem{theorem}{Theorem}[section]

\newcommand{\Var}{\text{Var}}

\title{\textbf{Optimal cohort Staircase designs}}

\author[]{Soumadeb Pain$^{*}$}
\author[]{Satya Prakash Singh}

\affil[]{
  \small Department of Statistics and Data Science, \\
  Indian Institute of Technology Kanpur, 208016, India \\
  $^{*}$Corresponding author: \texttt{soumadebp22@iitk.ac.in}
}

\date{}

\begin{document}
\maketitle

\label{firstpage}

\begin{abstract}
\noindent Staircase designs are a practical alternative to conventional stepped-wedge cluster randomised trials when collecting outcome measurements from every
cluster throughout the entire study is costly or burdensome. In a staircase design, each
cluster is observed only over a limited window surrounding the transition from control to
intervention, substantially reducing the measurement burden. However, existing works on
staircase designs have largely focused on equal cluster sizes and do not fully address
how clusters should be optimally allocated across treatment sequences.
In this work, we develop an optimal design framework for cohort staircase trials allowing
cluster sizes to vary across sequences, while maintaining equal cluster sizes within each
sequence. For several important classes of staircase designs, assuming equal cluster sizes, we further obtain analytical optimal allocations and establish symmetry properties of the optimal designs. We also introduce a Min–Min Design (MMD), which jointly optimises the allocation of clusters across sequences and the sequence-specific cluster sizes under a fixed total number of participants.  We further compare staircase and standard stepped-wedge designs in terms of statistical efficiency, participant burden and total cost. Results indicate that staircase designs can achieve large reductions in repeated measurements and participant observations while maintaining comparable power, particularly for longer trials.
\end{abstract}

\section{Introduction}

Cluster randomised trials (CRTs) are popular in many fields including medicine, education, and social sciences \citep{Donner_2000}. CRTs are pivotal in evaluating interventions where groups, rather than individuals, are randomised. Statistical methods for CRTs have been the focus of extensive research over the past several decades and are well-documented in various methodological reviews \citep{turner2017review1, turner2017review}. Among CRT designs, the stepped-wedge design (SWD) has garnered increasing attention, alongside traditional parallel and crossover CRTs \citep{brown2006stepped, mdege2011systematic}. While parallel designs randomize clusters to fixed intervention or control arms and crossover designs alternate clusters between arms over time, the SW-CRT \citep{Hussey_2007, hemming2015stepped} employs a unidirectional roll-out of all the clusters from control to intervention in sequential ``steps.'' The order in which the different individuals or clusters receive the intervention is random. The study continues until all the clusters are assigned to the intervention, with outcome data collected from each cluster throughout the study period. SWDs are particularly attractive when (i) withholding the intervention entirely is considered unethical, (ii) logistical constraints require a staggered rollout, or (iii) the intervention is complex and cannot be implemented in all clusters simultaneously. A substantial body of statistical methodology now exists for sample size calculations \citep{hooper2016sample, Hemming_2016, Baio_2015}, optimal design \citep{li2018optimal, Matthews_2020, Lawrie_2015}, and analysis of SWDs \citep{kasza2019impact, girling2016statistical, thompson2017optimal, li2021mixed}.

While the ethical and practical benefits of SWDs are significant, \citet{hemming2020reflection} discussed several key factors that should be considered when implementing a SW-CRT. In particular, a complete SWD requires every participating cluster to collect outcome data in every time period of the trial. In some trial settings, there may be sufficient time to stagger the introduction of the intervention across clusters, but it may nevertheless be burdensome or costly to collect individual-level data from each cluster throughout the entire trial duration. This repeated measurement burden can be especially important in settings where data collection is resource-intensive or logistically difficult. For example, researchers evaluating a catheter flushing education program deliberately kept measurement periods short to avoid overwhelming hospital wards with data collection \citep{keogh2020implementation}, while a trial in remote Aboriginal communities used a restricted measurement window because a complete SWD was considered too costly, geographically impractical, and burdensome for teachers and parents \citep{wagner2020school}. These practical considerations motivate the use of non-standard SWDs in which clusters are not observed throughout the complete trial.

To address this issue, \citet{grantham2024staircase} formally introduced the \textbf{staircase design} (SCD). Before the design had a formal definition, the statistical literature encountered it mostly as an example of an incomplete stepped wedge \citep{hemming2015stepped, unni2020variations}, as a high-efficiency sub-design running along the diagonal \citep{hooper2020hunt, kasza2021information}, or as a loose conceptual extension of the dog-leg design \citep{hooper2015cluster}. An SCD is a longitudinal CRT in which each cluster is measured only in a limited window of $R_0$ pre-switch control periods and $R_1$ post-switch intervention periods, rather than throughout the entire trial duration. Derived from the zigzag pattern of steps along the diagonal of the SWD schematic, the SCD retains much of the efficiency of the complete SWD while substantially reducing the measurement burden per cluster \citep{grantham2025relative}. The resulting design can therefore be attractive when repeated outcome measurement, rather than the availability of clusters or study duration, is the primary practical constraint.

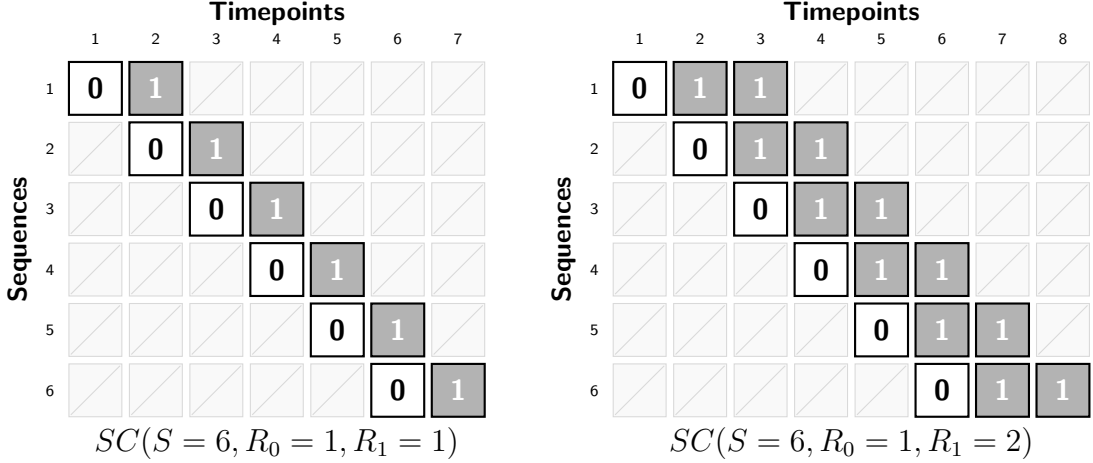
\begin{figure}
    \centering
    \begin{tikzpicture}[
    % --- Define Styles Once Globally ---
    cell base/.style={
        rectangle,
        minimum size=0.7cm,
        outer sep=0pt,
        font=\sffamily\bfseries\small,
        anchor=center
    },
    empty cell/.style={
        cell base,
        draw=gray!30,
        fill=gray!5,
        thin
    },
    control cell/.style={
        cell base,
        draw=black,
        thick,
        fill=white,
        text=black
    },
    treat cell/.style={
        cell base,
        draw=black,
        thick,
        fill=gray!60, 
        text=white
    },
    slash/.style={
        draw=gray!30,
        thin,
        shorten >=1pt, shorten <=1pt
    }
]

% ========================================================
% FIGURE 1: SC(6,1,1) (Left)
% ========================================================
\begin{scope}[shift={(0,0)}]
    \def\numrows{6}
    \def\numcols{7} 
    \def\colsep{0.8}
    \def\rowsep{0.8}

    % Background Empty & Slashed Cells
    \foreach \i in {1,...,\numrows} {
        \foreach \j in {1,...,\numcols} {
            \pgfmathsetmacro{\xcoord}{\j*\colsep}
            \pgfmathsetmacro{\ycoord}{-\i*\rowsep}
            \node[empty cell] (c1-\i-\j) at (\xcoord, \ycoord) {};
            \draw[slash] (c1-\i-\j.south west) -- (c1-\i-\j.north east);
        }
    }

    % Diagonal Control (0)
    \foreach \i in {1,...,\numrows} {
        \node[control cell] at (c1-\i-\i) {0};
    }
    
    % Diagonal Treatment (1) 
    \foreach \i in {1,...,\numrows} {
        \pgfmathtruncatemacro{\j}{\i+1}
        \ifnum\j>\numcols\else
            \node[treat cell] at (c1-\i-\j) {1};
        \fi
    }

    % X-Axis (Timepoints)
    \foreach \j in {1,...,\numcols} {
        \pgfmathsetmacro{\xcoord}{\j*\colsep}
        \node[font=\sffamily\tiny, anchor=south] at (\xcoord, -0.35) {\j};
    }
    \pgfmathsetmacro{\midx}{(\numcols+1)*\colsep/2}
    \node[font=\sffamily\footnotesize\bfseries] at (\midx, 0.2) {Timepoints};

    % Y-Axis (Sequences)
    \foreach \i in {1,...,\numrows} {
        \pgfmathsetmacro{\ycoord}{-\i*\rowsep}
        \node[font=\sffamily\tiny, anchor=east] at (0.4, \ycoord) {\i};
    }
    \pgfmathsetmacro{\midy}{(-\numrows-1)*\rowsep/2}
    \node[font=\sffamily\footnotesize\bfseries, rotate=90, anchor=south] at (0.1, \midy) {Sequences};

    % Bottom Label
    \pgfmathsetmacro{\bottomy}{(-\numrows*\rowsep) - 0.7}
    \node[font=\sffamily\normalsize] at (\midx, \bottomy) {$SC(S = 6,R_0 =1,R_1 = 1)$};
\end{scope}

% ========================================================
% FIGURE 2: SC(6,1,2) (Right)
% ========================================================
\begin{scope}[shift={(7.2,0)}] % Shifted safely to the right
    \def\numrows{6}
    \def\numcols{8} 
    \def\colsep{0.8}
    \def\rowsep{0.8}

    % Background Empty & Slashed Cells
    \foreach \i in {1,...,\numrows} {
        \foreach \j in {1,...,\numcols} {
            \pgfmathsetmacro{\xcoord}{\j*\colsep}
            \pgfmathsetmacro{\ycoord}{-\i*\rowsep}
            \node[empty cell] (c2-\i-\j) at (\xcoord, \ycoord) {};
            \draw[slash] (c2-\i-\j.south west) -- (c2-\i-\j.north east);
        }
    }

    % Diagonal Control (0)
    \foreach \i in {1,...,\numrows} {
        \node[control cell] at (c2-\i-\i) {0};
    }
    
    % Diagonal Treatment (1) - 2 periods long
    \foreach \i in {1,...,\numrows} {
        % Period 1
        \pgfmathtruncatemacro{\jone}{\i+1}
        \ifnum\jone>\numcols\else
            \node[treat cell] at (c2-\i-\jone) {1};
        \fi
        % Period 2
        \pgfmathtruncatemacro{\jtwo}{\i+2}
        \ifnum\jtwo>\numcols\else
            \node[treat cell] at (c2-\i-\jtwo) {1};
        \fi
    }

    % X-Axis (Timepoints)
    \foreach \j in {1,...,\numcols} {
        \pgfmathsetmacro{\xcoord}{\j*\colsep}
        \node[font=\sffamily\tiny, anchor=south] at (\xcoord, -0.35) {\j};
    }
    \pgfmathsetmacro{\midx}{(\numcols+1)*\colsep/2}
    \node[font=\sffamily\footnotesize\bfseries] at (\midx, 0.2) {Timepoints};

    % Y-Axis (Sequences)
    \foreach \i in {1,...,\numrows} {
        \pgfmathsetmacro{\ycoord}{-\i*\rowsep}
        \node[font=\sffamily\tiny, anchor=east] at (0.4, \ycoord) {\i};
    }
    \pgfmathsetmacro{\midy}{(-\numrows-1)*\rowsep/2}
    \node[font=\sffamily\footnotesize\bfseries, rotate=90, anchor=south] at (0.1, \midy) {Sequences};

    % Bottom Label
    \pgfmathsetmacro{\bottomy}{(-\numrows*\rowsep) - 0.7}
    \node[font=\sffamily\normalsize] at (\midx, \bottomy) {$SC(S = 6,R_0 =1,R_1 = 2)$};
\end{scope}

\end{tikzpicture}
    \caption{Different types of SCD showing control periods (white) and intervention periods (gray).}
    \label{fig:SCD}
\end{figure}

As depicted in Figure \ref{fig:SCD}, an SCD with $T$ time periods has $S=T-R_0-R_1+1$ sequences, where $R_0$ and $R_1$ are the numbers of control and intervention periods, respectively. In practice, equal numbers of clusters are often assigned to the different sequences. However, in the broader SWD literature, \citet{li2018optimal} showed that equal allocation across sequences is generally not optimal, even when all clusters have the same size. In particular, they showed that, under balanced cluster sizes, the extreme sequences receive equal but larger numbers of clusters than the intermediate sequences. More recently, \citet{Pain_2026} proposed optimal cluster allocation for SWDs with unequal cluster sizes. These results suggest that the allocation of clusters across sequences can have an important influence on statistical efficiency and that equal allocation in a SCD should not necessarily be taken for granted.

Despite the theoretical appeal and growing practical use of SCDs, several fundamental optimal-design questions remain unanswered. Existing work on SCDs has largely focused on equal cluster sizes across sequences. In practice, however, clusters are unlikely to be of equal size. We therefore consider a cohort SCD in which all clusters within a given sequence have the same size, while cluster sizes are allowed to vary across sequences. This assumption is reasonable in several settings. For example, geographic rollout phases may naturally group administrative areas of similar size. Clusters such as hospitals or schools are also frequently stratified by size prior to randomization, which can naturally result in balanced sizes within sequences. Furthermore, equal cluster sizes within sequences can simplify trial management by standardizing resource requirements during each step of the rollout. In addition, neither \citet{grantham2024staircase} nor \citet{rezaei2025inference} addresses the question of how clusters should be allocated across sequences to minimize the variance of the estimated treatment effect. 

\noindent Motivated by this gap, we develop a general optimal-design framework for cohort staircase trials that simultaneously addresses sequence allocation, sequence-specific cluster sizes, and the choice of observation window. We first derive the variance of the treatment-effect estimator under a general cohort covariance structure and establish that the resulting variance is a convex function of the sequence-allocation proportions. We then derive an equivalence theorem that provides a necessary and sufficient condition for verifying the optimality of a given allocation, together with a symmetry result that substantially reduces the dimension of the optimization problem for balanced SCDs. These results lead to exact analytical optimal allocations for several important SCDs. We further introduce a \emph{Min--Min Design} (MMD), which jointly optimises the allocation of clusters across sequences and the sequence-specific cluster sizes subject to a fixed total number of participants. Finally, we investigate the statistical and financial trade-offs between staircase and standard SWDs, study the choice of the observation window, and illustrate the practical implications of the proposed methodology using trial-based case studies. Together, these results provide a unified framework for designing cohort staircase trials when both measurement burden and cluster-size heterogeneity are important considerations.

\section{Model}
We have considered a cohort SCD trial denoted as $SC(S,R_0,R_1)$ where $S$ is the total number of sequences, $R_0$ and $R_1$ are the total number of control periods and intervention periods, respectively. For different values of $S, R_0 \ \text{and} \ R_1$, different SCDs are shown in Figure 1. An SCD is called a balanced SCD if $R_0 = R_1$ and for $R_0 = R_1 = 1$, it is a special case of balanced SCD called a basic SCD.  

Suppose sequence $s$ has $K_s$ clusters, each having size $m_s$, such that $\sum_{s=1}^SK_s = K$. Letting \( Y_{skti} \) represent the outcome for subject \( i = 1, \ldots, m_s \) in cluster \( k = 1, \ldots, K_s \) assigned to sequence \( s = 1, \ldots, S \) in period \( t = s, \ldots, s + R_0 + R_1 - 1 \). We considered the following mixed effects model for a continuous outcome:
\begin{equation}\label{Model}
\begin{aligned}
Y_{skti} &= \mathbf{Z}_{st} \boldsymbol{\beta} + X_{st} \theta + \delta_{sk} + CP_{skt} + u_{ski} + \epsilon_{skti} 
\end{aligned}
\end{equation}
where, \( \boldsymbol{\beta} \) is a \( p \)-dimensional column vector of fixed time effects, \( \mathbf{Z}_{st} \) is a \( p \)-dimensional row vector specifying the form for the fixed effects corresponding to period \( t \), \( X_{st} \) is the treatment indicator for sequence \( s \) in the period \( t \), \( \theta \) is the treatment effect, $\delta_{sk} \sim \mathcal{N}(0,\sigma_\delta^2)$ is cluster-specific random effect, \( CP_{skt} \sim \mathcal{N}(0, \sigma^2_C)\)  is the random effect for cluster k assigned to sequence $s$ in period $t$, $u_{ski} \sim \mathcal{N}(0,\sigma^2_u)$ is the subject-specific random effect shared across periods and \( \epsilon_{skti} \sim \mathcal{N}(0, \sigma^2_\epsilon)\) is the subject-level error term.

\noindent Note that $\mathbf{CP}_{sk} = (CP_{sks}, \ldots, CP_{sk,s+R_0+R_1-1})^T \sim \mathcal{N}_{R_0+R_1}(0, \mathbf{V})$ where $\mathbf{V}$ is the covariance matrix of the cluster-period random effects across the $R_0 + R_1$ periods of measurement under the assumption that all clusters assigned to all sequences have random effects with an identical distribution. A general form of $\mathbf{V}$ is given by $\text{Var}(CP_{skt}) = \sigma_C^2$ and $\text{Cov}(CP_{skt},CP_{skt'}) = r_{tt'}\sigma_C^2$ for $0 < r_{tt'} \leq 1$. Assuming $r_{tt'} = r \ \forall \ t \neq t'$, the block exchangeable or constant between-period intracluster correlation structure is obtained.  In longitudinal cluster trials, observations closer in time tend to be more highly correlated. Therefore an alternative correlation structure allows for these correlations to decrease as the time between subjects’ periods of measurement increases by assuming $r_{tt'} = r^{|t-t'|}$. 

The variance of $Y_{skti}$ is  $\sigma^2 = Var(Y_{skti})= \sigma_{\delta}^2 + \sigma_C^2 + \sigma_u^2 + \sigma_\epsilon^2$,  where $\sigma_{u}^2 ,   \sigma_C^2  \ge 0$ and $\sigma_{\delta}^2, \sigma_{\epsilon}^2 > 0$. The correlation between the responses of two distinct individuals within the same cluster and period is $\alpha_0=\operatorname{Corr}(Y_{skti},Y_{skti'})=\frac{\sigma_{\delta}^2+\sigma_C^2}{\sigma^2},\ i\neq i',$ which is referred to as the within-period correlation. Next, define $\alpha_1=\frac{\sigma_{\delta}^2}{\sigma^2}.$ For two distinct individuals in the same cluster observed at different periods $t\neq t'$, the inter-period correlation is $\alpha_{1,tt'}=\operatorname{Corr}(Y_{skti},Y_{skt'i'})=\alpha_1+r_{tt'}(\alpha_0-\alpha_1),\ i\neq i',$ so that $\alpha_1$ represents the persistent cluster-level contribution to the between-period dependence. Finally, define $\alpha_2=\frac{\sigma_{\delta}^2+\sigma_u^2}{\sigma^2}.$ The correlation between repeated responses from the same individual at two distinct periods is $\alpha_{2,tt'}=\operatorname{Corr}(Y_{skti},Y_{skt'i})=\alpha_2+r_{tt'}(\alpha_0-\alpha_1),\ t\neq t',$ which is referred to as the individual autocorrelation. We retain $\alpha_1$ and $\alpha_2$ as the fundamental model parameters rather than $\alpha_{1,tt'}$ and $\alpha_{2,tt'}$, because the latter are fully determined by $\alpha_1$, $\alpha_2$, and $r_{tt'}$. It can be observed that $0<\alpha_1\leq \min \{\alpha_0,\alpha_2\}<1$.

\noindent In Section \ref{Sec:Optimal.SCD}, we formally introduce the objective function or optimality criterion, used to determine the optimal allocation of clusters across sequences. The criterion is based on the variance of the generalized least squares estimator, $\widehat{\theta}$, of the treatment effect $\theta$. Deriving this variance in closed form is difficult due to its complex covariance structure induced by the cohort SCD. But we obtain a tractable analytical expression for the variance, which forms the basis of the proposed optimality criterion. A detailed derivation is provided in the Supplementary Material. There we have shown that
under block exchangeable correlation structure ($r_{tt'} = r \in (0,1] \ \ \forall \ t \neq t'$), the variance of the generalized least squares estimator $\hat{\theta}$ of $\theta$ is given by
\begin{equation}\label{Var:blockexchangeble}
    \text{Var}(\hat{\theta}) = \frac{1}{K}\left[c_\mathbf{p} - \mathbf{b}_\mathbf{p}^\top\mathrm{A}_\mathbf{p}^{-1}\mathbf{b}_\mathbf{p}\right]^{-1}
\end{equation}
where, $K$ is the total number of clusters, $\mathbf{p} = (p_1, \ldots, p_S)^{\top}; \ p_s = \frac{K_s}{K}:$ the proportion of total cluster that are in sequence $s$, $c_\mathbf{p} = \sum\limits_{s=1}^Sp_sg_s \ \  \text{with} \ \ g_s = \mathbf{X}^\top(\mathbf{V}_*^{s})^{-1}\mathbf{X}$, $\mathbf{b}_\mathbf{p} = \sum_{s=1}^Sp_s\mathbf{d}_s$ with $\mathbf{d}_s = \mathrm{Z}_s^\top(\mathbf{V}_*^{s})^{-1}\mathbf{X}$, $\mathrm{A}_\mathbf{p} = \sum_{s=1}^Sp_s\mathrm{B}_s$ with $\mathrm{B}_s = \mathrm{Z}_s^\top(\mathbf{V}_*^{s})^{-1} \mathrm{Z}_s$, $\mathrm{Z}_s = (\mathbf{Z}_{ss}, \mathbf{Z}_{ss+1}, \cdots, \mathbf{Z}_{ss+R_0+R_1-1} )^\top$ where $\mathbf{Z}_{st}$ is a $1 \times (S+R_0 +R_1-1)$ row vector with a 1 in column $t$ and 0s elsewhere, $\mathbf{X} = (\underbrace{0,\cdots0}_{R_0 \ \text{times}}, \underbrace{1,\cdots1}_{R_1 \ \text{times}})^\top$ is the treatment indicator vector and $\mathbf{V}_*^{s}$ is a $(R_0 + R_1) \times (R_0 + R_1)$ matrix given by:
\[
\left[\mathbf{V}_*^{(s)}\right]_{tt'} = 
\begin{cases} 
\frac{1}{m_s}\left( 1 + (m_s -1)\alpha_0\right) , & t = t' \\[10pt]
\alpha_1 + r(\alpha_0 - \alpha_1) + \frac{1}{m_s}(\alpha_2 - \alpha_1), & t \neq t'
\end{cases}
\]
Under the AR(1) (or equivalently, exponential decay) correlation structure, the
expression \eqref{Var:blockexchangeble} for $\mathrm{Var}(\hat{\theta})$ remains
structurally the same. The only modification is in the off-diagonal elements of
$\mathbf{V}_*^{(s)}$, which become $\alpha_1 +r_{tt'}(\alpha_0-\alpha_1) + 
(\alpha_2-\alpha_1)/m_s$ for $t \neq t'$, where $r_{tt'} = \exp(-\lambda |t - t'|)$ under exponential decay and  $r_{tt'} = r^{ |t - t'|}$, reflecting the time-decaying inter-period correlation.

\section{Optimal Design}\label{Sec:Optimal.SCD}

Our goal is to allocate the $K$ available clusters across the $S$ sequences so as to
minimize $\mathrm{Var}(\hat\theta)$ in \eqref{Var:blockexchangeble}. The objective
depends on the sequence sizes only through the proportions $p_s=K_s/K$ and $K$
itself enters \eqref{Var:blockexchangeble} only as the multiplicative factor $1/K$;
the search for an optimal design therefore reduces to a search over proportions
rather than integer cluster counts. This continuous relaxation is standard in
optimal design theory. Once the optimal proportion vector is found, an integer
allocation can be recovered with an efficient rounding algorithm such as the one
in \citet{Singh_eff_2019}. We work over the design space
\begin{equation}
\label{eq:design-space}
\Xi_S
=
\left\{
\mathbf{p}=(p_1,\ldots,p_S)^\top\in\mathbb{R}^S:
 p_s\geq 0,\;
\sum_{s=1}^S p_s=1
\right\},
\qquad
p_s=\frac{K_s}{K}.
\end{equation}
For fixed sequence-specific cluster sizes
$\mathbf{m}=(m_1,\ldots,m_S)^\top$, correlation parameters
$\boldsymbol{\alpha}=(\alpha_0,\alpha_1,\alpha_2,r)^\top$ and number of sequences
$S$, we define the estimable design space as
\begin{equation}
\label{eq:estimable-design-space}
\Xi_{\mathrm{est}}
= 
\left\{
\mathbf{p}\in\Xi_S:
\mathbf{A}_{\mathbf{p}}\succ0,\;
f(\mathbf{p})>0
\right\},
\qquad
f(\mathbf{p})
=
c_{\mathbf{p}}
-
\mathbf{b}_{\mathbf{p}}^\top
\mathbf{A}_{\mathbf{p}}^{-1}
\mathbf{b}_{\mathbf{p}}.
\end{equation}
The two conditions defining $\Xi_{\mathrm{est}}$ have a direct interpretation.
The condition $\mathbf{A}_{\mathbf{p}}\succ0$ ensures that all calendar-time
effects are estimable, which requires every trial period to be covered by at least
one sequence with positive allocation. The additional condition $f(\mathbf{p})>0$
ensures that the treatment effect is separately estimable from the time effects,
which requires at least two positively weighted sequences whose observation
windows overlap. A formal characterization of these conditions in terms of the support of $\mathbf{p}$ is given in Lemma~\ref{lem:est} in the Supplementary Material. Thus, $\Xi_{\mathrm{est}}$ contains the allocation vectors for which the time
effects and the treatment effect are estimable. Whenever a minimizer exists, $\mathbf p \in \Xi_{\mathrm{est}}$, we define the optimal allocation as
\begin{equation}
\label{eq:optimal}
\mathbf{p}^{*}
=
\operatorname*{arg\,min}_{\mathbf{p}\in\Xi_{\mathrm{est}}}
\Psi(\mathbf{p},\mathbf{m},\boldsymbol{\alpha},S),
\qquad
\Psi(\mathbf{p},\mathbf{m},\boldsymbol{\alpha},S)
=
\mathrm{Var}(\hat{\theta}).
\end{equation}

\noindent
Solving \eqref{eq:optimal} is complicated by the fact that $\Psi$ depends on
$\mathbf{p}$ nonlinearly through the matrix inverse
$\mathbf{A}_{\mathbf{p}}^{-1}$ in \eqref{Var:blockexchangeble}. However we are able to prove that for fixed $\{\mathbf{m}, \bm \alpha, S\}$, the variance
$\mathrm{Var}(\hat{\theta})$ is a convex function of the allocation vector
$\mathbf{p}$ over $\Xi_{\mathrm{est}}$, which implies that any local minimizer of
\eqref{eq:optimal} is a global minimizer. It does not, in general, imply uniqueness of the optimal allocation. A separate necessary and sufficient condition for verifying whether a particular allocation is globally optimal is
given by the following equivalence theorem.

\begin{theorem}[Equivalence Theorem]
\label{thm:equiv}
Let $\mathbf{p}^*\in\Xi_{\mathrm{est}}$. Then $\mathbf{p}^*$ minimizes
$\mathrm{Var}(\hat{\theta})$ over $\Xi_{\mathrm{est}}$ if and only if, for every
sequence $l=1,\ldots,S$,
\begin{equation}
\label{eq:eqv}
\delta_l(\mathbf{p}^*)
=
\left(
\mathbf{X}-\mathbf{Z}_l\mathbf{x}^*
\right)^\top
\left(\mathbf{V}_*^{(l)}\right)^{-1}
\left(
\mathbf{X}-\mathbf{Z}_l\mathbf{x}^*
\right)
\leq
f(\mathbf{p}^*),
\end{equation}
where $\mathbf{x}^*=\mathbf{A}_{\mathbf{p}^*}^{-1}\mathbf{b}_{\mathbf{p}^*}$ and
$f(\mathbf{p}^*)=c_{\mathbf{p}^*}-\mathbf{b}_{\mathbf{p}^*}^{\top}\mathbf{A}_{\mathbf{p}^*}^{-1}\mathbf{b}_{\mathbf{p}^*}$.
\end{theorem}

\begin{figure}[H]
    \centering
    \includegraphics[width=0.7\linewidth]{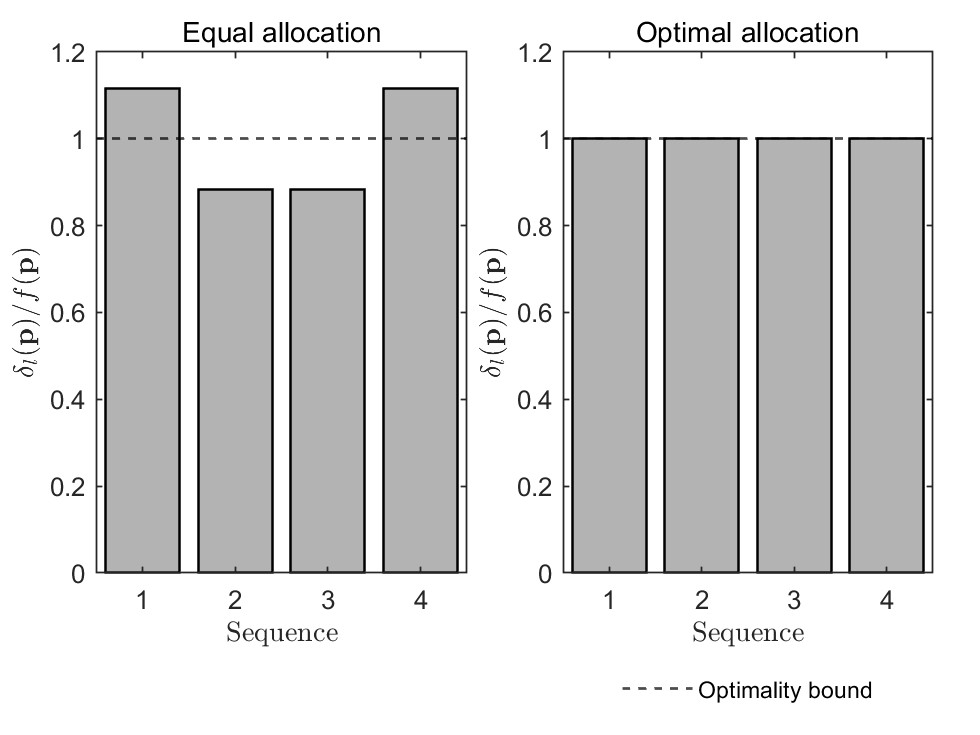}
    \caption{Application of Equivalence theorem for a basic $S=4$ SCD}
    \label{fig:equivalence_check}
\end{figure}

\noindent
Theorem~\ref{thm:equiv} gives a necessary and sufficient condition for optimality
that involves only quantities computed sequence by sequence: each sequence's contribution $\delta_l(\mathbf{p}^*)$ must not exceed the bound $f(\mathbf{p}^*)$. This is the
SCD analogue of the classical equivalence theorem. For any candidate allocation $\mathbf p\in\Xi_{\mathrm{est}}$,
define $h_l(\mathbf p) = \frac{\delta_l(\mathbf p)}{f(\mathbf p)},\ l=1,\ldots,S$. Then $\mathbf p$ is globally optimal if and only if $\max_{1\leq l\leq S} h_l(\mathbf p)\leq1$. Moreover, $\sum_{l=1}^S p_l\delta_l(\mathbf p)=f(\mathbf p)$, so that, at an optimal design, every sequence receiving positive allocation
must satisfy $\delta_l(\mathbf p^*)=f(\mathbf p^*)$ whereas sequences receiving zero allocation may satisfy the inequality strictly. Figure~\ref{fig:equivalence_check} illustrates the practical use of
Theorem~\ref{thm:equiv}. We consider a basic SCD with 4 sequences. Under equal allocation, the value of $h_1(\bm p)$ and $h_4(\bm p)$ exceed one, thereby violating the equivalence
condition. At the analytically optimal allocation (as derived in Section~\ref{S=4}), $h_l(\bm p)=1 \ \forall \ l$, providing an independent optimality check of the closed-form solution.

\noindent
The preceding theorem concerns optimization of the cluster allocation $\mathbf p$ for fixed sequence specific cluster sizes. We now
consider the broader problem in which both the allocation proportions and the cluster sizes are allowed to vary. Let
$\mathbf m=(m_1,\ldots,m_S)^\top$ denote the sequence-specific cluster-size
vector and fix the average cluster size $\bar m>0$. Define
\[
\Omega_{\bar m}
=
\left\{
(\mathbf p,\mathbf m)\in
\Delta_S\times\mathbb R_{>0}^{S}:
\sum_{s=1}^{S}p_sm_s=\bar m
\right\},
\qquad
\Delta_S
=
\left\{
\mathbf p\in\mathbb R_{>0}^{S}:
\sum_{s=1}^{S}p_s=1
\right\}.
\]
The MinMin Design (MMD) jointly minimizes the variance of the
treatment-effect estimator over $\Omega_{\bar m}$. Whenever a minimizer
exists, we denote a joint optimal design by
\begin{equation}
\label{eq:MMD}
(\mathbf p^{*},\mathbf m^{*})
\in
\operatorname*{arg\,min}_{(\mathbf p,\mathbf m)\in\Omega_{\bar m}}
\operatorname{Var}(\hat\theta).
\end{equation}
The following result characterizes the symmetry of a unique MMD solution
for balanced staircase designs.

\begin{theorem}[Symmetry of the MMD]
\label{thm:MMD_symmetry}
Consider a balanced closed-cohort SCD with $R_0=R_1=R$, $S$ sequences,
and $T=S+2R-1$ total periods. Suppose that $\theta$ is estimable and that
the covariance matrix for a cluster assigned to sequence $s$ depends on
$s$ only through its cluster size, so that $\mathbf V_*^{(s)}=\mathbf V_*(m_s)$. Assume further that, for every admissible cluster size $m$,
$\mathbf V_*(m)$ is persymmetric, i.e., $\mathbf J_{2R}\mathbf V_*(m)^{-1}\mathbf J_{2R}=\mathbf V_*(m)^{-1}$, where $\mathbf J_{2R}$ denotes the $2R\times2R$ anti-identity matrix. Suppose that
$(\mathbf p^{*},\mathbf m^{*})$ is the unique minimizer of $\operatorname{Var}(\hat\theta)$ over $\Omega_{\bar m}$. Then
$p_s^{*}=p_{S-s+1}^{*}$ and $m_s^{*}=m_{S-s+1}^{*}$ for all $s=1,\ldots,S$.
\end{theorem}

\noindent Theorem~\ref{thm:MMD_symmetry} shows that, whenever the joint MMD solution is
unique, both components of the optimal design are symmetric about the middle
sequence. Thus, reversing the order of the sequences leaves the optimal design
unchanged, so that the optimal allocation proportions and cluster sizes take the
palindromic forms
$\mathbf p^*=(p_1^*,\ldots,p_S^*)^\top$ with
$p_s^*=p_{S-s+1}^*$ and
$\mathbf m^*=(m_1^*,\ldots,m_S^*)^\top$ with
$m_s^*=m_{S-s+1}^*$. Consequently, the number of free parameters in the joint
optimization is substantially reduced. Without symmetry, the constraints
$\sum_{s=1}^S p_s=1$ and $\sum_{s=1}^S p_sm_s=\bar m$ leave $(2S-2)$ free
parameters, whereas under Theorem~\ref{thm:MMD_symmetry} only
$2\lceil S/2\rceil-2$ free parameters remain. A particularly important special case arises when the cluster sizes are fixed and equal, $m_s=m$ for all $s$. For a balanced SCD, define the reversed allocation by $\mathbf p^{R}=(p_S,p_{S-1},\ldots,p_1)^\top$. It can be shown that under a persymmetric covariance structure, $\operatorname{Var}(\hat\theta;\mathbf p^{R})= \operatorname{Var}(\hat\theta;\mathbf p)$. Since the variance function is convex in $\mathbf p$, the symmetrized allocation $\mathbf p^{\mathrm{sym}}=\frac{\mathbf p+\mathbf p^{R}}{2}$ satisfies $\operatorname{Var}(\hat\theta;\mathbf p^{\mathrm{sym}})
\leq
\frac{
\operatorname{Var}(\hat\theta;\mathbf p)
+
\operatorname{Var}(\hat\theta;\mathbf p^{R})
}{2}
=
\operatorname{Var}(\hat\theta;\mathbf p)$.
Hence, whenever a global minimizer exists, there is at least one symmetric
global minimizer. If the minimizer is unique, it must itself be symmetric.
A detailed proof is provided in the Supplementary Material. This result allows the equal-cluster-size optimization to be restricted to
symmetric allocations without assuming uniqueness. For $S=3$ we may write $\mathbf p^*=(b,1-2b,b)^\top$, while for $S=4$ we may write $\mathbf p^*=(b,0.5-b,0.5-b,b)^\top$. These one-dimensional parametrizations are used below to obtain the corresponding closed-form optimal allocations.

\subsection{Exact optimal allocations for equal cluster sizes}

\noindent
Throughout this subsection we specialize to equal cluster sizes, $m_s=m$ for all
$s=1,\ldots,S$. Under this restriction, only the allocation vector $\mathbf p$ remains to be optimised. By the symmetry result established
above, the optimal allocation is symmetric about the middle of the design, which reduces the optimization to a small number of free parameters and, for $S=3$ and $S=4$, permits explicit closed-form solutions. Under the block-exchangeable correlation structure, write
\begin{equation}\label{eq:psi-def}
\psi=\frac{c}{v},
\qquad
v=\frac{1}{m}\{1+(m-1)\alpha_0\},
\qquad
c=\alpha_1+r(\alpha_0-\alpha_1)
+\frac{1}{m}(\alpha_2-\alpha_1).
\end{equation}
Here $v$ and $c$ denote, respectively, the variance of a cluster-period
mean and the covariance between two cluster-period means from the same
cluster. Hence $\psi=c/v$ represents the correlation between two cluster-period means from the same cluster, under equal cluster-size. For every admissible covariance configuration considered here, $0<\psi<1$.

\subsubsection{Three-sequence designs}

\noindent For $S=3$, symmetry gives
$\mathbf p=(b,1-2b,b)^\top$, so that the original allocation problem reduces
to a one-dimensional optimization in the outer-sequence allocation $b$. The form of the resulting first-order condition depends on the observation window $R$.
\paragraph{Case $R=1$.}
For the basic SCD, $R_0=R_1=1$, the first-order condition is quadratic and yields
\begin{equation}\label{eq:opt3Basic}
p_1^*=p_3^*
=
\frac{2-\sqrt{2(1-\psi)}}{2(1+\psi)},
\qquad
p_2^*
=
\frac{\sqrt{2(1-\psi)}-(1-\psi)}{1+\psi}.
\end{equation}
\noindent The allocation between the inner and outer sequences is determined entirely
by $\psi$. In particular, $p_2^*\ge p_1^*$ if and only if $\psi\le 1/2$. Thus, when the correlation between cluster-period means is
relatively weak, more clusters are allocated to the middle sequence, whereas the outer sequences receive progressively greater weight as $\psi$ increases. This parameter $\psi$ also generalizes a quantity already in the SCD literature. Setting $\sigma_\delta^2=\sigma_u^2=0$ collapses $\psi$ to
$\psi \to \dfrac{mr\rho_0}{1+(m-1)\rho_0}$, which is precisely the correlation parameter used by \citet{grantham2024staircase} for the cross-sectional SCD.

\paragraph{Case $R\ge2$.}
When each sequence is observed for more than one period before and after crossover, the reduced first-order condition becomes cubic rather than
quadratic. It is convenient to define
\begin{equation}\label{eq:nu-def}
\nu=-\frac{\psi}{1+(2R-1)\psi},
\qquad
L=1+\nu.
\end{equation}
The optimal outer-sequence allocation is the admissible real solution
\begin{equation}\label{eq:opt3general}
p_1^*=p_3^*
=
\frac{5L-C_3-L^2/C_3}{6},
\end{equation}
where $C_3=
\sqrt[3]{\dfrac{\Delta_{1,3}+
\sqrt{\Delta_{1,3}^2-4\Delta_{0,3}^3}}{2}}$, $\Delta_{0,3}=L^2$ and $\Delta_{1,3}=110L^3-216L^2+108L$.
Consequently, $\mathbf p^*=
\bigl(p_1^*,\,1-2p_1^*,\,p_1^*\bigr)^\top.$

\subsubsection{Four-sequence designs}
\label{S=4}

\noindent For $S=4$, symmetry gives $\mathbf p=(b,0.5-b,0.5-b,b)^\top$. Hence the allocation problem again reduces to a single outer-sequence weight $b$. Unlike the three-sequence case, three distinct forms arise according
to the observation window: $R=1$, $R=2$, and $R\ge3$.

\paragraph{Case $R=1$.}
For the basic four-sequence SCD, the optimal allocation is
\begin{equation}\label{eq:opt4Basic}
p_1^*=p_4^*
=
\frac{2-(2+\psi)\sqrt{1-\psi}}{4\psi^2},
\qquad
p_2^*=p_3^*
=
\frac{(2+\psi)\sqrt{1-\psi}-2(1-\psi^2)}
{4\psi^2}.
\end{equation}
\noindent
The middle sequences receive at least as much weight as the outer sequences when $\psi\le\frac{\sqrt5-1}{2}\approx0.618$. This threshold is larger than the corresponding value $1/2$ for $S=3$, indicating that in the basic design, the preference for the middle
sequences persists over a wider range of correlation values when a fourth sequence is introduced.

\begin{figure}[H]
    \centering
 \includegraphics[width=1.0\linewidth]{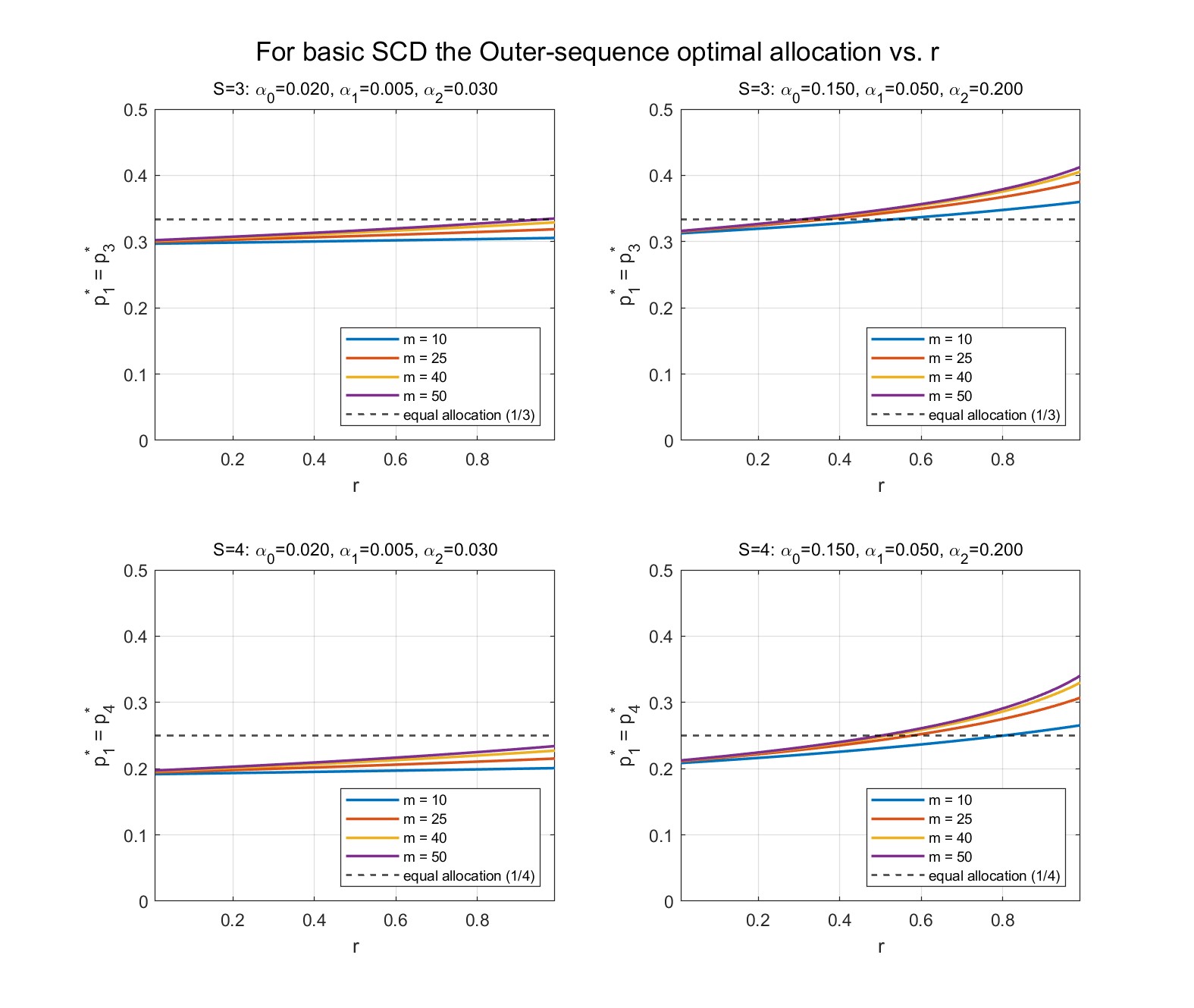}
    \caption{Optimal allocation of basic SCD for $S=3$ and $S=4$ over varying $r$ and cluster size $m$.}
    \label{Fig:S3S4optimal}
\end{figure}
\noindent Note that $\psi$ is a function of $\{\bm \alpha,r,m \}$. Figure~\ref{Fig:S3S4optimal} illustrates how the optimal outer-sequence allocation varies with the exchangeable correlation parameter $r$ for the basic $S=3$ and $S=4$ designs. The optimal allocation for outer sequences increases monotonically as $r$ grows. Furthermore, larger cluster sizes consistently demand a higher allocation to these outer sequences, an effect that amplifies significantly at higher correlations. Finally, standard equal allocation is rarely optimal as the optimal design under-allocates to outer sequences in low-correlation setting, but over-allocates to them in high-correlation setting and exchangeable correlation is moderate to high.

\paragraph{Case $R=2$.}
For $R=2$, define $\eta=-\dfrac{\psi}{1+2\psi}$. Note that $-\frac13<\eta<0$.

\noindent Under the symmetric parametrization
$\mathbf p=(0.5-b,b,b,0.5-b)^\top$, the first-order condition reduces to
the cubic equation
\begin{equation}\label{eq:opt4R2_cubic}
B_3b^3+B_2b^2+B_1b+B_0=0,
\end{equation}
where $B_3=128\eta^4-64\eta^3+64\eta^2, \ B_2=-176\eta^4+120\eta^3+8\eta^2+8\eta+8,\ B_1=80\eta^4-72\eta^3-40\eta^2+24\eta+8,\ B_0=-12\eta^4+15\eta^3+5\eta^2-6\eta-2.$

\noindent For every admissible value of $\eta$, \eqref{eq:opt4R2_cubic} has exactly
one root $b^*\in(0,0.5)$; hence this root determines the optimal allocation.
A proof of existence, uniqueness, and global optimality is given in the
Supplementary Material. To express $b^*$ explicitly, define
\[
\Delta_{0,4}^{(2)}
=
B_2^2-3B_3B_1,
\quad
\Delta_{1,4}^{(2)}
=
2B_2^3-9B_3B_2B_1+27B_3^2B_0, \quad \mathcal D_{4,2}
=
\left(\Delta_{1,4}^{(2)}\right)^2
-
4\left(\Delta_{0,4}^{(2)}\right)^3.
\]

\noindent The sign of $\mathcal D_{4,2}$ determines the number of real roots of the
cubic, but not the uniqueness of the admissible root.

\noindent If $\mathcal D_{4,2}>0$, the cubic has only one real root, which necessarily
lies in $(0,0.5)$. In this case,
\begin{equation}\label{eq:opt4R2_cardano}
b^*
=
-\frac{1}{3B_3}
\left(
B_2+C_{4,2}
+\frac{\Delta_{0,4}^{(2)}}{C_{4,2}}
\right) \ \ \text{where,} \ C_{4,2}
=\sqrt[3]{\dfrac{\Delta_{1,4}^{(2)}+ \sqrt{\mathcal D_{4,2}}}{2}}
\end{equation}

\noindent If $\mathcal D_{4,2}\leq0$, the cubic has three real roots, but exactly one
belongs to $(0,0.5)$. In this case, the admissible root is
\begin{equation}\label{eq:opt4R2_trig}
b^*
=
-\frac{1}{3B_3}
\left[
B_2+
2\sqrt{\Delta_{0,4}^{(2)}}
\cos\left(
\vartheta_{4,2}+\frac{2\pi}{3}
\right)
\right] \  \ \text{where,} \ \vartheta_{4,2}
=
\frac13
\arccos\left\{
\frac{\Delta_{1,4}^{(2)}}
{2\left(\Delta_{0,4}^{(2)}\right)^{3/2}}
\right\}.
\end{equation}

\noindent Hence the exact optimal
allocation is $\mathbf p^* =
(0.5-b^*,\,b^*,\,b^*,\,0.5-b^*)^\top$

\paragraph{Case $R\ge3$.}
For $R\ge3$, the first-order condition is cubic. Using the transformation
in \eqref{eq:nu-def}, define $A_0=-8L^4+L^3+2L^2+L, \ A_1=48L^4-56L^3+24L^2, \ A_2=-96L^4+212L^3-204L^2+92L-24, \ A_3=64L^4-208L^3+296L^2-208L+64$. Then let, $\Delta_{0,4}=A_2^2-3A_3A_1, \ \Delta_{1,4}=
2A_2^3-9A_3A_2A_1+27A_3^2A_0$ and $C_4
=
\sqrt[3]{
\dfrac{
\Delta_{1,4}
+
\sqrt{\Delta_{1,4}^2-4\Delta_{0,4}^3}
}{2}
}$.
The optimal outer-sequence allocation is the admissible real root
\begin{equation}\label{eq:opt4general}
p_1^*=p_4^*
=
-\frac{1}{3A_3}
\left(
A_2+C_4+\frac{\Delta_{0,4}}{C_4}
\right),
\end{equation}
with $p_2^*=p_3^*=0.5-p_1^*$.

\begin{figure}[H]
    \centering
    
    % First row (S=3)
    \begin{subfigure}{\textwidth}
        \centering
        % Limits height to 42% of page, scales width proportionally
        \includegraphics[width=\textwidth, height=0.42\textheight, keepaspectratio]{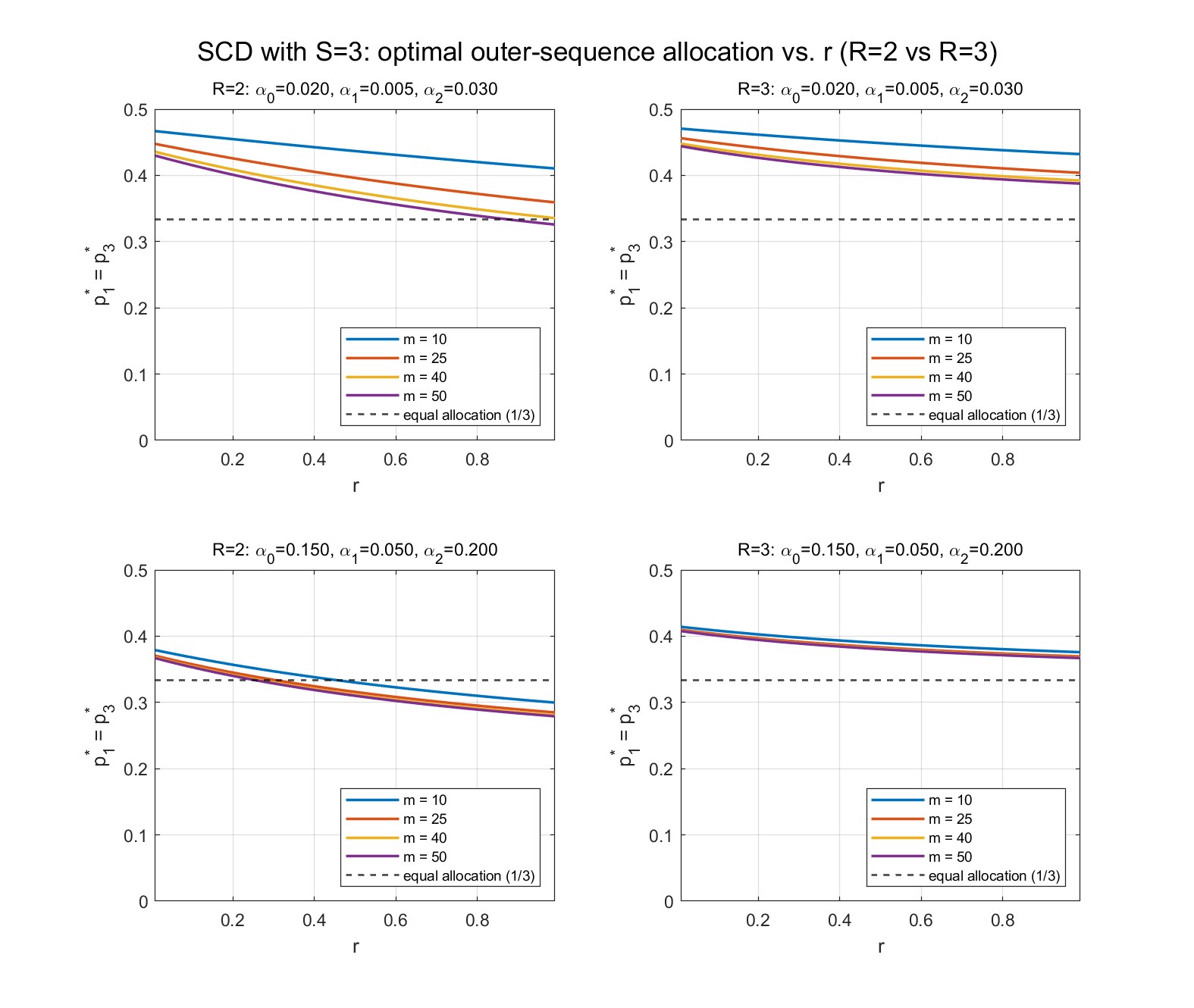}
        \caption{Three-sequence extended SCDs: $R=2$ and $R=3$.}
    \end{subfigure}
    
    \vspace{0.3cm} % Reduced from 0.5cm to save vertical space
    
    % Second row (S=4)
    \begin{subfigure}{\textwidth}
        \centering
        % Limits height to 42% of page, scales width proportionally
        \includegraphics[width=\textwidth, height=0.42\textheight, keepaspectratio]{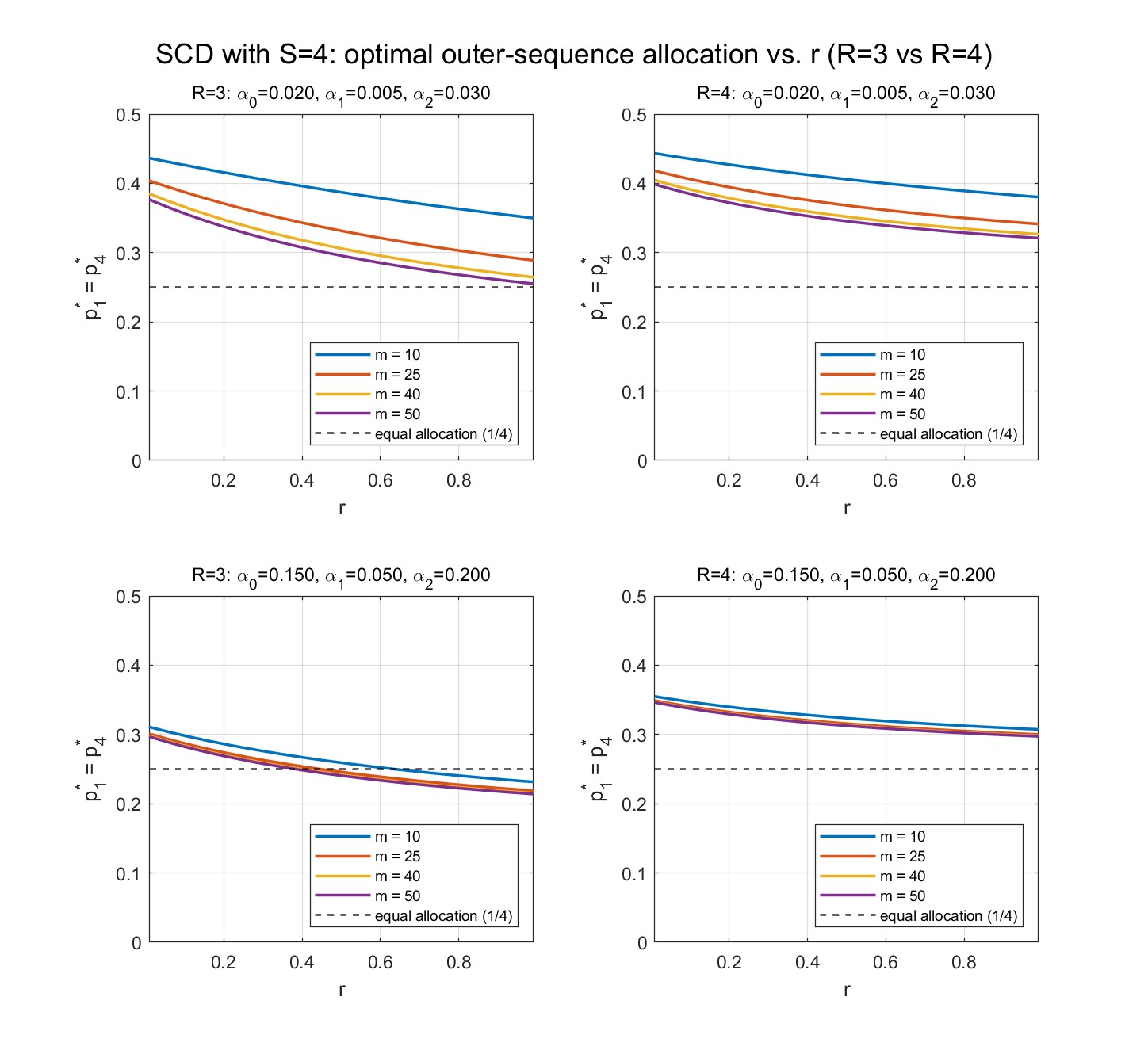}
        \caption{Four-sequence extended SCDs: $R=3$ and $R=4$.}
    \end{subfigure}
    
    \caption{Optimal outer-sequence allocation vs. exchangeable correlation ($r$) for extended SCDs.}
    \label{Fig:S3-S4optimal_generalR}
\end{figure}

\noindent Figure~\ref{Fig:S3-S4optimal_generalR} shows that the allocation pattern observed for the basic SCD does not generally carry over to wider observation windows. For the basic SCD with $R=1$, the optimal allocation to the outer sequences generally increases with the induced correlation between cluster-period means, $\psi$ and hence with $r$ when the remaining covariance parameters and cluster size are fixed. In contrast, for the extended designs considered in Figure~\ref{Fig:S3-S4optimal_generalR}, the optimal allocation in the outer sequences generally decreases as $r$ increases. For $S=3$, the decrease is particularly pronounced when $R=2$. A similar pattern is also observed for $S=4$. These numerical patterns are consistent with the exact analytical expressions derived above for $S=3$ and $S=4$. They also show that the effect of the correlation structure on the optimal allocation depends on the values of $R$.

\section{Numerical Analysis}

\subsection{Comparison between SWD and SCD}
\subsubsection{A cost analysis under block exchangeable correlation structure}

\noindent Traditionally, SWDs appear statistically superior because they observe every participating cluster continuously throughout the entire trial. However, in practice, this extended continuous observation comes at a high logistical and financial cost. To make a truly fair comparison between an SWD and a SCD, we cannot just look at the absolute variance. We must account for the data burden. For this reason we conduct a cost analysis between SWD and SCD.
Here first we fix the target statistical precision derived from the optimal SCD. Then we determine exactly how many clusters a standard Stepped Wedge Design requires to match it. To quantify the financial requirements of both designs, we use the budget framework from \cite{Liu_2024_SMMR}. Let $c_c$ denote the fixed cost of recruiting and setting up a cluster, $c_s$ the cost of enrolling a participant and $c_e$ the recurring cost per outcome measurement. In a standard Stepped Wedge Design with $K_{SWD}$ clusters and uniform cluster size $m$, every participant is measured across all $T$ periods, giving total budget $B_{SWD} = K_{SWD} \bigl( c_c + c_s m + c_e T m \bigr)$. The balanced Staircase Design $R_0 = R_1 = R$ restricts data collection to the crossover window spanning $2R$ periods rather than the entire trial duration $T$. This gives $B_{SCD} = K_{SCD} \bigl( c_c + c_s m + c_e (2R) m \bigr)$. The SCD restricts measurements to a narrow crossover window $(2R)$, which cuts longitudinal measurement costs. The SWD requires continuous observation throughout $T$, so it extracts more information per cluster and needs fewer clusters to reach target power. We map this trade-off to find when the SCD saves money and when it costs more than SWD.

To evaluate the financial trade-offs fairly, we force both designs to achieve identical statistical precision. We consider $T \in \{ 9,15\}$, cluster size to 20 participants and fix the SCD at 30 clusters. For any observation window and correlation scenario, we optimise the SCD to find its lowest variance—this becomes our target precision. We then optimise the SWD and calculate how many SWD clusters match that target. With cluster counts for both designs, we calculate total budget using $c_c = 3000$, $c_s = 250$ and $c_e = 175$. Finally we calculate the percentage difference in cost defined as $\Delta_B = \frac{B_{SWD}-B_{SCD}}{B_{SWD}}\times100$ for each setting. 

Tables \ref{tab:cost_analysis1} and \ref{tab:cost_analysis2} show that the SCD's cost advantage over the SWD depends on trial duration ($T$), observation window ($R$) and the underlying correlation structure. At 
$T=9$ with $R=1$, the SCD cuts total costs by roughly $11 \ \text{to}\ 24\%$. Expanding the window to $R=2$ under strong correlation reverses this: the SWD becomes up to 41\% cheaper by efficiently managing the secular time trend with its continuous data collection. But if we extend the trial duration to $T=15$, the scenarios change completely. At
$T=15$, the SCD is cheaper across all evaluated scenarios with savings reaching $33.11\%$. This change is primarily driven by the SWD's requirement for repeated measurements over time, which becomes financially demanding when the $T$ is large. Although the SWD achieves the target precision with substantially fewer clusters (requiring as few as $9$ clusters compared with $30$ for the SCD), the cost of collecting data from all participants over $15$ consecutive periods exceeds the cost of recruiting additional clusters in the SCD. This shift reflects a trade-off between trend estimation and measurement burden. When the trial duration is short ($T=9$), a narrow window ($R=1$) provides adequate overlap between sequences and requires fewer measurements. However, for a longer trial ($T=15$), a wider window ($R=2$) offers more longitudinal data to support estimation of the secular trend, while remaining less costly than the continuous measurements required in the SWD.

\begin{table}[ht]
\centering
\caption{Cost analysis and required SWD clusters to achieve identical statistical precision to the Staircase Design ($K_{SCD} = 30$, $m = 20$) for $T=9$. Total budget assumptions: Cluster setup ($c_c$)= 3,000, Enrollment ($c_s$)= 250, Measurement ($c_e$)= 175.}
\label{tab:cost_analysis1}
\resizebox{\textwidth}{!}{% Resizes table to fit page width if it gets too large
\begin{tabular}{lc ccc ccc}
\toprule
\multirow{2}{*}{\textbf{Scenario ($\alpha_0, \alpha_1, \alpha_2$)}} & \multirow{2}{*}{\textbf{$r$}} & \multicolumn{3}{c}{\textbf{$R = 1$}} & \multicolumn{3}{c}{\textbf{$R = 2$}} \\
\cmidrule(lr){3-5} \cmidrule(lr){6-8}
& & Target Var & $K_{SWD}$ & $\Delta_B$ & Target Var & $K_{SWD}$ & $\Delta_B$ \\
\midrule

% SCENARIO 1
\multirow{4}{*}{1. (0.05, 0.01, 0.05)} 
& 0.1 & 0.006093 & 13.09 & 12.98 & 0.004306 & 18.52 & 9.79 \\
& 0.3 & 0.005593 & 14.18 & 19.65 & 0.004203 & 18.87 & 11.44 \\
& 0.5 & 0.005089 & 14.79 & 22.95 & 0.004068 & 18.50 & 9.66 \\
& 0.7 & 0.004581 & 15.06 & 24.36 & 0.003896 & 17.71 & 5.65 \\
\midrule

% SCENARIO 2
\multirow{4}{*}{2. (0.10, 0.05, 0.10)} 
& 0.1 & 0.006939 & 15.04 & 24.23 & 0.005841 & 17.86 & 6.46 \\
& 0.3 & 0.006300 & 15.10 & 24.55 & 0.005586 & 17.03 & 1.87 \\
& 0.5 & 0.005653 & 14.99 & 24.01 & 0.005280 & 16.05 & -4.09 \\
& 0.7 & 0.004999 & 14.71 & 22.57 & 0.004911 & 14.98 & -11.56 \\
\midrule

% SCENARIO 3
\multirow{4}{*}{3. (0.15, 0.05, 0.30)} 
& 0.1 & 0.009457 & 14.98 & 23.96 & 0.007841 & 18.07 & 7.53 \\
& 0.3 & 0.008177 & 15.09 & 24.51 & 0.007334 & 16.83 & 0.69 \\
& 0.5 & 0.006878 & 14.81 & 23.07 & 0.006664 & 15.28 & -9.33 \\
& 0.7 & 0.005548 & 14.09  & 19.16 & 0.005771 & 13.55 & -23.33 \\
\midrule

% SCENARIO 4
\multirow{4}{*}{4. (0.25, 0.10, 0.40)} 
& 0.1 & 0.012737 & 15.10 & 24.56 & 0.011182 & 17.20 & 2.86 \\
& 0.3 & 0.010797 & 14.92 & 23.66 & 0.010250 & 15.72 & -6.30 \\
& 0.5 & 0.008816 & 14.33 & 20.51 & 0.009011 & 14.02 & -19.16 \\
& 0.7 & 0.006771 & 13.17 & 13.49 & 0.007326 & 12.17 & -37.28 \\
\midrule

% SCENARIO 5
\multirow{4}{*}{5. (0.30, 0.15, 0.30)} 
& 0.1 & 0.013486 & 15.04 &  24.25 & 0.012413 & 16.34 & -2.25 \\
& 0.3 & 0.011527 & 14.71 & 22.54 & 0.011332 & 14.96 & -11.68 \\
& 0.5 & 0.009525 & 14.04 & 18.84 & 0.009942  & 13.45 &  -24.24 \\
& 0.7 & 0.007459 & 12.88  & 11.53 & 0.008117 & 11.83 & -41.20 \\
% \midrule

\bottomrule
\end{tabular}
}
\end{table}

\begin{table}[ht]
\centering
\caption{Cost analysis and required SWD clusters to achieve identical statistical precision to the Staircase Design ($K_{SCD} = 30$, $m = 20$) for $T=15$. Total budget assumptions: Cluster setup ($c_c$)=3,000, Enrollment ($c_s$)=250, Measurement ($c_e$)=175.}
\label{tab:cost_analysis2}
\begin{tabular}{llcccccc}
\toprule
\multirow{2}{*}{\textbf{Scenario} ($\alpha_0, \alpha_1, \alpha_2$)} & \multirow{2}{*}{$r$} & \multicolumn{3}{c}{$R = 1$} & \multicolumn{3}{c}{$R = 2$} \\
\cmidrule(lr){3-5} \cmidrule(lr){6-8}
 & & Target Var & $K_{SWD}$ & $\Delta_B$ & Target Var & $K_{SWD}$ & $\Delta_B$ \\
\midrule
\multirow{4}{*}{1. (0.05, 0.01, 0.05)} & 0.1 & 0.005692 & 8.96 & 16.98 & 0.003287 & 15.51 & 29.68 \\
 & 0.3 & 0.005173 & 9.67 & 23.11 & 0.003085 & 16.22 & 32.74 \\
 & 0.5 & 0.004652 & 10.07 & 26.15 & 0.002873 & 16.31 & 33.11 \\
 & 0.7 & 0.004130 & 10.29 & 27.70 & 0.002651 & 16.03 & 31.94 \\
\midrule
\multirow{4}{*}{2. (0.10, 0.05, 0.10)} & 0.1 & 0.006272 & 10.26 & 27.51 & 0.003999 & 16.10 & 32.23 \\
 & 0.3 & 0.005618 & 10.36 & 28.20 & 0.003709 & 15.69 & 30.48 \\
 & 0.5 & 0.004961 & 10.38 & 28.36 & 0.003405 & 15.13 & 27.89 \\
 & 0.7 & 0.004302 & 10.33 & 28.00 & 0.003083 & 14.41 & 24.30 \\
\midrule
\multirow{4}{*}{3. (0.15, 0.05, 0.30)} & 0.1 & 0.008578 & 10.22 & 27.20 & 0.005416 & 16.18 & 32.57 \\
 & 0.3 & 0.007269 & 10.37 & 28.28 & 0.004838 & 15.58 & 29.99 \\
 & 0.5 & 0.005953 & 10.35 & 28.15 & 0.004214 & 14.62 & 25.41 \\
 & 0.7 & 0.004626 & 10.15 & 26.71 & 0.003528 & 13.31 & 18.02 \\
\midrule
\multirow{4}{*}{4. (0.25, 0.10, 0.40)} & 0.1 & 0.011391 & 10.35 & 28.12 & 0.007467 & 15.78 & 30.88 \\
 & 0.3 & 0.009420 & 10.37 & 28.30 & 0.006551 & 14.92 & 26.86 \\
 & 0.5 & 0.007434 & 10.22 & 27.25 & 0.005551 & 13.69 & 20.34 \\
 & 0.7 & 0.005423 & 9.83 & 24.31 & 0.004423 & 12.05 & 9.47 \\
\midrule
\multirow{4}{*}{5. (0.30, 0.15, 0.30)} & 0.1 & 0.011893 & 10.38 & 28.36 & 0.008069 & 15.30 & 28.71 \\
 & 0.3 & 0.009916 & 10.33 & 27.99 & 0.007112 & 14.40 & 24.24 \\
 & 0.5 & 0.007922 & 10.13 & 26.58 & 0.006069 & 13.22 & 17.50 \\
 & 0.7 & 0.005902 & 9.72 & 23.45 & 0.004900 & 11.70 & 6.79 \\
\bottomrule
\end{tabular}
\end{table}

\vspace{4cm}

\subsubsection{A power analysis under decay correlation structure}
\noindent To evaluate the performance of the SCD relative to SWD under an autoregressive correlation decay model, we conduct a power analysis based on the Wald test statistic. In large samples, the Wald statistic can be utilized to test the significance of the treatment effect. We test the null hypothesis $H_0:\theta = 0$ against the alternative hypothesis $H_1:\theta = \theta_1 (\neq 0)$. The Wald statistic is defined as: $Z = \dfrac{\widehat{\theta}}{\sqrt{\text{Var}(\widehat{\theta})}}$. Under the large sample approximation, the power function of the above test is given by:$$\pi = 1 - \mathrm{\Phi}\left(z_{\alpha/2} - \frac{\theta_1}{\sqrt{\text{Var}(\widehat{\theta})}}\right) + \mathrm{\Phi}\left(-z_{\alpha/2} - \frac{\theta_1}{\sqrt{\text{Var}(\widehat{\theta})}}\right)$$ where $\mathrm{\Phi}()$ denotes the cumulative distribution function of the standard normal distribution and $z_{\gamma}$ represents the $(1-\gamma)$ quantile of the standard normal distribution. To quantify the comparison between these two designs, the relative efficiency of power ($RE^{\pi}$) is calculated as: $RE^{\pi} = 100\left(\dfrac{\pi_{\rm SCD} - \pi_{\rm SWD}}{\pi_{\rm SWD}}\right)$. This numerical analysis incorporates three distinct scenarios for the correlation parameter vector $\boldsymbol{\alpha} = (\alpha_0, \alpha_1, \alpha_2)$: low $(0.05, 0.02, 0.20)$, moderate $(0.20, 0.10, 0.40)$ and high $(0.35, 0.20, 0.60)$. Within each configuration, the autoregressive correlation decay parameter is assessed at two levels, representing rapid decay ($r = 0.4$) and slow decay ($r = 0.8$). Finally, the comprehensive power analysis is computed across two different treatment effect sizes: $\theta_1 = 0.35$ and $\theta_1 = 0.50$. 

\begin{table}[ht]
    \centering
    \caption{Relative Efficiency of Power ($RE^{\pi}$) and Participant Reduction ($\Delta_N$) between SCD and SWD (assuming equal cluster size $m=10$) across different scenarios of $\boldsymbol{\alpha}=(\alpha_0, \alpha_1, \alpha_2)$ and effect sizes $\theta_1$.}
    \label{tab:power_combined}
    \begin{threeparttable}   
        \setlength{\tabcolsep}{4pt} 
        \begin{tabular}{cccc cc cc cc}
        \toprule
        \multicolumn{4}{c}{$\boldsymbol{\alpha}\to$} 
         & \multicolumn{2}{c}{$(0.05, 0.02, 0.20)$} 
         & \multicolumn{2}{c}{$(0.20, 0.10, 0.40)$} 
         & \multicolumn{2}{c}{$(0.35, 0.20, 0.60)$} \\
        \cmidrule(lr){5-6} \cmidrule(lr){7-8} \cmidrule(lr){9-10}
        
        $T$ & $R$ & $\Delta_N$ & $\theta_1$
        & $RE^\pi_{r=0.8}$ & $RE^\pi_{r=0.4}$ 
        & $RE^\pi_{r=0.8}$ & $RE^\pi_{r=0.4}$ 
        & $RE^\pi_{r=0.8}$ & $RE^\pi_{r=0.4}$ \\
        \midrule
        \multirow{4}{*}{7} 
        & \multirow{2}{*}{1} & \multirow{2}{*}{71.43} 
        & 0.35 & $-1.57$ & $-2.47$ & $-2.70$ & $-6.35$ & $-3.37$ & $-9.90$ \\
        & & & 0.50 & $-0.01$ & $-0.02$ & $-0.02$ & $-0.23$ & $-0.03$ & $-0.72$ \\
        \addlinespace
        & \multirow{2}{*}{2} & \multirow{2}{*}{42.86} 
        & 0.35 & $-2.46$ & $-2.86$ & $-4.37$ & $-10.76$ & $-1.60$ & $-13.56$ \\
        & & & 0.50 & $-0.01$ & $-0.02$ & $-0.06$ & $-0.65$ & $-0.01$ & $-1.30$ \\
        \midrule
        \multirow{4}{*}{12} 
        & \multirow{2}{*}{1} & \multirow{2}{*}{83.33} 
        & 0.35 & $-0.65$ & $-1.28$ & $-0.51$ & $-3.41$ & $-0.28$ & $-4.77$ \\
        & & & 0.50 & $-0.00$ & $-0.00$ & $-0.00$ & $-0.03$ & $-0.00$ & $-0.09$ \\
        \addlinespace
        & \multirow{2}{*}{2} & \multirow{2}{*}{66.67} 
        & 0.35 & $-0.07$ & $-0.15$ & $-0.41$ & $-2.02$ & $-0.27$ & $-4.24$ \\
        & & & 0.50 & $-0.00$ & $-0.00$ & $-0.00$ & $-0.01$ & $-0.00$ & $-0.07$ \\
        \midrule
        \multirow{4}{*}{17} 
        & \multirow{2}{*}{1} & \multirow{2}{*}{88.24} 
        & 0.35 & $-0.46$ & $-1.01$ & $-0.25$ & $-2.60$ & $-0.08$ & $-3.58$ \\
        & & & 0.50 & $-0.00$ & $-0.00$ & $-0.00$ & $-0.02$ & $-0.00$ & $-0.04$ \\
        \addlinespace
        & \multirow{2}{*}{2} & \multirow{2}{*}{76.47} 
        & 0.35 & $-0.01$ & $-0.04$ & $-0.08$ & $-0.83$ & $-0.06$ & $-2.07$ \\
        & & & 0.50 & $-0.00$ & $-0.00$ & $-0.00$ & $-0.00$ & $-0.00$ & $-0.01$ \\
        \bottomrule
        \end{tabular}  
    \end{threeparttable}
\end{table}

\noindent Table \ref{tab:power_combined} compares between SCD and SWD in terms of the percentage reduction in relative efficiency of power ($RE^{\pi}$) alongside the percentage reduction in total participants ($\Delta_N$). Across all evaluated scenarios, the SCD achieves substantial reductions in required participant observations ($\Delta_N$ ranging from $42.86\%$ up to $88.24\%$) while sustaining only marginal losses in statistical power. In the majority of configurations, the percentage reduction in relative efficiency of power remains well below $5\%$. Also the rate at which within-cluster correlation decays over time directly impacts the efficiency of the SCD. A slower decay rate ($r=0.8$) consistently preserves more statistical power compared with a faster decay rate ($r=0.4$). For example, at $T=7$, $R=2$, $\theta_1=0.35$ and $\boldsymbol{\alpha}=(0.35, 0.20, 0.60)$, moving from $r=0.8$ to $r=0.4$ increases the power penalty from $-1.60\%$ to $-13.56\%$. This indicates that the SCD is more useful than SWD in such environments where cluster effects are highly stable. Lastly as $T$ increases, the relative efficiency of the SCD improves drastically and also simultaneously maximizing participant reduction. For instance, extending the trial to $T=17$ with $R=1$ yields an $88.24\%$ reduction in sample size with maximum power penalties remaining under $4\%$. Additionally, widening the staircase window from $R=1$ to $R=2$ improves relative power in many of the longer trials, but not uniformly and also it inherently decreases the magnitude of the participant reduction.

\subsection{Interaction between trial duration ($T$) and observation-window width ($R$)}
A practical advantage of the SCD over the SWD is that the researcher controls how long clusters are observed. In a standard SWD every cluster contributes data across all $T$ periods. The SCD restricts measurement to a $2R$-period crossover window and $R$ is a free design parameter. The choice of $R$ is not trivial. Extending the window raises within-cluster information but reduces the number of sequences, which weakens the design's ability to separate the treatment effect from the time trend. We examine this trade-off numerically, restricting attention to the balanced case $R_0 = R_1 = R$. To study how $R$ interacts with trial duration, we fix $K=30$ clusters and sweep $T$ from $6$ to $14$. At each $T$ we evaluate all valid observation windows, where $S=T-2R+1$ must satisfy $2 \leq S \leq T-1$. For each $(T,R)$ pair we optimise the cluster allocation $\mathbf{p}$ and record the resulting minimum variance. The window $R^*$ is whichever $R$ produces the lowest variance at that $T$. We repeat this for cluster sizes $m \in \{5, 20\}$ under three ICC scenarios, giving the six columns in Table~\ref{tab:optimal_R_combined}. From this table while it is clear that the optimal observation window $R^*$ generally behaves as a monotonically non-decreasing step function of the trial duration $T$, but a universal theoretical formula for $R^*$ is quite hard to find as it depends heavily on the underlying correlation structures.  
\begin{table}[htbp]
\centering
\caption{Optimal observation window ($R^*$) across varying trial durations ($T$) and cluster sizes ($m$). The correlation scenarios are parameterized as $(\alpha_0, \alpha_1, \alpha_2, r)$.}
\label{tab:optimal_R_combined}
\begin{tabular}{c cc cc cc}
\toprule
 & \multicolumn{2}{c}{\textbf{Low Correlation}} & \multicolumn{2}{c}{\textbf{Moderate Correlation}} & \multicolumn{2}{c}{\textbf{High Correlation}} \\
 & \multicolumn{2}{c}{\footnotesize $(0.05, 0.01, 0.05, 0.1)$} & \multicolumn{2}{c}{\footnotesize $(0.18, 0.10, 0.15, 0.4)$} & \multicolumn{2}{c}{\footnotesize $(0.30, 0.15, 0.30, 0.7)$} \\
\cmidrule(lr){2-3} \cmidrule(lr){4-5} \cmidrule(lr){6-7}
\textbf{$T$} & $m=5$ & $m=20$ & \quad $m=5$ & $m=20$ & $m=5$ & $m=20$ \\
\midrule
6  & 2 & 1 & 1 & 1 & 1 & 1 \\
7  & 2 & 2 & 1 & 1 & 1 & 1 \\
8  & 2 & 2 & 2 & 1 & 1 & 3 \\
9  & 2 & 2 & 2 & 2 & 1 & 3 \\
10 & 2 & 2 & 2 & 2 & 2 & 4 \\
11 & 3 & 2 & 2 & 2 & 2 & 4 \\
12 & 3 & 3 & 2 & 2 & 2 & 4 \\
13 & 3 & 3 & 2 & 2 & 2 & 5 \\
14 & 3 & 3 & 2 & 2 & 2 & 5 \\
\bottomrule
\end{tabular}
\end{table}

\subsection{Comparison between MMD vs EOD}
\subsubsection{Efficiency comparison under block exchangeable correlation structure}
The EOD fixes every cluster to the same size $m_s = 20 \ \forall \ s=1,2,\ldots,S$ and optimises only the proportion of clusters allocated to each sequence. On the other hand, the MMD treats both the allocation proportions $\mathbf{p}$ and the cluster sizes $\mathbf{m} = (m_1,\dots,m_S)^\top$ as free variables and optimises them simultaneously. To make the comparison fair, every scenario holds fixed the total trial duration ($T = 9$ periods), total clusters ($K = 30$) and total participants ($N = K\bar{m} = 600$). The participant constraint is enforced in MMD through $\sum_{s=1}^S p_s m_s = \bar{m}$, so both designs use exactly the same budget.

For each combination of observation window
$R \in \{1, 2, 3\}$ (which determines the total number of sequences $S = T - 2R + 1$) and
cluster-period correlation $r \in \{0.1, 0.3, 0.7\}$, we evaluated all valid ICC triples $(\alpha_0, \alpha_1, \alpha_2)$ drawn from the grid $\{0.01, 0.05, 0.10, 0.20, 0.30, 0.40,\\ 0.50\}^3$, subject to the constraints
$\alpha_1 \le \alpha_0$ and $\alpha_1 \le \alpha_2$. For each valid configuration, the numerically optimised variance is obtained using a multi-start \texttt{fmincon} (MATLAB, interior-point algorithm) function under both EOD and MMD. For every configuration we recorded the percentage efficiency gain of MMD over EOD:
\begin{equation}
  \text{Gain}(\%) =
    \frac{\text{Var}_{\text{EOD}} - \text{Var}_{\text{MMD}}}{\text{Var}_{\text{EOD}}} \times 100.
  \label{eq:gain}
\end{equation}
In table \ref{tab:gain_summary1} we report six summary statistics of the efficiency gain (\%) across all valid ICC configurations $(\alpha_0, \alpha_1, \alpha_2)$ for each combination of $R$ and $r$. The \textbf{Mean} and \textbf{Median} together characterise the centre of the gain distribution. The \textbf{Max} represents the largest gain observed over the evaluated parameter grid, representing the practical ceiling of what MMD can deliver under those design conditions. Finally, the $\mathbf{>5\%}$ and $\mathbf{>10\%}$ columns report the proportion of configurations for which the gain crosses practically meaningful thresholds.

\begin{table}[htbp]
  \centering
  \caption{Summary of efficiency gains (\%) of MMD over EOD across 140 ICC parameter combinations. Fixed: $T=9$, $K=30$, $N=600$.}
  \label{tab:gain_summary1}
  \begin{tabular}{cccccccc}
  \toprule
  $R$ & $r$ & Mean & Median & Max & SD & $>5\%$ & $>10\%$ \\
  \midrule
 & $0.1$ & $2.72$ & $0.06$ & $31.64$ & $6.92$ & $15.0\%$ & $10.7\%$ \\
$1$ & $0.3$ & $2.78$ & $0.09$ & $31.64$ & $6.89$ & $15.0\%$ & $10.7\%$ \\
 & $0.7$ & $3.41$ & $0.91$ & $31.64$ & $6.71$ & $15.0\%$ & $10.7\%$ \\
 \hline
 & $0.1$ & $2.13$ & $0.17$ & $16.41$ & $4.31$ & $15.0\%$ & $11.4\%$ \\
$2$ & $0.3$ & $2.16$ & $0.22$ & $16.41$ & $4.29$ & $15.0\%$ & $11.4\%$ \\
 & $0.7$ & $2.54$ & $0.59$ & $16.41$ & $4.18$ & $15.0\%$ & $11.4\%$ \\
 \hline
 & $0.1$ & $0.67$ & $0.06$ & $7.19$ & $1.37$ & $2.9\%$ & $0.0\%$ \\
$3$ & $0.3$ & $0.65$ & $0.05$ & $7.19$ & $1.38$ & $2.9\%$ & $0.0\%$ \\
 & $0.7$ & $0.66$ & $0.03$ & $7.19$ & $1.38$ & $2.9\%$ & $0.0\%$ \\
  \bottomrule
  \end{tabular}
\end{table}

\begin{table}[htbp]
  \centering
  \caption{Summary of efficiency gains (\%) of MMD over EOD across 140 ICC parameter combinations. Fixed: $T=15$, $K=30$, $N=600$.}
  \label{tab:gain_summary2}
  \begin{tabular}{cccccccc}
  \toprule
  $R$ & $r$ & Mean & Median & Max & SD & $>5\%$ & $>10\%$ \\
  \midrule
 & $0.1$ & $1.65$ & $0.02$ & $25.36$ & $4.65$ & $10.7\%$ & $7.1\%$ \\
$1$ & $0.3$ & $1.68$ & $0.03$ & $25.36$ & $4.64$ & $10.7\%$ & $7.1\%$ \\
 & $0.7$ & $1.99$ & $0.38$ & $25.36$ & $4.56$ & $10.7\%$ & $7.1\%$ \\
  \hline
 & $0.1$ & $2.53$ & $0.13$ & $21.55$ & $5.59$ & $15.0\%$ & $12.9\%$ \\
$2$ & $0.3$ & $2.59$ & $0.19$ & $21.55$ & $5.57$ & $15.0\%$ & $12.9\%$ \\
 & $0.7$ & $3.16$ & $0.98$ & $21.55$ & $5.39$ & $15.0\%$ & $12.9\%$ \\
  \hline
 & $0.1$ & $2.01$ & $0.21$ & $16.70$ & $3.99$ & $17.1\%$ & $8.6\%$ \\
$3$ & $0.3$ & $2.02$ & $0.23$ & $16.70$ & $3.98$ & $17.1\%$ & $7.9\%$ \\
 & $0.7$ & $2.33$ & $0.47$ & $16.70$ & $3.91$ & $17.1\%$ & $8.6\%$ \\
  \bottomrule
  \end{tabular}
\end{table}

From table \ref{tab:gain_summary1} and \ref{tab:gain_summary2}, the relative efficiency of the MMD compared with the EOD is fundamentally governed by the interplay between the total trial duration ($T$) and the active observation window ($R$). This relationship dictates a critical trade-off between the total number of sequences and the longitudinal depth of the data collected per sequence. A wider observation window gives each sequence more longitudinal data but shrinks the total number of sequences, since $S= T - 2R + 1$. Fewer sequences means fewer cluster sizes to optimise and that is where MMD lacks. To identify the threshold where the MMD yields maximum efficiency over the EOD, we define the quantity, $\kappa = \dfrac{S}{2R} = \dfrac{T - 2R + 1}{2R}$ which is a ratio mathematically formalizes the tension between the sequence dimensionality ($S$) and the total measurement burden per cluster ($2R$).

When $\kappa$ is too small $(\kappa \leq 1)$, the observation window has consumed most of the trial. At $T=9, R =3 \implies \kappa \approx 0.67$, the trial contains more active measurement periods per cluster than total sequences. In this constrained space, the MMD optimiser lacks the degrees of freedom required to overcome the EOD. For this setting, only four sequences survive and the mean gain collapses to $0.65\%$ and no ICC configuration in that block crosses $10\%$. When $\kappa$ is too large $(\kappa \gg 4)$, the opposite problem occurs. At $T=15, R=1 \implies \kappa = 7.0$ yields $14$ sequences but each spans just two periods. Here the design is highly flexible but information-poor. As a result the mean gains stay below $2\%$. Across the parameter configurations considered here, the largest gains occurred for approximately $1.5 \leq \kappa \le 4.5$. In this range, the optimiser is supplied with both sufficient sequence heterogeneity to justify asymmetric sizing and sufficient longitudinal depth to anchor the generalized least squares estimator. For instance, at $T=9$ and $R=1$, the ratio stabilizes at $\kappa = 4.0$, yielding the highest mean efficiency gains across the parameter sweep. But this range should be regarded as an empirical diagnostic rather than a universal threshold. By evaluating this aspect ratio $\kappa$ during the planning phase, investigators can assess whether the mathematical complexity of deploying an MMD protocol is justified by the proposed trial structure or not.

\subsubsection{Robustness to correlation misspecification}
\label{subsec:correlation_robustness}

The optimal design depends on the correlation parameters, which in practice must be specified before the trial is conducted and may therefore be subject to misspecification. We assess the sensitivity of the proposed MMD to such uncertainty by comparing its performance with the EOD when the correlation parameters specified at the design stage differ from their true values. Throughout this analysis, we consider the balanced SCD with $R_0=R_1=1$ and average cluster size $\bar m=20$, and examine $S\in\{4,6,10\}$. Both the EOD and MMD are first optimised under the correlation configuration $(\alpha_0,\alpha_1,\alpha_2,r)=(0.20,0.10,0.40,0.40)$. The resulting optimal designs are then held fixed. To represent misspecification, the true correlation parameters are obtained by multiplying all four correlation parameters by a common perturbation factor $\lambda\in\{0.8,0.9,1.0,1.1,1.2\}$. For example, $\lambda=0.8$ represents a setting in which the true correlations are uniformly lower than assumed at the design stage, whereas $\lambda=1.2$ represents uniformly higher correlations. For each value of $\lambda$, the exact analytical variances of the fixed EOD and MMD are evaluated under the corresponding true correlation structure. Their relative efficiency is calculated according to (\ref{eq:gain}). Positive values therefore indicate an efficiency advantage of the MMD. 
%Since the analysis is based on continuous designs, the total number of clusters $K$ only appears as a common multiplicative factor $1/K$ in both variances and consequently cancels from the relative efficiency measure.
\begin{figure}[H]
    \centering
    \includegraphics[width=0.6\linewidth]{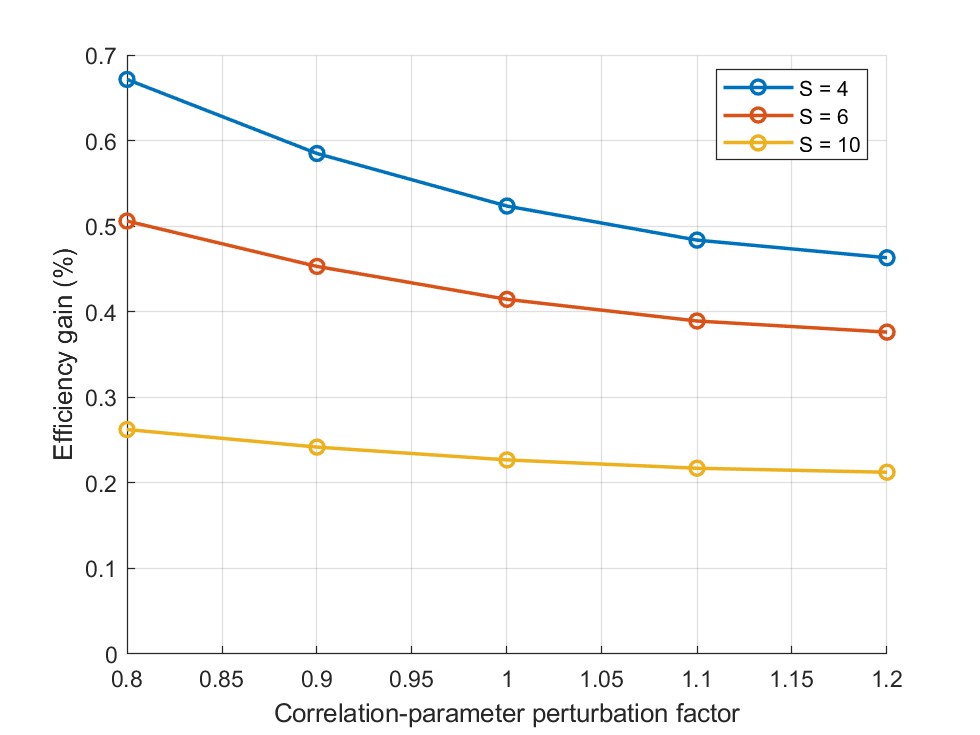}
    \caption{Efficiency gain of the MMD over the EOD under perturbation of the assumed correlation parameters for $S=4,6, 10$.}
    \label{fig:correlation_misspecification}
\end{figure}

\noindent Figure~\ref{fig:correlation_misspecification} shows that the MMD retains a positive efficiency advantage throughout the entire range of correlation misspecification considered. At the nominal specification, $\lambda=1$, the variance reduction is approximately $0.52\%$, $0.41\%$, and $0.23\%$ for $S=4$, $6$, and $10$, respectively. The gain increases when the true correlations are lower than assumed and decreases gradually as the true correlations become stronger. For example, when $\lambda=0.8$, the gain is approximately $0.67\%$ for $S=4$, compared with approximately $0.47\%$ when $\lambda=1.2$. A similar pattern is observed for $S=6$ and $S=10$. The efficiency advantage also becomes smaller as the number of sequences increases. Across the entire perturbation range, the largest gains occur for $S=4$, followed by $S=6$ and the gains for $S=10$ are comparatively modest. Nevertheless, the gain remains positive even under a $\pm20\%$ perturbation of all correlation parameters. These results indicate that, although the absolute improvement of MMD over EOD is modest under this correlation configuration, the advantage is not sensitive to moderate misspecification of the correlation parameters and does not reverse within the range examined.

\subsection{Illustrative trial-design case studies}
\subsubsection{PROMPT Trial}
\noindent To show the practical implications of our closed-cohort framework, we adapt the clinical context of the PROMPT trial \citep{turner2017tiered}. PROMPT was a SW-CRT evaluating a tiered psychosocial intervention for depression in
adult cancer patients across regional treatment centres. We adapt its geometry to a closed-cohort setting with $S = 5$ sequences, cluster size $m = 20$. Covariance parameters follow the published trial: within-period ICC $\alpha_0 = 0.032$ and cluster autocorrelation $r = 0.93$. We additionally consider $r = 0.80$ to assess sensitivity to the autocorrelation assumption. The closed-cohort structure requires two additional parameters: inter-period correlation $\alpha_1$ and individual autocorrelation $\alpha_2$. To make the comparison between SWD and SCD, we set $\alpha_1 = 0.020$ and then create $5$ different scenarios based on $5$ different values of $\alpha_2 \in [0.35, 0.5]$. The SWD spans $T = 6$ periods while the SCD restricts data collection
to the two periods only ($R_0 = R_1 = 1$). Comparing both designs at the same cluster count would be
misleading, since the SWD collects three times as many observations per cluster. We instead hold the total budget fixed at \$1160000, using
the cost framework of \cite{Liu_2024_SMMR} with per-cluster setup cost $c=\$3000$, per-participant enrollment cost $s = \$250$ and per-period measurement cost $e=\$175$.
As per the original PROMPT Trial, the SWD uses this budget to recruit 40 clusters, producing
$40 \times 6 \times 20 = 4800$ observations in total. The SCD, limited to two periods per cluster, needs only $1600$ observations for the same 40 clusters; the saved measurement expenditure funds 37 additional clusters, bringing the SCD total clusters
to 77. \\
Table \ref{tab:prompt_budget_comparison} illustrates the relative reduction in variance achieved by fixing equal budget among SCD and SWD across varying levels of individual auto-correlation ($\alpha_2$) and cluster autocorrelation ($r$). The relative efficiency of the SCD increases with a lower cluster autocorrelation. For example, at $\alpha_2 = 0.5$, the variance reduction improves from $6.63\%$ at $r = 0.93$ to $8.66\%$ at $r = 0.8$. This happens because the SWD tracks clusters across all $T$ periods. It uses this extended data to estimate the secular time trend and the treatment effect. Now, when the cluster autocorrelation ($r$) is low, measurements taken far apart lose their correlation which causes severe diminishing returns for the SWD's distant observations. The SCD avoids this by restricting data collection to the periods immediately before and after the intervention switch. By dropping distant observations, the SCD frees up resources to recruit more independent clusters. This strategy makes the SCD superior whenever $r$ is low. \\
Another finding from Table \ref{tab:prompt_budget_comparison} is that the SCD loses its relative advantage over the SWD as individual auto-correlation ($\alpha_2$) increases. For example, when $\alpha_2$ rises from $0.35$ to $0.5$, the variance reduction at $r = 0.93$ drops sharply from $14.33\%$ to $6.63\%$. A higher $\alpha_2$ means a patient's measurements remain highly stable over time. In the closed-cohort framework, the SWD tracks the exact same individuals across all six periods. It uses this prolonged observation to build a precise baseline for every patient which effectively filters out individual-level noise. But the SCD cannot exploit this deep stability because it only observes patients for two periods and that is why the efficiency gap becomes smaller when $\alpha_2$ increases.

\begin{table}[htbp]
\centering
\caption{Comparison of SWD and SCD under a fixed budget constraint across varying individual auto-correlation ($\alpha_2$) scenarios in PROMPT Trial for $r = 0.93$ and $r = 0.8$.}
\label{tab:prompt_budget_comparison}
\begin{tabular}{c ccc ccc}
\toprule
 & \multicolumn{3}{c}{$r = 0.93$} & \multicolumn{3}{c}{$r = 0.8$} \\
\cmidrule(lr){2-4} \cmidrule(lr){5-7}
$\alpha_2$ & $\Var(\hat{\theta})_{\mathrm{SWD}}$ & $\Var(\hat{\theta})_{\mathrm{SCD}}$ & Reduction (\%) & $\Var(\hat{\theta})_{\mathrm{SWD}}$ & $\Var(\hat{\theta})_{\mathrm{SCD}}$ & Reduction (\%) \\
\midrule
0.350 & 0.001634 & 0.001400 & 14.33 & 0.001703 & 0.001441 & 15.40 \\
0.387 & 0.001549 & 0.001351 & 12.83 & 0.001620 & 0.001392 & 14.09 \\
0.425 & 0.001463 & 0.001301 & 11.09 & 0.001535 & 0.001342 & 12.56 \\
0.463 & 0.001376 & 0.001251 &  9.04 & 0.001449 & 0.001293 & 10.77 \\
0.500 & 0.001287 & 0.001202 &  6.63 & 0.001361 & 0.001243 &  8.66 \\
\bottomrule
\end{tabular}
\end{table}

\subsubsection{Enhancing Recruitment Using Teleconference and Commitment Contract (ERUTECC) Trial}

\noindent To evaluate the practical implications of the proposed optimal allocation strategy, we present a case study based on the ERUTECC trial (Enhancing Recruitment Using Teleconference and Commitment Contract; \citet{lundstrom2018enhancing}), a cluster randomised trial embedded within the EFFECTS stroke trial. Although ERUTECC was described as a SWD, participating centres contributed observations only within a limited window around their intervention date, producing a measurement pattern that is closer to a
SCD than to a complete SWD. We therefore adapt its footprint for our analysis with $K=22$ clusters distributed across $S=11$ sequences, with $R_0=R_1=2$, giving a trial duration of $T=14$ periods. Because different patients contribute recruitment outcomes in different periods, we treat the design as repeated cross-sectional and set $\sigma_u^2=0$, so that $\alpha_2=\alpha_1$. For this illustration, we further assume an exchangeable correlation structure by setting $\alpha_0=\alpha_1=\alpha_2=\rho$. We compare two allocation strategies under a fixed total of $K=22$ clusters: a uniform allocation, which was used in the original study by assigning exactly 2 clusters to each of the 11 sequences ($p_s=1/11$) and an optimal allocation, which distributes clusters across sequences to minimize $\mathrm{Var}(\hat\theta)$. Table~\ref{tab:erutecc_variance_reduction} reports the variance under each strategy ($V_U$ and $V_O$) and the resulting percentage reduction $\Delta
=100\dfrac{V_U-V_O}{V_U}$ across $\rho \in [0.05, 0.40]$ and $m \in \{2, 5, 10, 20\}$.

\noindent Table~\ref{tab:erutecc_variance_reduction} shows that the benefit of optimizing the sequence allocation depends strongly on both the cluster size and the correlation parameter. At $m=2$, increasing \(\rho\) mostly reduces the gain from optimal design, whereas at $m=10$ and $m=20$, the gain eventually increases strongly with $\rho$. For small cluster sizes, the improvement is generally modest. When
$m=2$, the variance reduction decreases from $2.30\%$ at $\rho=0.05$ to
around $0.4\%$ for the largest values of $\rho$. This indicates that when clusters contain relatively few observations, the uniform allocation is already close to optimal over much of the parameter range. Across all configurations, the optimal allocation improves on the uniform design, although the magnitude
of the gain varies from about $0.4\%$ to $5.26\%$.  
\begin{table}[htbp]
    \centering
    \caption{Variance Reduction using Optimal Allocation in the Modified ERUTECC Trial (SCD with $S=11, R=2, K=22$). Here, $\text{V}_{\text{U}}$ and $\text{V}_{\text{O}}$ denote the variance under uniform and optimal allocations, respectively, while $\Delta$ represents the percentage reduction.}
        \label{tab:erutecc_variance_reduction}
    \begin{threeparttable}
        \setlength{\tabcolsep}{4pt} 
        \begin{tabular}{c ccc ccc ccc ccc}
        \toprule
        \multirow{2}{*}{$\rho$} 
        & \multicolumn{3}{c}{$m=2$} 
        & \multicolumn{3}{c}{$m=5$} 
        & \multicolumn{3}{c}{$m=10$} 
        & \multicolumn{3}{c}{$m=20$} \\
        \cmidrule(lr){2-4} \cmidrule(lr){5-7} \cmidrule(lr){8-10} \cmidrule(lr){11-13}
        
        & $\text{V}_{\text{U}}$ & $\text{V}_{\text{O}}$ & $\Delta$ 
        & $\text{V}_{\text{U}}$ & $\text{V}_{\text{O}}$ & $\Delta$ 
        & $\text{V}_{\text{U}}$ & $\text{V}_{\text{O}}$ & $\Delta$ 
        & $\text{V}_{\text{U}}$ & $\text{V}_{\text{O}}$ & $\Delta$ \\
        \midrule
        
        0.05 & 0.0264 & 0.0258 & 2.30 & 0.0111 & 0.0110 & 1.57 & 0.0060 & 0.0059 & 0.83 & 0.0034 & 0.0033 & 0.41 \\
        0.10 & 0.0260 & 0.0256 & 1.74 & 0.0114 & 0.0113 & 0.77 & 0.0064 & 0.0064 & 0.41 & 0.0039 & 0.0038 & 1.18 \\
        0.15 & 0.0255 & 0.0252 & 1.26 & 0.0116 & 0.0116 & 0.44 & 0.0068 & 0.0068 & 0.77 & 0.0042 & 0.0041 & 2.39 \\
        0.20 & 0.0250 & 0.0248 & 0.88 & 0.0118 & 0.0117 & 0.44 & 0.0071 & 0.0070 & 1.45 & 0.0045 & 0.0043 & 3.45 \\
        0.25 & 0.0244 & 0.0243 & 0.61 & 0.0119 & 0.0118 & 0.69 & 0.0073 & 0.0071 & 2.22 & 0.0046 & 0.0044 & 4.25 \\
        0.30 & 0.0238 & 0.0237 & 0.45 & 0.0119 & 0.0117 & 1.11 & 0.0074 & 0.0072 & 2.98 & 0.0047 & 0.0045 & 4.80 \\
        0.35 & 0.0231 & 0.0230 & 0.41 & 0.0118 & 0.0116 & 1.63 & 0.0075 & 0.0072 & 3.67 & 0.0047 & 0.0045 & 5.12 \\
        0.40 & 0.0224 & 0.0223 & 0.48 & 0.0117 & 0.0114 & 2.22 & 0.0074 & 0.0071 & 4.25 & 0.0046 & 0.0044 & 5.26 \\
        
        \bottomrule
        \end{tabular}  
    \end{threeparttable}
\end{table}

\section{Discussion}

This paper develops an optimal-design framework for cohort staircase designs, with particular emphasis on how clusters and participants should be distributed across treatment sequences. Existing methodological work on staircase designs has primarily focused on variance, power, and finite-sample inference, while the sequence allocation itself has generally been treated as fixed. Our results show that this need not be optimal. By deriving the variance of the generalized least squares estimator under a cohort covariance structure, characterizing the estimable design space, establishing convexity of the allocation criterion, and deriving an equivalence theorem, we provide a general framework for identifying and verifying optimal sequence allocations. For the equal-cluster-size designs considered here, reversal invariance of the criterion together with its convexity allows the search to be restricted to symmetric allocations, substantially simplifying the
optimization problem. This leads to exact analytical solutions for several three- and four sequence designs.

One of the main implications is that equal allocation across sequences should not be assumed automatically, even when all clusters have the same size. The optimal weights depend on the correlation structure and on the observation window. For the basic three- and four-sequence designs, the exact solutions show how the balance between inner and outer sequences changes with the induced correlation between cluster-period means. In particular, weaker correlation can favour greater weight on the middle sequences, whereas stronger correlation shifts allocation towards the outer sequences. The pattern changes once the observation window is widened. For the extended designs shown in Figure~\ref{Fig:S3-S4optimal_generalR}, the optimal outer-sequence allocation generally decreases as $r$ increases, in contrast to the basic case. Thus, the allocation of clusters into sequences depends heavily on the observation window width ($R$) and this should be taken into account at the design stage.

% The symmetry results provide a useful structural simplification. In a balanced staircase design, sequences at equal distances from the centre of the design play equivalent roles under reversal. Consequently, an optimal fixed-cluster-size design can be taken to be symmetric, while a unique joint MMD solution must have both symmetric allocation proportions and symmetric sequence-specific cluster sizes. This is useful not only for deriving the closed-form results presented here, but also for numerical optimization when the number of sequences is larger.

The Min--Min Design extends the allocation problem by allowing cluster proportions and sequence-specific cluster sizes to be optimised simultaneously under a fixed participant constraint. The numerical results show that the resulting gains over the equal-cluster-size optimal design are not uniform across settings. In many configurations the gain is modest, but substantially larger improvements are possible for particular combinations of trial duration, observation window, and correlation parameters. This suggests that MMD is most relevant when recruitment or cluster sizes can be varied flexibly across rollout phases. The results should therefore be interpreted as identifying when such flexibility can be used efficiently, rather than as implying that unequal sequence-specific cluster sizes are always preferable.

The comparison with standard SWDs highlights the practical motivation for SCDs. A SWD collects more information from each cluster by observing it throughout the trial, whereas a SCD deliberately reduces the number of repeated measurements by concentrating data collection around the treatment transition. Our numerical results show that the preferred design depends on the relative value of clusters, participants, and repeated measurements. In shorter trials or when extended follow-up is inexpensive, the additional information from the SWD can be worthwhile. In longer trials, however, the repeated-measurement burden can become substantial and a SCD may achieve comparable precision at lower cost. The power comparisons similarly show that large reductions in participant-period observations can often be achieved with relatively small losses in power. This reinforces the idea that the observation window $R$ should itself be regarded as a design parameter rather than a fixed feature of the SCD.

Because the optimal allocation depends on different correlation parameters specified at the planning stage, sensitivity to misspecification is also important. In the settings examined here, the MMD retained an efficiency advantage over the EOD. This provides some reassurance that the proposed design is not excessively sensitive to small planning errors. At the same time, previous work of \cite{rezaei2025inference} has shown that misspecification of the correlation structure at the analysis stage can affect inference in staircase trials. Hence careful modelling of the dependence structure remains important.

Several limitations of the present work suggest directions for future research. The theoretical development assumes a continuous outcome and a linear mixed model, with covariance parameters treated as known at the design stage. Extending the framework to binary or count outcomes, and studying finite-sample inference under the proposed optimal designs, would broaden its practical relevance. The current formulation also assumes equal cluster sizes within each sequence; fully heterogeneous or random cluster sizes would be a natural extension. The exact analytical allocations are derived only for balanced designs with small numbers of sequences, and further work is needed for imbalanced designs, more general observation windows, and other correlation structures. For the MMD problem, existence and uniqueness of the joint optimum are assumed rather than established. Unlike \cite{singh2024bayesian} or \cite{Pain_2026} robust or Bayesian design approaches could account directly for uncertainty in the correlation parameters in staircase settings.

Overall, the results suggest that staircase trials should be viewed as flexible designs whose sequence allocation, cluster sizes, and observation window can all be chosen deliberately. Optimizing these components can improve statistical efficiency while preserving the main practical advantage of the SCD that is reducing unnecessary measurement burden.

\section*{Acknowledgements}
The work of Soumadeb Pain is funded through an IIT Kanpur assistantship, funded by the Ministry of Education (MoE), Govt. of India. The work of Satya Prakash Singh is supported by Science and Engineering Research Board, India grant no. MTR/2022/000627 and Indian Institute of Technology Kanpur, India grant no. IITK/MATH/2021277. We are gratefully acknowledged.

\newpage
\bibliographystyle{apalike}
\bibliography{reference}

\end{document}

\maketitle
Equations numbered with the prefix ``S'' refer to the Supplementary
Material, whereas equation numbers without this prefix refer to the
main manuscript.

\section{Derivation of the variance of the GLS estimator of the treatment effect}
\label{sec:supp_variance}

\noindent Let $n=R_0+R_1$. For cluster $k$ assigned to sequence $s$, define the
cluster-period mean $\bar Y_{skt}
=
\frac{1}{m_s}\sum_{i=1}^{m_s}Y_{skti}$. Averaging the individual-level model over the $m_s$ gives
\begin{equation}\label{eq:supp_mean_model}
\bar Y_{skt}
=
\mathbf Z_{st}\boldsymbol\beta
+
X_{st}\theta
+
\delta_{sk}
+
CP_{skt}
+
\bar u_{sk}
+
\bar\epsilon_{skt},
\end{equation}
where
\[
\bar u_{sk}
=
\frac{1}{m_s}\sum_{i=1}^{m_s}u_{ski}
\sim
\mathcal N\left(0,\frac{\sigma_u^2}{m_s}\right),
\qquad
\bar\epsilon_{skt}
=
\frac{1}{m_s}\sum_{i=1}^{m_s}\epsilon_{skti}
\sim
\mathcal N\left(0,\frac{\sigma_\epsilon^2}{m_s}\right).
\]

Let $\bar{\mathbf Y}_{sk}=
(\bar Y_{sks},\ldots,\bar Y_{sk,s+n-1})^\top$ and let $\mathbf Z_s$ denote the $n\times q$ matrix obtained by stacking the corresponding rows $\mathbf Z_{st}$. The treatment vector is common
across sequences and is given by $\mathbf X=(\mathbf 0_{R_0}^\top,\mathbf 1_{R_1}^\top)^\top$. Further, define
\[
\mathbf{CP}_{sk}
=
(CP_{sks},\ldots,CP_{sk,s+n-1})^\top,
\qquad
\bar{\boldsymbol\epsilon}_{sk}
=
(\bar\epsilon_{sks},\ldots,\bar\epsilon_{sk,s+n-1})^\top.
\]
Then \eqref{eq:supp_mean_model} can be written in vector form as
\begin{equation}\label{eq:supp_vector_model}
\bar{\mathbf Y}_{sk}
=
\mathbf Z_s\boldsymbol\beta
+
\mathbf X\theta
+
\delta_{sk}\mathbf 1_n
+
\mathbf{CP}_{sk}
+
\bar u_{sk}\mathbf 1_n
+
\bar{\boldsymbol\epsilon}_{sk}.
\end{equation}
Assume $\mathbf{CP}_{sk} \sim \mathcal N_n(\mathbf 0,\mathbf V_{CP})$, where the diagonal elements of $\mathbf V_{CP}$ are $\sigma_C^2$ and for $t\neq t'$, the corresponding off-diagonal element is
$r_{tt'}\sigma_C^2$. Since $\delta_{sk}$ and $\bar u_{sk}$ are common
to all periods observed for the same cohort, whereas the components of
$\bar{\boldsymbol\epsilon}_{sk}$ are independent across periods, the
covariance matrix of \eqref{eq:supp_vector_model} is
\begin{equation}\label{eq:supp_Vstar}
\mathbf V_*^{(s)}
= \operatorname{Cov}(\bar{\mathbf Y}_{sk})
= \mathbf V_{CP}+\left(
\sigma_\delta^2+\frac{\sigma_u^2}{m_s}
\right)
\mathbf 1_n\mathbf 1_n^\top
+
\frac{\sigma_\epsilon^2}{m_s}\mathbf I_n.
\end{equation}
Hence,
\begin{equation}\label{eq:supp_Vstar_elements}
\left[\mathbf V_*^{(s)}\right]_{tt'}
=
\begin{cases}
\displaystyle
\sigma_\delta^2+\sigma_C^2
+
\frac{\sigma_u^2+\sigma_\epsilon^2}{m_s},
& t=t',\\[10pt]
\displaystyle
\sigma_\delta^2+r_{tt'}\sigma_C^2
+
\frac{\sigma_u^2}{m_s},
& t\neq t'.
\end{cases}
\end{equation}
Let $\sigma^2= \sigma_\delta^2+\sigma_C^2+\sigma_u^2+\sigma_\epsilon^2$ denote the marginal variance of an individual response. Then define 
\[
\alpha_0
=
\frac{\sigma_\delta^2+\sigma_C^2}{\sigma^2},
\qquad
\alpha_1
=
\frac{\sigma_\delta^2}{\sigma^2},
\qquad
\alpha_2
=
\frac{\sigma_\delta^2+\sigma_u^2}{\sigma^2}.
\]
Note that $0<\alpha_1\leq\min\{\alpha_0,\alpha_2\}<1$. Under the normalization $\sigma^2=1$, \eqref{eq:supp_Vstar_elements}
becomes
\begin{equation}\label{eq:supp_Vstar_alpha}
\left[\mathbf V_*^{(s)}\right]_{tt'}
=
\begin{cases}
\displaystyle
v_s
:=
\frac{1+(m_s-1)\alpha_0}{m_s},
& t=t',\\[10pt]
\displaystyle
c_s:= \alpha_1+r_{tt'}(\alpha_0-\alpha_1)
+
\frac{\alpha_2-\alpha_1}{m_s},
& t\neq t'.
\end{cases}
\end{equation}
For the block-exchangeable correlation structure, $r_{tt'}=r$ for all
$t\neq t'$ and hence $c_s =\alpha_1+r(\alpha_0-\alpha_1)+\frac{\alpha_2-\alpha_1}{m_s}$.
Therefore,
\begin{equation}\label{eq:supp_Vstar_BE}
\mathbf V_*^{(s)}
=
(v_s-c_s)\mathbf I_n
+
c_s\mathbf 1_n\mathbf 1_n^\top.
\end{equation}
We now derive the variance of the GLS estimator of the treatment effect. Stack the cluster-period mean vectors over all
clusters and sequences into $\bar{\mathbf Y}$, and define $\boldsymbol\gamma= (\boldsymbol\beta\ \ 
\theta)^\top$. The covariance matrix of $\bar{\mathbf Y}$ is block diagonal:
\begin{equation}\label{eq:supp_Sigma}
\boldsymbol\Sigma
=
\operatorname{Cov}(\bar{\mathbf Y})
=
\bigoplus_{s=1}^S
\bigoplus_{k=1}^{K_s}
\mathbf V_*^{(s)},
\end{equation}
where $K_s$ denotes the number of clusters assigned to sequence $s$.

\noindent Let $\mathbf H$ denote the corresponding stacked fixed-effects design
matrix, whose block for each cluster assigned to sequence $s$ is
$[\mathbf Z_s\ \mathbf X]$. The GLS estimator is
\[
\widehat{\boldsymbol\gamma}
=
(\mathbf H^\top\boldsymbol\Sigma^{-1}\mathbf H)^{-1}
\mathbf H^\top\boldsymbol\Sigma^{-1}\bar{\mathbf Y} \quad \text{with} \ \operatorname{Cov}(\widehat{\boldsymbol\gamma})
=
(\mathbf H^\top\boldsymbol\Sigma^{-1}\mathbf H)^{-1}.
\]
Because all clusters assigned to the same sequence have the same covariance
block,
\begin{align}
\mathbf H^\top\boldsymbol\Sigma^{-1}\mathbf H
&=
\sum_{s=1}^S K_s
\begin{pmatrix}
\mathbf Z_s^\top(\mathbf V_*^{(s)})^{-1}\mathbf Z_s
&
\mathbf Z_s^\top(\mathbf V_*^{(s)})^{-1}\mathbf X
\\
\mathbf X^\top(\mathbf V_*^{(s)})^{-1}\mathbf Z_s
&
\mathbf X^\top(\mathbf V_*^{(s)})^{-1}\mathbf X
\end{pmatrix}
=
K\begin{pmatrix}\mathbf A_{\mathbf p} & \mathbf b_{\mathbf p}\\
\mathbf b_{\mathbf p}^\top & c_{\mathbf p}
\end{pmatrix},
\label{eq:supp_information}
\end{align}
where $K=\sum_{s=1}^S K_s, \ p_s=\frac{K_s}{K} \ \ \mathbf A_{\mathbf p}=
\sum_{s=1}^S p_s\mathbf B_s,
\ \
\mathbf b_{\mathbf p}
=
\sum_{s=1}^S p_s\mathbf d_s,
\ \
c_{\mathbf p}
=
\sum_{s=1}^S p_s g_s,$ and 
\[ \mathbf B_s=\mathbf Z_s^\top(\mathbf V_*^{(s)})^{-1}\mathbf Z_s,
\ \
\mathbf d_s
=
\mathbf Z_s^\top(\mathbf V_*^{(s)})^{-1}\mathbf X,
\ \
g_s
=
\mathbf X^\top(\mathbf V_*^{(s)})^{-1}\mathbf X.
\]
Provided $\mathbf A_{\mathbf p}\succ0$ and $f(\mathbf p)=c_{\mathbf p}
-\mathbf b_{\mathbf p}^{\top}\mathbf A_{\mathbf p}^{-1}\mathbf b_{\mathbf p}>0$, finally we have
\begin{equation}\label{eq:supp_var_theta}
\operatorname{Var}(\hat\theta)
=
\frac{1}{K}
\left[
c_{\mathbf p}
-
\mathbf b_{\mathbf p}^\top
\mathbf A_{\mathbf p}^{-1}
\mathbf b_{\mathbf p}
\right]^{-1}
\end{equation}

\section{Estimability and convexity of the variance criterion}

For $\mathbf{p}=(p_1,\ldots,p_S)^\top$ in the simplex
\[
\Xi_S=\left\{\mathbf{p}\in\mathbb{R}^S:p_s\geq0,\ \sum_{s=1}^S p_s=1\right\},
\]
define $\mathbf{A}_{\mathbf{p}}=\sum_{s=1}^S p_s\mathbf{B}_s$, $\mathbf{b}_{\mathbf{p}}=\sum_{s=1}^S p_s\mathbf{d}_s$, and $c_{\mathbf{p}}=\sum_{s=1}^S p_sg_s$, where $\mathbf{B}_s=\mathbf{Z}_s^\top(\mathbf{V}_*^{(s)})^{-1}\mathbf{Z}_s$, $\mathbf{d}_s=\mathbf{Z}_s^\top(\mathbf{V}_*^{(s)})^{-1}\mathbf{X}$, and $g_s=\mathbf{X}^\top(\mathbf{V}_*^{(s)})^{-1}\mathbf{X}$. Whenever $\mathbf{A}_{\mathbf{p}}\succ0$, the Schur-complement information for the treatment effect is $f(\mathbf{p})= c_{\mathbf{p}} - \mathbf{b}_{\mathbf{p}}^\top\mathbf{A}_{\mathbf{p}}^{-1}\mathbf{b}_{\mathbf{p}}$. For allocations satisfying $\mathbf{A}_{\mathbf{p}}\succ0$ and $f(\mathbf{p})>0$, we have $\operatorname{Var}(\hat{\theta})=\frac{1}{Kf(\mathbf{p})}.$
\begin{lemma}[Estimability Condition]\label{lem:est}
Let $n=R_0+R_1$. The time-effect information matrix $\mathbf{A}_{\mathbf{p}}$ is positive definite if and only if the following conditions hold:
\begin{enumerate}
    \item[(a)] $p_1>0$ and $p_S>0$;
    \item[(b)] no $n$ consecutive indices $s,\ldots,s+n-1$ satisfy
    $p_s=\cdots=p_{s+n-1}=0$.
\end{enumerate}
Moreover, the full information matrix
\[
\mathbf{M}_{\mathbf{p}}
=
\begin{pmatrix}
\mathbf{A}_{\mathbf{p}} & \mathbf{b}_{\mathbf{p}}\\
\mathbf{b}_{\mathbf{p}}^\top & c_{\mathbf{p}}
\end{pmatrix}
\]
is positive definite if and only if, in addition to (a) and (b), the following condition holds:
\begin{enumerate}
    \item[(c)] there exist two active sequences $s<l$ such that $l-s<n$.
\end{enumerate}
Consequently, under conditions (a)--(c), $f(\mathbf{p})>0$.
\end{lemma}

\begin{proof}
Because the single-cluster covariance matrix $\mathbf{V}_*^{(s)}$ is positive definite, its inverse is also positive definite. For any $\mathbf{a}\in\mathbb{R}^T\setminus\{\mathbf{0}\}$,
\[
\mathbf{a}^\top\mathbf{A}_{\mathbf{p}}\mathbf{a}
=
\sum_{s=1}^S
p_s
(\mathbf{Z}_s\mathbf{a})^\top
(\mathbf{V}_*^{(s)})^{-1}
(\mathbf{Z}_s\mathbf{a})
\geq0,
\]
with equality if and only if $\mathbf{Z}_s\mathbf{a}=\mathbf{0}$ for every active sequence, that is, every sequence $s$ for which $p_s>0$.

Period $t$ is observed only by sequences
$s\in C_t:=\{\max(1,t-n+1),\ldots,\min(S,t)\}$. Therefore,
$\mathbf{A}_{\mathbf{p}}\succ0$ if and only if every calendar period is observed by
at least one active sequence, or equivalently,
$C_t\cap\operatorname{supp}(\mathbf{p})\neq\emptyset$ for every
$t=1,\ldots,T$, where
$\operatorname{supp}(\mathbf{p})=\{s:p_s>0\}$. Since $C_1=\{1\}$ and
$C_T=\{S\}$, conditions $p_1>0$ and $p_S>0$ are necessary. For an interior
period $t$ for which $|C_t|=n$, the set $C_t$ forms a run of $n$ consecutive
indices. Thus, a zero-run of length $n$ leaves at least one calendar period
unobserved. Conversely, if $p_1>0$, $p_S>0$, and no $n$ consecutive sequences
have zero allocation, then every $C_t$ contains at least one active sequence.
Hence, if $\mathbf{Z}_s\mathbf{a}=\mathbf{0}$ for every active sequence, then
$a_t=0$ for every calendar period $t$, which implies $\mathbf{a}=\mathbf{0}$.
Therefore, $\mathbf{A}_{\mathbf{p}}\succ0$.

Next consider the full information matrix $\mathbf{M}_{\mathbf{p}}$. For any
\[
\mathbf{q}=
\begin{pmatrix}
\mathbf{a}\\ b
\end{pmatrix}
\in\mathbb{R}^{T+1}\setminus\{\mathbf{0}\},
\]
we have
\[
\mathbf{q}^\top\mathbf{M}_{\mathbf{p}}\mathbf{q}
=
\sum_{s=1}^S
p_s
(\mathbf{Z}_s\mathbf{a}+\mathbf{X}b)^\top
(\mathbf{V}_*^{(s)})^{-1}
(\mathbf{Z}_s\mathbf{a}+\mathbf{X}b)
\geq0,
\]
with equality if and only if
$\mathbf{Z}_s\mathbf{a}+\mathbf{X}b=\mathbf{0}$ for every active sequence.

Suppose conditions (a)--(c) hold. By condition (c), there exist active sequences
$s<l$ such that $l-s<n$, so their observation windows overlap. More precisely,
there exists a common period $t$ satisfying
\[
\max(l,s+R_0)
\leq t
\leq
\min(s+R_0+R_1-1,l+R_0-1).
\]
At such a period, sequence $s$ is under treatment whereas sequence $l$ is still
under control; hence $X_{s,t}=1$ and $X_{l,t}=0$. If
$\mathbf{Z}_s\mathbf{a}+\mathbf{X}b=\mathbf{0}$ for both sequences, then the
corresponding period-$t$ components give $a_t+b=0$ and $a_t=0$, respectively.
Thus $b=0$. It then follows that
$\mathbf{Z}_s\mathbf{a}=\mathbf{0}$ for every active sequence, and by conditions
(a) and (b), $\mathbf{a}=\mathbf{0}$. Hence
$\mathbf{q}^\top\mathbf{M}_{\mathbf{p}}\mathbf{q}>0$ for every
$\mathbf{q}\neq\mathbf{0}$, so $\mathbf{M}_{\mathbf{p}}\succ0$.

Conversely, suppose conditions (a) and (b) hold but no two active sequences have
overlapping observation windows. Then $l-s\geq n$ for every pair of active
sequences $s<l$, and the corresponding observation windows are disjoint. Hence,
at each calendar period, the treatment indicator is constant across the active
observations. The treatment column is therefore in the column space of the
calendar-time effects, so the stacked design matrix has rank at most $T$ rather
than $T+1$. Consequently, $\mathbf{M}_{\mathbf{p}}$ is singular and cannot be
positive definite. Thus condition (c) is also necessary.

Finally, under conditions (a)--(c), both $\mathbf{A}_{\mathbf{p}}$ and
$\mathbf{M}_{\mathbf{p}}$ are positive definite. By the Schur complement
theorem,
\[
f(\mathbf{p})
=
c_{\mathbf{p}}
-
\mathbf{b}_{\mathbf{p}}^\top
\mathbf{A}_{\mathbf{p}}^{-1}
\mathbf{b}_{\mathbf{p}}
>0.
\]
\end{proof}
\noindent The estimable allocation space is therefore
$\Xi_{\mathrm{est}}
=
\left\{
\mathbf{p}\in\Xi_S:
\mathbf{A}_{\mathbf{p}}\succ0,\;
f(\mathbf{p})>0
\right\},$ which is characterized by conditions (a)--(c) in Lemma~\ref{lem:est}.

\begin{lemma}
\label{lem:domain_convex}
The set $\Xi_+
=
\left\{
\mathbf{p}\in\Xi_S:
\mathbf{A}_{\mathbf{p}}\succ0
\right\}$ is convex.
\end{lemma}

\begin{proof}
Since each $\mathbf{B}_s\succeq0$ and $p_s\geq0$,
\[
\mathbf{A}_{\mathbf{p}}
=
\sum_{s=1}^S p_s\mathbf{B}_s
\succeq0
\]
for every $\mathbf{p}\in\Xi_S$. Let $\mathbf{p}^{(1)},\mathbf{p}^{(2)}\in\Xi_+$
and $\lambda\in[0,1]$. Since $\mathbf{p}\mapsto\mathbf{A}_{\mathbf{p}}$ is affine,
\[
\mathbf{A}_{\lambda\mathbf{p}^{(1)}+(1-\lambda)\mathbf{p}^{(2)}}
=
\lambda\mathbf{A}_{\mathbf{p}^{(1)}}
+
(1-\lambda)\mathbf{A}_{\mathbf{p}^{(2)}}.
\]
For $\lambda\in(0,1)$, the right-hand side is a positive combination of two
positive definite matrices and is therefore positive definite. The cases
$\lambda=0$ and $\lambda=1$ are immediate. Hence
$\lambda\mathbf{p}^{(1)}+(1-\lambda)\mathbf{p}^{(2)}\in\Xi_+$, proving that
$\Xi_+$ is convex.
\end{proof}

\begin{theorem}\label{Th:convexity}
For fixed sequence-specific cluster sizes $\mathbf m=(m_1,\ldots,m_S)^\top$, $f(\mathbf p)$ is concave on $\mathbf\Xi_+$, and $\operatorname{Var}(\hat\theta)$ is convex in $\mathbf p$ over $\mathbf\Xi_{\mathrm{est}}$.
\end{theorem}

\begin{proof}
Both $\mathbf A_{\mathbf p}$ and $\mathbf b_{\mathbf p}$ are affine in $\mathbf p$, and the matrix-fractional function $h(\mathbf A,\mathbf b)=\mathbf b^\top\mathbf A^{-1}\mathbf b$ is jointly convex on $\{\mathbf A\succ0\}$ (Boyd \& Vandenberghe, 2004, \S3.1.5). For $\mathbf p^{(1)},\mathbf p^{(2)}\in\mathbf\Xi_+$ and $\lambda\in[0,1]$, $\mathbf p^{(\lambda)}=\lambda\mathbf p^{(1)}+(1-\lambda)\mathbf p^{(2)}\in\mathbf\Xi_+$ by Lemma~\ref{lem:domain_convex}. Further
\[
g(\mathbf p^{(\lambda)}):=h(\mathbf A_{\mathbf p^{(\lambda)}},\mathbf b_{\mathbf p^{(\lambda)}})\le\lambda h(\mathbf A_{\mathbf p^{(1)}},\mathbf b_{\mathbf p^{(1)}})+(1-\lambda)h(\mathbf A_{\mathbf p^{(2)}},\mathbf b_{\mathbf p^{(2)}})=\lambda g(\mathbf p^{(1)})+(1-\lambda)g(\mathbf p^{(2)}),
\]
so $g$ is convex on $\mathbf\Xi_+$. Since $c_{\mathbf p}$ is linear, $f(\mathbf p)=c_{\mathbf p}-g(\mathbf p)$ is concave on $\mathbf\Xi_+$, proving the first claim.

\noindent Since $\mathbf\Xi_{\mathrm{est}}\subseteq\mathbf\Xi_+$, $f$ restricted to $\mathbf\Xi_{\mathrm{est}}$ is concave there too. The scalar map $u(x)=x^{-1}$ is convex and non-increasing on $x>0$, and the composition of a convex non-increasing function with a positive concave function is convex; since by definition $f(\mathbf p)>0$ on $\mathbf\Xi_{\mathrm{est}}$, $f(\mathbf p)^{-1}$ is convex there. Multiplying by $1/K>0$ preserves convexity, so $\operatorname{Var}(\hat\theta)=[Kf(\mathbf p)]^{-1}$ is convex in $\mathbf p$ over $\mathbf\Xi_{\mathrm{est}}$.
\end{proof}

\begin{corollary}\label{cor:est_convex}
$\mathbf\Xi_{\mathrm{est}}$ is convex.
\end{corollary}

\begin{proof}
By definition $\mathbf\Xi_{\mathrm{est}}=\{\mathbf p\in\mathbf\Xi_+:f(\mathbf p)>0\}$, a strict superlevel set of $f$ restricted to $\mathbf\Xi_+$. By Lemma~\ref{lem:domain_convex}, $\mathbf\Xi_+$ is convex; by Theorem~\ref{Th:convexity}, $f$ is concave on $\mathbf\Xi_+$. Superlevel sets of a concave function on a convex domain are convex, so $\mathbf\Xi_{\mathrm{est}}$ is convex.
\end{proof}

\section{Equivalence Theorem}

\begin{theorem}\label{thm:equiv}
Let $\mathbf p^*\in\mathbf\Xi_{\mathrm{est}}$. Then $\mathbf p^*$ minimizes $\operatorname{Var}(\hat\theta)$ over $\mathbf\Xi_{\mathrm{est}}$ if and only if, for every $l=1,\ldots,S$,
\[
\boxed{\delta_l(\mathbf p^*):=(\mathbf X-\mathbf Z_l\mathbf x^*)^\top(\mathbf V_*^{(l)})^{-1}(\mathbf X-\mathbf Z_l\mathbf x^*)\le f(\mathbf p^*)},\qquad \mathbf x^*=\mathbf A_{\mathbf p^*}^{-1}\mathbf b_{\mathbf p^*}.
\]
\end{theorem}

\begin{proof}
Since $\operatorname{Var}(\hat\theta)=[Kf(\mathbf p)]^{-1}$ and $f>0$ on $\mathbf\Xi_{\mathrm{est}}$, minimizing $\operatorname{Var}(\hat\theta)$ is equivalent to maximizing $f$, which is concave on $\mathbf\Xi_{\mathrm{est}}$ by Theorem~\ref{Th:convexity}.

For $\mathbf p,\mathbf p^1\in\mathbf\Xi_{\mathrm{est}}$, let $\mathbf p(\epsilon)=\mathbf p+\epsilon(\mathbf p^1-\mathbf p)$ and $f'(\mathbf p;\mathbf p^1)=\lim_{\epsilon\to0^+}\epsilon^{-1}[f(\mathbf p(\epsilon))-f(\mathbf p)]$. Writing $\dot c=\sum_s(p_s^1-p_s)g_s$, $\dot{\mathbf A}=\sum_s(p_s^1-p_s)\mathbf B_s$, $\dot{\mathbf b}=\sum_s(p_s^1-p_s)\mathbf d_s$, and using $\frac{d}{d\epsilon}\mathbf A_{\mathbf p(\epsilon)}^{-1}=-\mathbf A_{\mathbf p(\epsilon)}^{-1}\dot{\mathbf A}\mathbf A_{\mathbf p(\epsilon)}^{-1}$, we obtain, with $\mathbf x=\mathbf A_{\mathbf p}^{-1}\mathbf b_{\mathbf p}$,
\[
\left.\frac{d}{d\epsilon}\mathbf b_{\mathbf p(\epsilon)}^\top\mathbf A_{\mathbf p(\epsilon)}^{-1}\mathbf b_{\mathbf p(\epsilon)}\right|_{\epsilon=0}=2\dot{\mathbf b}^\top\mathbf x-\mathbf x^\top\dot{\mathbf A}\mathbf x=\sum_{s=1}^S(p_s^1-p_s)\bigl(2\mathbf d_s^\top\mathbf x-\mathbf x^\top\mathbf B_s\mathbf x\bigr).
\]
Hence $f'(\mathbf p;\mathbf p^1)=(\mathbf p^1-\mathbf p)^\top\boldsymbol\delta(\mathbf p)$, where $\delta_s(\mathbf p)=g_s-2\mathbf d_s^\top\mathbf x+\mathbf x^\top\mathbf B_s\mathbf x=(\mathbf X-\mathbf Z_s\mathbf x)^\top(\mathbf V_*^{(s)})^{-1}(\mathbf X-\mathbf Z_s\mathbf x)$, the last equality following by substituting the definitions of $g_s,\mathbf d_s,\mathbf B_s$. A direct computation using $\mathbf A_{\mathbf p}\mathbf x=\mathbf b_{\mathbf p}$ gives the identity
\begin{equation}\label{eq:delta_identity}
\sum_{s=1}^Sp_s\delta_s(\mathbf p)=c_{\mathbf p}-2\mathbf b_{\mathbf p}^\top\mathbf x+\mathbf x^\top\mathbf A_{\mathbf p}\mathbf x=c_{\mathbf p}-\mathbf b_{\mathbf p}^\top\mathbf A_{\mathbf p}^{-1}\mathbf b_{\mathbf p}=f(\mathbf p).
\end{equation}

$(\Rightarrow)$ Suppose $\mathbf p^*$ maximizes $f$ over $\mathbf\Xi_{\mathrm{est}}$. Fix $l$ and let $\mathbf e_l$ be the $l$th unit vector; since $\mathbf A_{\mathbf p^*}\succ0$ and $f(\mathbf p^*)>0$ are open conditions, $\mathbf p^*(\epsilon)=(1-\epsilon)\mathbf p^*+\epsilon\mathbf e_l\in\mathbf\Xi_{\mathrm{est}}$ for sufficiently small $\epsilon>0$, even though $\mathbf e_l$ itself need not belong to $\mathbf\Xi_{\mathrm{est}}$. Optimality gives $f'(\mathbf p^*;\mathbf e_l)\le0$, i.e.\ $(\mathbf e_l-\mathbf p^*)^\top\boldsymbol\delta(\mathbf p^*)\le0$, which by \eqref{eq:delta_identity} equals $\delta_l(\mathbf p^*)-f(\mathbf p^*)$. Hence $\delta_l(\mathbf p^*)\le f(\mathbf p^*)$ for every $l=1,\ldots,S$.

$(\Leftarrow)$ Suppose $\delta_l(\mathbf p^*)\le f(\mathbf p^*)$ for all $l$. Any $\mathbf p^1\in\mathbf\Xi_{\mathrm{est}}\subset\mathbf\Xi_S$ can be written $\mathbf p^1=\sum_{l=1}^Sc_l\mathbf e_l$ with $c_l\ge0$, $\sum_lc_l=1$, so by \eqref{eq:delta_identity},
\[
f'(\mathbf p^*;\mathbf p^1)=(\mathbf p^1-\mathbf p^*)^\top\boldsymbol\delta(\mathbf p^*)=\sum_{l=1}^Sc_l\delta_l(\mathbf p^*)-f(\mathbf p^*)\le\sum_{l=1}^Sc_lf(\mathbf p^*)-f(\mathbf p^*)=0.
\]
Since $\mathbf\Xi_{\mathrm{est}}$ is convex (Corollary~\ref{cor:est_convex}) and $f$ is concave there, $f'(\mathbf p^*;\mathbf p^1)\le0$ for every $\mathbf p^1\in\mathbf\Xi_{\mathrm{est}}$ implies $f(\mathbf p^1)\le f(\mathbf p^*)$ for every such $\mathbf p^1$. Thus $\mathbf p^*$ maximizes $f$, and therefore minimizes $\operatorname{Var}(\hat\theta)$, over $\mathbf\Xi_{\mathrm{est}}$.
\end{proof}

\section{Symmetry of the MMD}

\noindent We now prove Theorem~\ref{thm:MMD_symmetry}. Let
$n=2R$ and $T=S+2R-1$, and denote by $\mathbf J_n$ and
$\mathbf J_T$ the corresponding anti-identity matrices. For any
$\mathbf p=(p_1,\ldots,p_S)^\top$ and
$\mathbf m=(m_1,\ldots,m_S)^\top$, define their reversals by
\[
p_s^{R}=p_{S-s+1},
\qquad
m_s^{R}=m_{S-s+1},
\qquad s=1,\ldots,S.
\]

\begin{lemma}
\label{lem:Z_reversal}
For a balanced staircase design, $\mathbf Z_{S-s+1}=\mathbf J_n\mathbf Z_s\mathbf J_T,\ \  s=1,\ldots,S$. Moreover, $\mathbf J_n\mathbf X=\mathbf 1_n-\mathbf X.$
\end{lemma}

\begin{proof}
Sequence $s$ observes the $n=2R$ consecutive periods
$s,\ldots,s+n-1$. Reversing both the rows within the observation
window and the $T$ calendar periods maps this window to that of
sequence $S-s+1$, which gives
$\mathbf Z_{S-s+1}=\mathbf J_n\mathbf Z_s\mathbf J_T$.
Since
$\mathbf X=(\mathbf 0_R^\top,\mathbf 1_R^\top)^\top$,
reversing its entries yields
$\mathbf J_n\mathbf X
=(\mathbf 1_R^\top,\mathbf 0_R^\top)^\top
=\mathbf 1_n-\mathbf X$.
\end{proof}

\noindent For fixed $(\mathbf p,\mathbf m)$, write
\[
\mathbf A_{\mathbf p}(\mathbf m)
=
\sum_{s=1}^{S}
p_s\mathbf Z_s^\top
\mathbf V_*(m_s)^{-1}
\mathbf Z_s\ , \ \mathbf b_{\mathbf p}(\mathbf m)=\sum_{s=1}^{S}p_s\mathbf Z_s^\top
\mathbf V_*(m_s)^{-1}
\mathbf X \ ,\ c_{\mathbf p}(\mathbf m)
=
\sum_{s=1}^{S}
p_s\mathbf X^\top
\mathbf V_*(m_s)^{-1}
\mathbf X
\]
Also define
\[
\mathbf X^{R}=\mathbf J_n\mathbf X= \mathbf 1_n-\mathbf X \quad \text{and} \quad \mathbf b_{\mathbf p}^{R}(\mathbf m)
=
\sum_{s=1}^{S}
p_s\mathbf Z_s^\top
\mathbf V_*(m_s)^{-1}
\mathbf X^{R}
\]
\begin{lemma}
\label{lem:dual_b}
Suppose $\mathbf V_*(m)$ is persymmetric for every admissible $m$.
Then
\[
\left\{
\mathbf b_{\mathbf p}^{R}(\mathbf m)
\right\}^{\top}
\mathbf A_{\mathbf p}(\mathbf m)^{-1}
\mathbf b_{\mathbf p}^{R}(\mathbf m)
=
\mathbf b_{\mathbf p}(\mathbf m)^{\top}
\mathbf A_{\mathbf p}(\mathbf m)^{-1}
\mathbf b_{\mathbf p}(\mathbf m).
\]
\end{lemma}

\begin{proof}
Since $\mathbf Z_s\mathbf 1_T=\mathbf 1_n$,
\[
\begin{aligned}
\mathbf b_{\mathbf p}^{R}(\mathbf m)
&=
\sum_{s=1}^{S}
p_s\mathbf Z_s^\top
\mathbf V_*(m_s)^{-1}
(\mathbf 1_n-\mathbf X)\\
&=
\mathbf A_{\mathbf p}(\mathbf m)\mathbf 1_T
-
\mathbf b_{\mathbf p}(\mathbf m).
\end{aligned}
\]
By persymmetry,
\[
\mathbf J_n\mathbf V_*(m_s)^{-1}\mathbf J_n
=
\mathbf V_*(m_s)^{-1},
\]
and hence $\mathbf J_n$ commutes with
$\mathbf V_*(m_s)^{-1}$. Since
$\mathbf J_n\mathbf 1_n=\mathbf 1_n$,
\[
\begin{aligned}
\mathbf 1_n^\top
\mathbf V_*(m_s)^{-1}\mathbf X
&=
\mathbf 1_n^\top
\mathbf V_*(m_s)^{-1}
\mathbf J_n\mathbf X\\
&=
\mathbf 1_n^\top
\mathbf V_*(m_s)^{-1}
(\mathbf 1_n-\mathbf X).
\end{aligned}
\]
Therefore,
\[
2\mathbf 1_n^\top
\mathbf V_*(m_s)^{-1}\mathbf X
=
\mathbf 1_n^\top
\mathbf V_*(m_s)^{-1}\mathbf 1_n.
\]
Multiplying by $p_s$ and summing over $s$ gives
\[
\mathbf 1_T^\top
\mathbf A_{\mathbf p}(\mathbf m)\mathbf 1_T
=
2\mathbf 1_T^\top
\mathbf b_{\mathbf p}(\mathbf m).
\]
Consequently,
\[
\begin{aligned}
&
\left\{
\mathbf b_{\mathbf p}^{R}(\mathbf m)
\right\}^{\top}
\mathbf A_{\mathbf p}(\mathbf m)^{-1}
\mathbf b_{\mathbf p}^{R}(\mathbf m)
\\
=
&\left\{
\mathbf A_{\mathbf p}(\mathbf m)\mathbf 1_T
-\mathbf b_{\mathbf p}(\mathbf m)
\right\}^{\top}
\mathbf A_{\mathbf p}(\mathbf m)^{-1}
\left\{
\mathbf A_{\mathbf p}(\mathbf m)\mathbf 1_T
-\mathbf b_{\mathbf p}(\mathbf m)
\right\}
\\
=
&\mathbf 1_T^\top
\mathbf A_{\mathbf p}(\mathbf m)\mathbf 1_T
-
2\mathbf 1_T^\top
\mathbf b_{\mathbf p}(\mathbf m)
+
\mathbf b_{\mathbf p}(\mathbf m)^\top
\mathbf A_{\mathbf p}(\mathbf m)^{-1}
\mathbf b_{\mathbf p}(\mathbf m)
\\
=
&\mathbf b_{\mathbf p}(\mathbf m)^\top
\mathbf A_{\mathbf p}(\mathbf m)^{-1}
\mathbf b_{\mathbf p}(\mathbf m).
\end{aligned}
\]
\end{proof}

\begin{proof}[Proof of Theorem~\ref{thm:MMD_symmetry}]
Let
$(\mathbf p,\mathbf m)\in\Omega_{\bar m}$ and define its simultaneous
reversal $(\mathbf p^{R},\mathbf m^{R})$. Since reversal preserves
positivity,
\[
\sum_{s=1}^{S}p_s^{R}
=
\sum_{s=1}^{S}p_s
=
1,
\]
and, after re-indexing,
\[
\sum_{s=1}^{S}p_s^{R}m_s^{R}
=
\sum_{s=1}^{S}
p_{S-s+1}m_{S-s+1}
=
\sum_{s=1}^{S}p_sm_s
=
\bar m.
\]
Thus
$(\mathbf p^{R},\mathbf m^{R})\in\Omega_{\bar m}$.

We next show that the variance criterion is invariant under simultaneous
reversal. Since the covariance matrix depends on the sequence only through
the corresponding cluster size,
\[
\mathbf V_*^{(s)}(\mathbf m^{R})
=
\mathbf V_*(m_{S-s+1}).
\]
Using Lemma~\ref{lem:Z_reversal}, re-indexing with
$k=S-s+1$, and the persymmetry of $\mathbf V_*(m_k)$,
\[
\begin{aligned}
\mathbf A_{\mathbf p^{R}}(\mathbf m^{R})
&=
\sum_{s=1}^{S}
p_{S-s+1}
\mathbf Z_s^\top
\mathbf V_*(m_{S-s+1})^{-1}
\mathbf Z_s\\
&=
\mathbf J_T
\mathbf A_{\mathbf p}(\mathbf m)
\mathbf J_T.
\end{aligned}
\]
Similarly,
\[
\begin{aligned}
\mathbf b_{\mathbf p^{R}}(\mathbf m^{R})
&=
\sum_{k=1}^{S}
p_k
\mathbf Z_{S-k+1}^{\top}
\mathbf V_*(m_k)^{-1}
\mathbf X\\
&=
\mathbf J_T
\sum_{k=1}^{S}
p_k\mathbf Z_k^\top
\mathbf V_*(m_k)^{-1}
\mathbf J_n\mathbf X\\
&=
\mathbf J_T
\mathbf b_{\mathbf p}^{R}(\mathbf m).
\end{aligned}
\]
Moreover,
\[
c_{\mathbf p^{R}}(\mathbf m^{R})
=
c_{\mathbf p}(\mathbf m).
\]
Since
$\mathbf J_T^{-1}=\mathbf J_T^\top=\mathbf J_T$,
\[
\begin{aligned}
&
\mathbf b_{\mathbf p^{R}}(\mathbf m^{R})^\top
\mathbf A_{\mathbf p^{R}}(\mathbf m^{R})^{-1}
\mathbf b_{\mathbf p^{R}}(\mathbf m^{R})
\\
&\quad=
\left\{
\mathbf b_{\mathbf p}^{R}(\mathbf m)
\right\}^{\top}
\mathbf A_{\mathbf p}(\mathbf m)^{-1}
\mathbf b_{\mathbf p}^{R}(\mathbf m)
\\
&\quad=
\mathbf b_{\mathbf p}(\mathbf m)^\top
\mathbf A_{\mathbf p}(\mathbf m)^{-1}
\mathbf b_{\mathbf p}(\mathbf m),
\end{aligned}
\]
where the final equality follows from
Lemma~\ref{lem:dual_b}. Hence $f(\mathbf p^{R},\mathbf m^{R})=f(\mathbf p,\mathbf m)$ and therefore
\[
\operatorname{Var}
\!\left(
\hat\theta;
\mathbf p^{R},\mathbf m^{R}
\right)
=
\operatorname{Var}
\!\left(
\hat\theta;
\mathbf p,\mathbf m
\right).
\]
Now let
$(\mathbf p^{*},\mathbf m^{*})$
be the unique minimiser over $\Omega_{\bar m}$.
The simultaneous reversal
$\bigl((\mathbf p^{*})^{R},(\mathbf m^{*})^{R}\bigr)$
is also feasible and, by the preceding invariance result, attains the
same minimum variance. Uniqueness therefore implies
\[
(\mathbf p^{*})^{R}
=
\mathbf p^{*},
\qquad
(\mathbf m^{*})^{R}
=
\mathbf m^{*}.
\]
Equivalently,
\[
p_s^{*}
=
p_{S-s+1}^{*},
\qquad
m_s^{*}
=
m_{S-s+1}^{*},
\qquad
s=1,\ldots,S.
\]
Thus the optimal cluster-allocation proportions and the
sequence-specific cluster sizes are simultaneously symmetric about the
centre of the staircase design.
\end{proof}

\section{Symmetrization under fixed equal cluster sizes}
\label{sec:supp_fixed_m_symmetry}

\begin{lemma}[Fixed-cluster-size symmetrization]
\label{lem:fixed_m_symmetry}
Consider a balanced staircase design with $R_0=R_1=R$ and fixed equal
cluster sizes $m_s=m$. Suppose the covariance matrix is persymmetric and let
$\Phi(\mathbf p)=\operatorname{Var}(\hat\theta;\mathbf p)$ denote the variance criterion defined on the estimable design space. Then for $\mathbf p^{R}=(p_S,p_{S-1},\ldots,p_1)^\top$, we have
\[
\Phi(\mathbf p^{R})=\Phi(\mathbf p).
\]
Consequently, $\mathbf p^{\mathrm{sym}}=\frac{\mathbf p+\mathbf p^{R}}{2}$ is symmetric and satisfies $\Phi(\mathbf p^{\mathrm{sym}})\leq\Phi(\mathbf p)$.
Thus, whenever a global minimiser exists, there exists a symmetric global
minimiser. If the global minimiser is unique, it must satisfy
$\mathbf p^*=\mathbf p^{*R}$.
\end{lemma}

\begin{proof}
Let $n=2R$ and $T=S+2R-1$, and let $\mathbf J_q$ denote the
$q\times q$ reversal matrix. For sequence $s$, define its reversed sequence $s^{R}=S-s+1$. 

\noindent Then the balanced staircase gives $\mathbf Z_{s^{R}}=
\mathbf J_n\mathbf Z_s\mathbf J_T$ and $\mathbf J_n\mathbf X=\mathbf 1_n-\mathbf X$. Also for fixed equal cluster sizes, due to persymmetry we have $\mathbf J_n\mathbf V_*^{-1}\mathbf J_n =
\mathbf V_*^{-1}$.

\noindent Recall that
\[
\mathbf B_s
=
\mathbf Z_s^\top\mathbf V_*^{-1}\mathbf Z_s,
\qquad
\mathbf d_s
=
\mathbf Z_s^\top\mathbf V_*^{-1}\mathbf X,
\qquad
g_s
=
\mathbf X^\top\mathbf V_*^{-1}\mathbf X.
\]
Using the preceding identities we have, $\mathbf B_{s^{R}}=\mathbf J_T\mathbf B_s\mathbf J_T$, $\mathbf d_{s^{R}}=\mathbf J_T\left(\mathbf B_s\mathbf 1_T-\mathbf d_s\right)$.

\noindent For the reversed allocation $\mathbf p^{R}$, it follows that $\mathbf A_{\mathbf p^{R}}=\mathbf J_T\mathbf A_{\mathbf p}\mathbf J_T$ and $\mathbf b_{\mathbf p^{R}}=\mathbf J_T
\left(\mathbf A_{\mathbf p}\mathbf 1_T-\mathbf b_{\mathbf p}\right)$. Moreover, $c_{\mathbf p^{R}}=c_{\mathbf p}$.

Because $\mathbf V_*^{-1}$ is persymmetric and
$\mathbf J_n\mathbf X=\mathbf 1_n-\mathbf X$,
\[
\mathbf 1_n^\top\mathbf V_*^{-1}\mathbf X
=
\frac12
\mathbf 1_n^\top\mathbf V_*^{-1}\mathbf 1_n.
\]
Consequently,
\[
\mathbf 1_T^\top\mathbf A_{\mathbf p}\mathbf 1_T
=
2\mathbf 1_T^\top\mathbf b_{\mathbf p}.
\]
Therefore,
\begin{align*}
\mathbf b_{\mathbf p^{R}}^\top
\mathbf A_{\mathbf p^{R}}^{-1}
\mathbf b_{\mathbf p^{R}}
&=
\left(
\mathbf A_{\mathbf p}\mathbf 1_T-\mathbf b_{\mathbf p}
\right)^\top
\mathbf A_{\mathbf p}^{-1}
\left(
\mathbf A_{\mathbf p}\mathbf 1_T-\mathbf b_{\mathbf p}
\right) =
\mathbf b_{\mathbf p}^\top
\mathbf A_{\mathbf p}^{-1}
\mathbf b_{\mathbf p}.
\end{align*}
Hence $\Phi(\mathbf p^{R})=\Phi(\mathbf p)$.

\noindent The estimable design space is convex, so $\mathbf p^{\mathrm{sym}}=\frac{\mathbf p+\mathbf p^{R}}{2}$ is admissible whenever both $\mathbf p$ and $\mathbf p^{R}$ are admissible. Since $\Phi$ is convex,
\[
\Phi(\mathbf p^{\mathrm{sym}})
\leq
\frac12\Phi(\mathbf p)
+
\frac12\Phi(\mathbf p^{R})
=
\Phi(\mathbf p).
\]
Thus any global minimiser can be replaced by a symmetric global minimiser
without increasing the variance. If the global minimiser is unique, this
replacement must coincide with the original design, so that
$\mathbf p^*=\mathbf p^{*R}$.
\end{proof}

\section{Derivation of the exact optimal allocation for a three-sequence basic staircase design}
\label{sec:supp_s3_basic}

\noindent
Consider a balanced closed-cohort staircase design with $S=3$ sequences and
$R_0=R_1=1$, so that the trial spans $T=4$ periods. Throughout this derivation,
all clusters are assumed to have the same size $m$. Under the block-exchangeable
correlation structure, the covariance matrix of the two cluster-period means
contributed by a cluster is common to all sequences and can be written as
\[
\mathbf V_*
=
\begin{pmatrix}
v & c\\
c & v
\end{pmatrix},
\]
where, under the normalization $\sigma^2=1$ adopted in the main text,
\[
v=\frac{1+(m-1)\alpha_0}{m},
\qquad
c=\alpha_1+r(\alpha_0-\alpha_1)+\frac{\alpha_2-\alpha_1}{m}.
\]
Equivalently, in terms of the underlying variance components,
\[
v=\sigma_\delta^2+\sigma_C^2
+\frac{\sigma_u^2+\sigma_\epsilon^2}{m},
\qquad
c=\sigma_\delta^2+r\sigma_C^2+\frac{\sigma_u^2}{m},
\]
where
$\operatorname{Cov}(CP_{skt},CP_{skt'})=r\sigma_C^2$ for $t\neq t'$
within the same cluster, and the remaining random-effect components are
mutually independent. Define
\[
\psi=\frac{c}{v}\in(0,1),
\qquad
\kappa=1-\psi^2>0,
\qquad
w=\frac{v}{v^2-c^2}>0.
\]
Then
\[
\mathbf V_*^{-1}
=
w
\begin{pmatrix}
1 & -\psi\\
-\psi & 1
\end{pmatrix}.
\]

\noindent
The treatment indicator within the two-period observation window is
$\mathbf X=(0,1)^\top$. Under categorical time effects, the sequence-specific
time-effect design matrices are
\[
\mathbf Z_1=
\begin{pmatrix}
1&0&0&0\\
0&1&0&0
\end{pmatrix},
\qquad
\mathbf Z_2=
\begin{pmatrix}
0&1&0&0\\
0&0&1&0
\end{pmatrix},
\qquad
\mathbf Z_3=
\begin{pmatrix}
0&0&1&0\\
0&0&0&1
\end{pmatrix}.
\]
For $s=1,2,3$, let
$\mathbf B_s=\mathbf Z_s^\top\mathbf V_*^{-1}\mathbf Z_s$ and
$\mathbf d_s=\mathbf Z_s^\top\mathbf V_*^{-1}\mathbf X$. Since
$\mathbf V_*^{-1}\mathbf X=w(-\psi,1)^\top$, direct calculation gives
\[
\mathbf B_1=w
\begin{pmatrix}
1&-\psi&0&0\\
-\psi&1&0&0\\
0&0&0&0\\
0&0&0&0
\end{pmatrix},
\qquad
\mathbf d_1=w
\begin{pmatrix}
-\psi\\1\\0\\0
\end{pmatrix},
\]
\[
\mathbf B_2=w
\begin{pmatrix}
0&0&0&0\\
0&1&-\psi&0\\
0&-\psi&1&0\\
0&0&0&0
\end{pmatrix},
\qquad
\mathbf d_2=w
\begin{pmatrix}
0\\-\psi\\1\\0
\end{pmatrix},
\]
and
\[
\mathbf B_3=w
\begin{pmatrix}
0&0&0&0\\
0&0&0&0\\
0&0&1&-\psi\\
0&0&-\psi&1
\end{pmatrix},
\qquad
\mathbf d_3=w
\begin{pmatrix}
0\\0\\-\psi\\1
\end{pmatrix}.
\]

\noindent
Let $\mathbf p=(p_1,p_2,p_3)^\top$ denote the sequence-allocation proportions.
With
$\mathbf A_{\mathbf p}=\sum_{s=1}^3p_s\mathbf B_s$ and
$\mathbf b_{\mathbf p}=\sum_{s=1}^3p_s\mathbf d_s$, define
$\widetilde{\mathbf A}_{\mathbf p}=\mathbf A_{\mathbf p}/w$ and
$\widetilde{\mathbf b}_{\mathbf p}=\mathbf b_{\mathbf p}/w$. Then
\[
\widetilde{\mathbf A}_{\mathbf p}
=
\begin{pmatrix}
p_1 & -p_1\psi & 0 & 0\\
-p_1\psi & p_1+p_2 & -p_2\psi & 0\\
0 & -p_2\psi & p_2+p_3 & -p_3\psi\\
0 & 0 & -p_3\psi & p_3
\end{pmatrix},
\]
and
\[
\widetilde{\mathbf b}_{\mathbf p}
=
\begin{pmatrix}
-p_1\psi\\
p_1-p_2\psi\\
p_2-p_3\psi\\
p_3
\end{pmatrix}.
\]
Moreover, since
$\mathbf X^\top\mathbf V_*^{-1}\mathbf X=w$ for every sequence,
$c_{\mathbf p}=w$. Therefore
\begin{equation}\label{eq:supp_var_s3_basic}
\operatorname{Var}(\hat\theta)
=
\frac{1}{Kw}\{1-W(\mathbf p)\}^{-1},
\qquad
W(\mathbf p)
=
\widetilde{\mathbf b}_{\mathbf p}^{\top}
\widetilde{\mathbf A}_{\mathbf p}^{-1}
\widetilde{\mathbf b}_{\mathbf p}.
\end{equation}

%\paragraph{Reduction of the GLS system.}
To evaluate $W(\mathbf p)$, solve
$\widetilde{\mathbf A}_{\mathbf p}\mathbf x
=\widetilde{\mathbf b}_{\mathbf p}$.
For an estimable interior allocation, $p_1,p_3>0$, and the first and fourth
rows give
\[
x_1=-\psi(1-x_2),
\qquad
x_4=1+\psi x_3.
\]
Substituting these expressions into the second and third rows yields
\[
\begin{pmatrix}
p_1\kappa+p_2 & -p_2\psi\\
-p_2\psi & p_2+p_3\kappa
\end{pmatrix}
\begin{pmatrix}
x_2\\x_3
\end{pmatrix}
=
\begin{pmatrix}
p_1\kappa-p_2\psi\\
p_2
\end{pmatrix}.
\]
Its determinant is
$\kappa\{p_2+\kappa p_1p_3\}>0$, and hence
\[
x_2
=
\frac{p_1p_2+p_1p_3\kappa-p_2p_3\psi}
{p_2+\kappa p_1p_3},
\qquad
x_3
=
\frac{p_1p_2(1+\psi)+p_2^2}
{p_2+\kappa p_1p_3}.
\]

By Lemma~\ref{lem:fixed_m_symmetry}, we can write
\[
p_1=p_3=p,
\qquad
p_2=1-2p,
\qquad
0<p<\frac12.
\]
Under this parametrization, let
\[
M=1-p(1-\psi),
\qquad
D=1-2p+\kappa p^2.
\]
Then
\[
x_2=\frac{p(1-\psi)M}{D},
\qquad
x_3=\frac{(1-2p)M}{D}.
\]
Using
\[
W(\mathbf p)
=
\begin{pmatrix}
-p\psi &
p-(1-2p)\psi &
(1-2p)-p\psi &
p
\end{pmatrix}
\begin{pmatrix}
-\psi(1-x_2)\\
x_2\\
x_3\\
1+\psi x_3
\end{pmatrix},
\]
we obtain, after simplification,
\[
W(p)
=
p(1+\psi^2)
+x_2\{p\kappa-\psi(1-2p)\}
+x_3(1-2p),
\]
and consequently
\begin{equation}\label{eq:supp_g_s3_basic}
g(p):=1-W(p)
=
\frac{
2\kappa p(1-2p)\{1-(1-\psi)p\}
}{
1-2p+\kappa p^2
}.
\end{equation}
For $0<p<1/2$, all factors in the numerator and denominator of
\eqref{eq:supp_g_s3_basic} are positive, so $g(p)>0$. Furthermore,
\[
\lim_{p\downarrow0}g(p)
=
\lim_{p\uparrow1/2}g(p)
=
0.
\]
Thus minimizing \eqref{eq:supp_var_s3_basic} is equivalent to maximizing
$g(p)$ over $0<p<1/2$.

%\paragraph{Optimality condition.}
Write
\[
g(p)=2\kappa\frac{N(p)}{D(p)},
\qquad
N(p)=p(1-2p)\{1-(1-\psi)p\},
\qquad
D(p)=1-2p+\kappa p^2.
\]
The first-order condition $g'(p)=0$ is equivalent to
$N'(p)D(p)-N(p)D'(p)=0$. Expanding and factorizing gives
\begin{equation}\label{eq:supp_foc_s3_basic}
\bigl\{(1-\psi)p-1\bigr\}^2
\left\{
2(1+\psi)p^2-4p+1
\right\}
=0.
\end{equation}
The repeated root of the first factor is
$p=1/(1-\psi)>1$ and is therefore infeasible. The quadratic factor gives
\[
p_{\pm}
=
\frac{
2\pm\sqrt{2(1-\psi)}
}{
2(1+\psi)
}.
\]
For $0<\psi<1$, $p_+>1/2$, whereas $0<p_-<1/2$. Hence
\eqref{eq:supp_foc_s3_basic} has exactly one stationary point in the admissible
interval. Since $g(p)>0$ in the interior and its continuous extension to
$[0,1/2]$ is zero at both endpoints, its global maximum must occur in the
interior. The unique admissible stationary point is therefore the global
maximizer of $g(p)$ and consequently minimizes
$\operatorname{Var}(\hat\theta)$.

\noindent
Thus the exact optimal allocation is
\begin{equation}\label{eq:supp_opt3Basic}
\boxed{
p_1^*=p_3^*
=
\frac{2-\sqrt{2(1-\psi)}}{2(1+\psi)},
\qquad
p_2^*
=
\frac{\sqrt{2(1-\psi)}-(1-\psi)}{1+\psi}
}
\end{equation}

\section{Derivation of the exact optimal allocation for a three-sequence staircase design with $R\geq2$}
\label{sec:supp_s3_general}

\noindent
Consider a balanced closed-cohort staircase design with $S=3$ sequences and
$R_0=R_1=R\geq2$. Each sequence is therefore observed over $2R$ consecutive
periods, and the total trial duration is $T=2R+2$. Throughout this derivation,
all clusters are assumed to have the same size $m$. Under the block-exchangeable
correlation structure, the covariance matrix of the $2R$ cluster-period means
from a single cluster has diagonal entry $v$ and common off-diagonal entry $c$,
so that
\[
\mathbf V_*
=
(v-c)\mathbf I_{2R}
+
c\,\mathbf 1_{2R}\mathbf 1_{2R}^{\top}.
\]
Under the normalization $\sigma^2=1$,
\[
v=\frac{1+(m-1)\alpha_0}{m},
\qquad
c=\alpha_1+r(\alpha_0-\alpha_1)
+\frac{\alpha_2-\alpha_1}{m}.
\]
Define $\psi=c/v\in(0,1)$. The inverse covariance matrix is
$\mathbf V_*^{-1}
=
w\mathbf I_{2R}
+
u\,\mathbf 1_{2R}\mathbf 1_{2R}^{\top}$,
where
\[
w=\frac{1}{v-c},
\qquad
u=-\frac{c}{(v-c)\{v+(2R-1)c\}}.
\]
It is convenient to write
\begin{equation}\label{eq:supp_nu_s3_general}
\nu=\frac{u}{w}
=
-\frac{c}{v+(2R-1)c}
=
-\frac{\psi}{1+(2R-1)\psi},
\qquad
L=1+\nu.
\end{equation}
Since $0<\psi<1$ and $R\geq2$,
\[
-\frac{1}{2R}<\nu<0,
\qquad
1-\frac{1}{2R}<L<1,
\]
and hence $3/4<L<1$.

\noindent By Lemma~\ref{lem:fixed_m_symmetry}, we can write
\[
p_1=p_3=p,
\qquad
p_2=1-2p,
\qquad
0<p\leq\frac{1}{2}.
\]

\noindent Let $\mathbf{X} = (\mathbf 0_R, \mathbf 1_R)^\top$ be the treatment indicator within a sequence-specific observation window, and
let $\mathbf Z_s$ be the $2R\times T$ matrix selecting the $2R$ calendar
periods observed by sequence $s$. Define $\mathbf u^{(s)} = \mathbf Z_s^\top\mathbf 1_{2R}$, $\mathbf x^{(s)}= \mathbf Z_s^\top\mathbf X$ and collect the sequence-window indicators in $\mathbf U=
\begin{bmatrix}
\mathbf u^{(1)}&
\mathbf u^{(2)}&
\mathbf u^{(3)}
\end{bmatrix}.$

\noindent Also let
\[
\mathbf P=\operatorname{diag}(p,1-2p,p).
\]
Since
$\mathbf Z_s^\top\mathbf Z_s
=\operatorname{diag}(\mathbf u^{(s)})$, the weighted time-effect information
matrix can be written as
\begin{equation}
\mathbf A_{\mathbf p}
=
\sum_{s=1}^3p_s
\mathbf Z_s^\top\mathbf V_*^{-1}\mathbf Z_s 
=
w\mathbf D_{\mathbf p}
+
u\mathbf U\mathbf P\mathbf U^\top
=
w\left(
\mathbf D_{\mathbf p}
+
\nu\mathbf U\mathbf P\mathbf U^\top
\right),
\label{eq:supp_A_s3_general}
\end{equation}
where
\[
\mathbf D_{\mathbf p}
=
\sum_{s=1}^3p_s\operatorname{diag}(\mathbf u^{(s)})
=
\operatorname{diag}
\left(
p,\,
1-p,\,
\underbrace{1,\ldots,1}_{2R-2\ {\rm entries}},\,
1-p,\,
p
\right).
\]
Thus $(\mathbf D_{\mathbf p})_{tt}$ is the total allocation proportion observed
in calendar period $t$.

\noindent Similarly,
\begin{align}
\mathbf b_{\mathbf p}
=
\sum_{s=1}^3p_s
\mathbf Z_s^\top\mathbf V_*^{-1}\mathbf X 
=
w\sum_{s=1}^3p_s\mathbf x^{(s)}
+
uR\sum_{s=1}^3p_s\mathbf u^{(s)}
=
w\mathbf V_X
+
w\nu R\,\mathbf D_{\mathbf p}\mathbf 1_T,
\label{eq:supp_b_s3_general}
\end{align}
where
\[
\mathbf V_X
=
\sum_{s=1}^3p_s\mathbf x^{(s)}
=
\left(
\underbrace{0,\ldots,0}_{R\ {\rm entries}},
p,\,
1-p,\,
\underbrace{1,\ldots,1}_{R-2\ {\rm entries}},
1-p,\,
p
\right)^\top.
\]
For $R=2$, the block containing $R-2$ entries is empty. Finally, because
$\mathbf X^\top\mathbf X=R$ and
$\mathbf 1_{2R}^\top\mathbf X=R$, the treatment-information scalar is
\begin{equation}\label{eq:supp_cp_s3_general}
c_{\mathbf p}
=
\mathbf X^\top\mathbf V_*^{-1}\mathbf X
=
wR+uR^2
=
wR(1+R\nu),
\end{equation}
which is independent of $\mathbf p$.

\noindent Let $W(\mathbf p)=\mathbf b_{\mathbf p}^{\top}\mathbf A_{\mathbf p}^{-1}\mathbf b_{\mathbf p}$. To evaluate this quantity, define
\[
\mathbf\Omega
=
\mathbf U^\top\mathbf D_{\mathbf p}^{-1}\mathbf U,
\qquad
\mathbf H
=
\frac{1}{\nu}\mathbf P^{-1}
+
\mathbf\Omega.
\]
Applying the Woodbury identity to \eqref{eq:supp_A_s3_general} gives
\begin{equation}\label{eq:supp_woodbury_s3_general}
\mathbf A_{\mathbf p}^{-1}
=
\frac{1}{w}
\left[
\mathbf D_{\mathbf p}^{-1}
-
\mathbf D_{\mathbf p}^{-1}\mathbf U
\mathbf H^{-1}
\mathbf U^\top\mathbf D_{\mathbf p}^{-1}
\right].
\end{equation}
Therefore,
\begin{equation}\label{eq:supp_Wsplit_s3_general}
W(\mathbf p)
=
\frac{1}{w}
\left[
\mathbf b_{\mathbf p}^{\top}\mathbf D_{\mathbf p}^{-1}\mathbf b_{\mathbf p}
-
\mathbf y^\top\mathbf H^{-1}\mathbf y
\right],
\qquad
\mathbf y
=
\mathbf U^\top\mathbf D_{\mathbf p}^{-1}\mathbf b_{\mathbf p}.
\end{equation}

From \eqref{eq:supp_b_s3_general},
\[
\mathbf b_{\mathbf p}^{\top}\mathbf D_{\mathbf p}^{-1}\mathbf b_{\mathbf p}
=
w^2
\left[
\mathbf V_X^\top\mathbf D_{\mathbf p}^{-1}\mathbf V_X
+
2R\nu\,\mathbf V_X^\top\mathbf 1_T
+
R^2\nu^2\mathbf 1_T^\top\mathbf D_{\mathbf p}\mathbf 1_T
\right].
\]
Direct calculation gives
\[
\mathbf V_X^\top\mathbf D_{\mathbf p}^{-1}\mathbf V_X
=
2p^2-2p+R,
\qquad
\mathbf V_X^\top\mathbf 1_T=R,
\qquad
\mathbf 1_T^\top\mathbf D_{\mathbf p}\mathbf 1_T=2R,
\]
and hence
\begin{equation}\label{eq:supp_firstterm_s3_general}
\mathbf b_{\mathbf p}^{\top}\mathbf D_{\mathbf p}^{-1}\mathbf b_{\mathbf p}
=
w^2
\left[
2p^2-2p+R+2R^2\nu(1+R\nu)
\right].
\end{equation}
Next define
\[
\mathbf S_E
=
\mathbf U^\top
\mathbf D_{\mathbf p}^{-1}\mathbf V_X.
\]
Summing over the three sequence windows gives
\[
\mathbf S_E
=
(R-1,R,R+1)^\top,
\qquad
\mathbf U^\top\mathbf 1_T
=
2R\,\mathbf 1_3,
\]
so that
\begin{equation}\label{eq:supp_y_s3_general}
\mathbf y
=
w\left(
\mathbf S_E+2R^2\nu\,\mathbf 1_3
\right).
\end{equation}

To simplify the second term in \eqref{eq:supp_Wsplit_s3_general}, define
\[
\mathbf z
=
\mathbf P\mathbf U^\top
\mathbf A_{\mathbf p}^{-1}\mathbf b_{\mathbf p}.
\]
Using \eqref{eq:supp_woodbury_s3_general} and
$\mathbf H=\nu^{-1}\mathbf P^{-1}+\mathbf\Omega$ gives
\[
\mathbf z
=
\frac{1}{w\nu}\mathbf H^{-1}\mathbf y,
\]
or equivalently,
\begin{equation}\label{eq:supp_zsystem_s3_general}
\left(
\mathbf I_3+\nu\mathbf P\mathbf\Omega
\right)\mathbf z
=
\mathbf P
\left(
\mathbf S_E+2R^2\nu\,\mathbf 1_3
\right).
\end{equation}

Premultiplying \eqref{eq:supp_zsystem_s3_general} by
$\mathbf 1_3^\top$ yields $\mathbf 1_3^\top\mathbf z=R$. Indeed,
\[
\mathbf p^\top\mathbf\Omega
=
(\mathbf U\mathbf p)^\top
\mathbf D_{\mathbf p}^{-1}\mathbf U
=
\mathbf 1_T^\top\mathbf U
=
2R\,\mathbf 1_3^\top,
\]
and
$\mathbf p^\top\mathbf S_E=R$, where
$\mathbf p=(p,1-2p,p)^\top$. Since $1+2R\nu>0$,
\[
(1+2R\nu)\mathbf 1_3^\top\mathbf z
=
R(1+2R\nu),
\]
which establishes the claim.

It remains to determine $z_3-z_1$. From the sequence-window geometry, $\Omega_{12}=\Omega_{32}$, while
\[
\Omega_{11}-\Omega_{31}
=
\frac{1}{p}+\frac{1}{1-p}
=
\frac{1}{p(1-p)},
\]
and
\[
\Omega_{13}-\Omega_{33}
=
-\frac{1}{p(1-p)}.
\]
Subtracting the third row of \eqref{eq:supp_zsystem_s3_general} from the
first therefore gives
\[
(z_1-z_3)
\left(
1+\frac{\nu}{1-p}
\right)
=
-2p,
\]
and hence
\begin{equation}\label{eq:supp_zdiff_s3_general}
z_3-z_1
=
\frac{2p(1-p)}{1-p+\nu}.
\end{equation}

From the definition of $\mathbf z$,
$\mathbf H^{-1}\mathbf y=w\nu\mathbf z$. Therefore, using
\eqref{eq:supp_y_s3_general},
\begin{align}
\mathbf y^\top\mathbf H^{-1}\mathbf y
&=
w^2\nu
\left(
\mathbf S_E+2R^2\nu\mathbf 1_3
\right)^\top\mathbf z \nonumber\\
&=
w^2\nu
\left[
R^2+(z_3-z_1)+2R^3\nu
\right],
\label{eq:supp_secondterm_s3_general}
\end{align}
where we used
$\mathbf S_E^\top\mathbf z
=R\mathbf 1_3^\top\mathbf z+(z_3-z_1)
=R^2+(z_3-z_1)$.
Substituting \eqref{eq:supp_firstterm_s3_general},
\eqref{eq:supp_zdiff_s3_general}, and
\eqref{eq:supp_secondterm_s3_general} into
\eqref{eq:supp_Wsplit_s3_general} gives
\begin{equation}\label{eq:supp_Wfinal_s3_general}
W(p)
=
w\left[
R+\nu R^2
-
\frac{2p(1-p)(1-p+2\nu)}
{1-p+\nu}
\right].
\end{equation}

Combining \eqref{eq:supp_cp_s3_general} and
\eqref{eq:supp_Wfinal_s3_general}, the Schur-complement information for
$\theta$ is
\begin{equation}\label{eq:supp_f_s3_general}
f(p)
=
c_{\mathbf p}-W(p)
=
\frac{2w\,p(1-p)(1-p+2\nu)}
{1-p+\nu}.
\end{equation}
Consequently,
\[
\operatorname{Var}(\hat\theta)
=
\frac{1}{Kf(p)}.
\]
Since $w>0$, minimizing the treatment-effect variance is equivalent to
maximizing
\[
h(p)
=
\frac{2p(1-p)(1-p+2\nu)}
{1-p+\nu},
\qquad
0<p<\frac{1}{2}.
\]
The bounds $-1/(2R)<\nu<0$ and $R\geq2$ ensure that both
$1-p+\nu$ and $1-p+2\nu$ are positive throughout this interval.

\noindent Differentiating $h(p)$ and simplifying gives
\begin{equation}\label{eq:supp_hprime_s3_general}
h'(p)
=
-\frac{2q(p)}{(1-p+\nu)^2},
\end{equation}
where
\begin{equation}\label{eq:supp_cubic_s3_general}
q(p)
=
2p^3
-
5(1+\nu)p^2
+
4(1+\nu)^2p
-
(1+\nu)(1+2\nu).
\end{equation}
Writing $L=1+\nu$, the first-order condition becomes
\begin{equation}\label{eq:supp_cubic_L_s3_general}
2p^3-5Lp^2+4L^2p-(2L^2-L)=0.
\end{equation}

The cubic in \eqref{eq:supp_cubic_L_s3_general} has exactly one real root.
To see this, for
\[
a=2,\qquad
b=-5L,\qquad
c_0=4L^2,\qquad
d=-(2L^2-L),
\]
the Cardano quantities are
\[
\Delta_{0,3}=b^2-3ac_0=L^2
\]
and
\[
\Delta_{1,3}
=
2b^3-9abc_0+27a^2d
=
110L^3-216L^2+108L.
\]
Moreover,
\[
\Delta_{1,3}^2-4\Delta_0^3
=
432L^2(L-1)^2(28L^2-54L+27)>0,
\]
because $3/4<L<1$ and
$28L^2-54L+27>0$ for all real $L$. Thus the cubic has one real root and
two complex conjugate roots. In addition,
\[
q(0)=-L(2L-1)<0,
\qquad
q\!\left(\frac{1}{2}\right)=\frac{1-L}{4}>0,
\]
so its unique real root lies in $(0,1/2)$. By
\eqref{eq:supp_hprime_s3_general}, $h'(p)>0$ before this root and
$h'(p)<0$ after it. Hence the unique admissible stationary point is the
global maximizer of $h(p)$ and therefore minimizes
$\operatorname{Var}(\hat\theta)$.

Using Cardano's formula, define
\[
C_3
=
\sqrt[3]{
\frac{
\Delta_{1,3}+\sqrt{\Delta_{1,3}^2-4\Delta_{0,3}^3}
}{2}
},
\qquad
\Delta_{0,3}=L^2,
\qquad
\Delta_{1,3}=110L^3-216L^2+108L,
\]
where the real cube root is taken. The unique admissible solution of
\eqref{eq:supp_cubic_L_s3_general} is therefore
\begin{equation}\label{eq:supp_opt3general}
\boxed{
p_1^*=p_3^*
=
\frac{5L-C_3-\dfrac{L^2}{C_3}}{6}
}.
\end{equation}
Accordingly, the exact optimal allocation vector is $\mathbf p^*=\left(p_1^*,\,1-2p_1^*,\,p_1^*\right)^\top$, with $L$ and $\nu$ defined in \eqref{eq:supp_nu_s3_general}. 

\section{Derivation of the exact optimal allocation for a four-sequence basic staircase design}
\label{sec:supp_s4_basic}

\noindent
Consider a balanced closed-cohort staircase design with $S=4$ sequences and
$R_0=R_1=1$, so that the trial spans $T=5$ periods. Throughout this derivation,
all clusters are assumed to have the same size $m$. Each sequence is observed
in two consecutive periods, with treatment indicator
$\mathbf X=(0,1)^\top$ within its observation window.

\noindent Under the block-exchangeable correlation structure, the covariance matrix of
the two cluster-period means from a single cluster is common to all sequences
and is given by
\[
\mathbf V_*
=
\begin{pmatrix}
v & c\\
c & v
\end{pmatrix},
\]
where, under the normalization $\sigma^2=1$,
\[
v=\frac{1+(m-1)\alpha_0}{m},
\qquad
c=\alpha_1+r(\alpha_0-\alpha_1)
+\frac{\alpha_2-\alpha_1}{m}.
\]
Define
\[
\psi=\frac{c}{v}\in(0,1),
\qquad
\kappa=1-\psi^2>0,
\qquad
\omega=\frac{v}{v^2-c^2}>0.
\]
Then
\begin{equation}\label{eq:supp_Vinv_s4_basic}
\mathbf V_*^{-1}
=
\omega
\begin{pmatrix}
1 & -\psi\\
-\psi & 1
\end{pmatrix}.
\end{equation}

\noindent
Under categorical time effects
$\boldsymbol{\beta}=(\beta_1,\ldots,\beta_5)^\top$, the sequence-specific
design matrices are
\[
\mathbf Z_s
=
\begin{pmatrix}
\mathbf e_s^\top\\
\mathbf e_{s+1}^\top
\end{pmatrix},
\qquad s=1,\ldots,4,
\]
where $\mathbf e_j$ is the $j$th standard basis vector in $\mathbb R^5$.
Let
\[
\mathbf A_{\mathbf p}
=
\sum_{s=1}^4p_s\mathbf Z_s^\top\mathbf V_*^{-1}\mathbf Z_s,
\qquad
\mathbf b_{\mathbf p}
=
\sum_{s=1}^4p_s\mathbf Z_s^\top\mathbf V_*^{-1}\mathbf X.
\]
Since
$\mathbf X^\top\mathbf V_*^{-1}\mathbf X=\omega$ for every sequence,
$c_{\mathbf p}=\omega$. Define
\[
\widetilde{\mathbf A}_{\mathbf p}
=
\frac{\mathbf A_{\mathbf p}}{\omega},
\qquad
\widetilde{\mathbf b}_{\mathbf p}
=
\frac{\mathbf b_{\mathbf p}}{\omega}.
\]
Using \eqref{eq:supp_Vinv_s4_basic}, direct calculation gives
\[
\widetilde{\mathbf A}_{\mathbf p}
=
\begin{pmatrix}
p_1 & -p_1\psi & 0 & 0 & 0\\
-p_1\psi & p_1+p_2 & -p_2\psi & 0 & 0\\
0 & -p_2\psi & p_2+p_3 & -p_3\psi & 0\\
0 & 0 & -p_3\psi & p_3+p_4 & -p_4\psi\\
0 & 0 & 0 & -p_4\psi & p_4
\end{pmatrix}
\]
and
\[
\widetilde{\mathbf b}_{\mathbf p}
=
\begin{pmatrix}
-p_1\psi \ \
p_1-p_2\psi\ \
p_2-p_3\psi\ \
p_3-p_4\psi\ \
p_4
\end{pmatrix}^\top
\]
Therefore,
\begin{equation}\label{eq:supp_var_s4_basic}
\operatorname{Var}(\hat\theta)
=
\frac{1}{K\omega}
\{g(\mathbf p)\}^{-1},
\qquad
g(\mathbf p)
=
1-W(\mathbf p),
\end{equation}
where
\[
W(\mathbf p)
=
\widetilde{\mathbf b}_{\mathbf p}^{\top}
\widetilde{\mathbf A}_{\mathbf p}^{-1}
\widetilde{\mathbf b}_{\mathbf p}.
\]

\noindent By Lemma~\ref{lem:fixed_m_symmetry}, we can write
\[
p_1=p_4=a,
\qquad
p_2=p_3=b.
\]
Since $\sum_{s=1}^4p_s=1$,
\[
a+b=\frac12,
\qquad
a=\frac12-b,
\qquad
0<b<\frac12.
\]
Thus the allocation problem reduces to one free parameter $b$.

\noindent To evaluate $W(\mathbf p)$, solve
\[
\widetilde{\mathbf A}_{\mathbf p}\mathbf x
=
\widetilde{\mathbf b}_{\mathbf p}.
\]
Under the symmetric parametrization, the first and fifth rows give
\[
x_1=-\psi(1-x_2),
\qquad
x_5=1+\psi x_4.
\]
Substitution into the remaining three equations yields
\[
\begin{pmatrix}
a\kappa+b & -b\psi & 0\\
-b\psi & 2b & -b\psi\\
0 & -b\psi & a\kappa+b
\end{pmatrix}
\begin{pmatrix}
x_2\\x_3\\x_4
\end{pmatrix}
=
\begin{pmatrix}
a\kappa-b\psi\\
b(1-\psi)\\
b
\end{pmatrix}.
\]
Solving this reduced system gives
\begin{equation}\label{eq:supp_x_s4_basic}
x_2
=
\frac{2a\kappa-b\psi}{2(a\kappa+b)},
\qquad
x_3=\frac12,
\qquad
x_4
=
\frac{b(2+\psi)}{2(a\kappa+b)}.
\end{equation}

Substituting \eqref{eq:supp_x_s4_basic}, together with the boundary
expressions for $x_1$ and $x_5$, into
$W(\mathbf p)=\widetilde{\mathbf b}_{\mathbf p}^\top\mathbf x$ and simplifying
gives
\begin{equation}\label{eq:supp_g_s4_basic}
g(b)
=
\frac{
\kappa b\{a(5+4\psi)+b\}
}{
2(a\kappa+b)
},
\qquad
a=\frac12-b.
\end{equation}
Since $K\omega>0$, minimizing
$\operatorname{Var}(\hat\theta)$ in \eqref{eq:supp_var_s4_basic} is
equivalent to maximizing $g(b)$ over $0<b<1/2$.

\noindent Differentiating \eqref{eq:supp_g_s4_basic} and simplifying gives
\begin{equation}\label{eq:supp_gprime_s4_basic}
g'(b)
=
-\frac{
8(1-\psi)(1+\psi)^2
}{
\{1-\psi^2+2\psi^2b\}^2
}
\,Q(b),
\end{equation}
where
\begin{equation}\label{eq:supp_Q_s4_basic}
Q(b)
=
\psi^2b^2
+
(1-\psi^2)b
-
\frac{(1-\psi)(5+4\psi)}{16}.
\end{equation}
Hence the first-order condition $g'(b)=0$ is equivalent to
\begin{equation}\label{eq:supp_foc_s4_basic}
\psi^2b^2
+
(1-\psi^2)b
-
\frac{(1-\psi)(5+4\psi)}{16}
=
0.
\end{equation}

For $0<\psi<1$,
\[
Q(0)
=
-\frac{(1-\psi)(5+4\psi)}{16}<0,
\]
whereas
\[
Q\!\left(\frac12\right)
=
\frac{3+\psi}{16}>0.
\]
Moreover, the product of the two roots of $Q$ is
\[
-\frac{(1-\psi)(5+4\psi)}{16\psi^2}<0,
\]
so the quadratic has exactly one positive root and one negative root.
Consequently, the positive root is the unique solution of
\eqref{eq:supp_foc_s4_basic} in $(0,1/2)$. Since the prefactor multiplying
$Q(b)$ in \eqref{eq:supp_gprime_s4_basic} is strictly negative,
\[
g'(b)>0
\quad\text{before the admissible root},
\qquad
g'(b)<0
\quad\text{after the admissible root}.
\]
Thus the unique admissible stationary point is the global maximizer of $g(b)$
over the symmetric allocation class and therefore minimizes
$\operatorname{Var}(\hat\theta)$.

Solving \eqref{eq:supp_foc_s4_basic} by the quadratic formula gives the
admissible root
\begin{equation}\label{eq:supp_bstar_s4_basic}
b^*
=
\frac{
(2+\psi)\sqrt{1-\psi}
-
2(1-\psi^2)
}{
4\psi^2
}.
\end{equation}
Since $a^*=1/2-b^*$,
\[
a^*
=
\frac{
2-(2+\psi)\sqrt{1-\psi}
}{
4\psi^2
}.
\]
Therefore, the exact optimal allocation is
\begin{equation}\label{eq:supp_opt4Basic}
\boxed{
p_1^*=p_4^*
=
\frac{2-(2+\psi)\sqrt{1-\psi}}{4\psi^2},
\qquad
p_2^*=p_3^*
=
\frac{(2+\psi)\sqrt{1-\psi}-2(1-\psi^2)}
{4\psi^2}
}
\end{equation}

\section{Derivation of the exact optimal allocation for a four-sequence staircase design with $R=2$}
\label{sec:supp_s4_R2}

\noindent
Consider a balanced closed-cohort staircase design with $S=4$ sequences and
$R_0=R_1=2$. Each sequence is observed over four consecutive periods, and
the total trial duration is $T=7$. Throughout this derivation, all clusters
are assumed to have the same size $m$. The treatment indicator within a
sequence-specific observation window is
\[
\mathbf X=(0,0,1,1)^\top.
\]
Under categorical time effects, the sequence-specific design matrices are
\[
\mathbf Z_s
=
\begin{pmatrix}
\mathbf e_s^\top\\
\mathbf e_{s+1}^\top\\
\mathbf e_{s+2}^\top\\
\mathbf e_{s+3}^\top
\end{pmatrix}
\in\mathbb R^{4\times7},
\qquad s=1,\ldots,4,
\]
where $\mathbf e_j$ is the $j$th standard basis vector in $\mathbb R^7$.

Under the block-exchangeable correlation structure, the covariance matrix of
the four cluster-period means from a single cluster is
\[
\mathbf V_*
=
(v-c)\mathbf I_4
+
c\,\mathbf 1_4\mathbf 1_4^\top,
\]
where, under the normalization $\sigma^2=1$ adopted in the main manuscript,
\[
v=\frac{1+(m-1)\alpha_0}{m},
\qquad
c=\alpha_1+r(\alpha_0-\alpha_1)
+\frac{\alpha_2-\alpha_1}{m}.
\]
Writing $\psi=c/v\in(0,1)$, the inverse covariance matrix is
\[
\mathbf V_*^{-1}
=
\frac{1}{v-c}\mathbf I_4
-
\frac{c}{(v-c)(v+3c)}
\mathbf 1_4\mathbf 1_4^\top.
\]
Equivalently,
\begin{equation}\label{eq:supp_Vinv_s4_R2}
\mathbf V_*^{-1}
=
\omega
\begin{pmatrix}
1&\eta&\eta&\eta\\
\eta&1&\eta&\eta\\
\eta&\eta&1&\eta\\
\eta&\eta&\eta&1
\end{pmatrix},
\end{equation}
where
\begin{equation}\label{eq:supp_eta_s4_R2}
\omega
=
\frac{v+2c}{(v-c)(v+3c)},
\qquad
\eta
=
-\frac{c}{v+2c}
=
-\frac{\psi}{1+2\psi}.
\end{equation}
Since $0<\psi<1$,
\[
-\frac13<\eta<0.
\]
Moreover,
\[
\frac{\mathbf V_*^{-1}\mathbf X}{\omega}
=
(2\eta,\,2\eta,\,1+\eta,\,1+\eta)^\top,
\]
and hence
\begin{equation}\label{eq:supp_cp_s4_R2}
c_{\mathbf p}
=
\mathbf X^\top\mathbf V_*^{-1}\mathbf X
=
2\omega(1+\eta),
\end{equation}
which is independent of the sequence allocation.

By Lemma~\ref{lem:fixed_m_symmetry}, we can write
\[
p_1=p_4=a,
\qquad
p_2=p_3=b,
\qquad
a=\frac12-b,
\qquad
0\leq b<\frac12.
\]
Define
\[
\widetilde{\mathbf A}_{\mathbf p}
=
\frac{\mathbf A_{\mathbf p}}{\omega},
\qquad
\widetilde{\mathbf b}_{\mathbf p}
=
\frac{\mathbf b_{\mathbf p}}{\omega}.
\]
The treatment-effect variance can then be written as
\begin{equation}\label{eq:supp_var_s4_R2}
\operatorname{Var}(\hat\theta)
=
\frac{1}{K\omega}\{g(b)\}^{-1},
\qquad
g(b)
=
2(1+\eta)-W(b),
\end{equation}
where
\[
W(b)
=
\widetilde{\mathbf b}_{\mathbf p}^{\top}
\widetilde{\mathbf A}_{\mathbf p}^{-1}
\widetilde{\mathbf b}_{\mathbf p}.
\]

Let $\mathbf x$ solve
\[
\widetilde{\mathbf A}_{\mathbf p}\mathbf x
=
\widetilde{\mathbf b}_{\mathbf p}.
\]
The first and seventh equations give
\begin{align}
x_1
&=
\eta(2-x_2-x_3-x_4),
\label{eq:supp_x1_s4_R2}\\
x_7
&=
1+\eta-\eta(x_4+x_5+x_6).
\label{eq:supp_x7_s4_R2}
\end{align}
Under the symmetric allocation,
$\widetilde{\mathbf A}_{\mathbf p}$ is centrosymmetric with respect to the
$7\times7$ reversal matrix. Introduce the symmetric and antisymmetric
combinations
\[
u_2=x_2+x_6,
\qquad
u_3=x_3+x_5,
\qquad
u_4=x_4,
\]
and
\[
v_2=x_2-x_6,
\qquad
v_3=x_3-x_5.
\]
Adding each interior equation to its reflected counterpart yields
\begin{equation}\label{eq:supp_us_s4_R2}
u_2=1,
\qquad
u_3=1,
\qquad
u_4=\frac12.
\end{equation}
Thus all design dependence is contained in the antisymmetric component.
Subtracting the reflected equations and using $a=1/2-b$ gives
\begin{equation}\label{eq:supp_vsystem_s4_R2}
\begin{pmatrix}
\frac12-a\eta^2 & a\eta(1-\eta)\\
a\eta & 1-a(1-\eta)
\end{pmatrix}
\begin{pmatrix}
v_2\\v_3
\end{pmatrix}
=
\begin{pmatrix}
(1-\eta)(a\eta-\frac12)\\
a(3+\eta)-1
\end{pmatrix}.
\end{equation}
Its determinant is
\[
\frac12\left\{1-a(1-\eta+2\eta^2)\right\}>0
\]
for $0<a<1/2$ and $-1/3<\eta<0$, so the system is nonsingular.
Solving \eqref{eq:supp_vsystem_s4_R2} gives
\begin{align}
v_2
&=
\frac{
(\eta-1)
\left[
16b^2\eta-2b(5\eta-1)+(\eta+1)
\right]
}{
2b(2\eta^2-\eta+1)
-(\eta-1)(2\eta+1)
},
\label{eq:supp_v2_s4_R2}\\[6pt]
v_3
&=
\frac{
-16b^2\eta^2
+2b(7\eta^2-2\eta-3)
-(\eta-1)(3\eta+1)
}{
2b(2\eta^2-\eta+1)
-(\eta-1)(2\eta+1)
}.
\label{eq:supp_v3_s4_R2}
\end{align}

Using \eqref{eq:supp_us_s4_R2},
\[
x_2=\frac{1+v_2}{2},
\qquad
x_6=\frac{1-v_2}{2},
\qquad
x_3=\frac{1+v_3}{2},
\qquad
x_5=\frac{1-v_3}{2},
\qquad
x_4=\frac12,
\]
while $x_1$ and $x_7$ follow from
\eqref{eq:supp_x1_s4_R2}--\eqref{eq:supp_x7_s4_R2}.
Substitution into
$W(b)=\widetilde{\mathbf b}_{\mathbf p}^{\top}\mathbf x$
and simplification yields
\begin{equation}\label{eq:supp_g_s4_R2}
g(b)
=
\frac{
(\eta-1)N(b)
}{
4D(b)
},
\end{equation}
where
\begin{align}
N(b)
={}&
128\eta^2b^3
-32(5\eta^2-2\eta-1)b^2 \nonumber\\
&\quad
+2(21\eta^2-16\eta-9)b
+(3\eta+1)(\eta-1),
\label{eq:supp_N_s4_R2}
\end{align}
and
\begin{equation}\label{eq:supp_D_s4_R2}
D(b)
=
2b(2\eta^2-\eta+1)
-(2\eta+1)(\eta-1).
\end{equation}
For $-1/3<\eta<0$ and $0<b<1/2$,
\[
D(b)
=
2b(2\eta^2-\eta+1)
+(1-\eta)(2\eta+1)>0.
\]
All four allocation proportions are positive on this interval, so the
treatment effect is estimable and $g(b)>0$.

Differentiating \eqref{eq:supp_g_s4_R2} gives
\begin{equation}\label{eq:supp_gprime_s4_R2}
g'(b)
=
\frac{
2(\eta-1)P_\eta(b)
}{
\left[
2b(2\eta^2-\eta+1)
-(2\eta+1)(\eta-1)
\right]^2
},
\end{equation}
where
\begin{equation}\label{eq:supp_P_s4_R2}
P_\eta(b)
=
B_3b^3+B_2b^2+B_1b+B_0
\end{equation}
with
\begin{align*}
B_3&=128\eta^4-64\eta^3+64\eta^2,\\
B_2&=-176\eta^4+120\eta^3+8\eta^2+8\eta+8,\\
B_1&=80\eta^4-72\eta^3-40\eta^2+24\eta+8,\\
B_0&=-12\eta^4+15\eta^3+5\eta^2-6\eta-2.
\end{align*}
Thus the stationary points are precisely the roots of $P_\eta(b)=0$.

At the left endpoint,
\[
P_\eta(0)
=
(1-\eta)
\left(
12\eta^3-3\eta^2-8\eta-2
\right)<0.
\]
Indeed, the cubic factor equals $-1/9$ at $\eta=-1/3$ and is strictly
decreasing on $(-1/3,0)$. At the other endpoint,
\[
P_\eta\!\left(\frac12\right)
=
\eta^3-5\eta^2+8\eta+4>0,
\]
because this expression is increasing on $(-1/3,0)$ and equals $20/27$ at
$\eta=-1/3$. Hence, by continuity, $P_\eta$ has at least one root in
$(0,1/2)$. Since $\eta-1<0$, \eqref{eq:supp_gprime_s4_R2} also shows that
$g'(b)>0$ near $b=0$ and $g'(b)<0$ near $b=1/2$.

Now let
\[
V(b)
=
\operatorname{Var}
\left(
\hat\theta;
\frac12-b,b,b,\frac12-b
\right).
\]
The map from $b$ to the allocation vector is affine. Therefore, by convexity
of the variance criterion in the allocation proportions, $V(b)$ is convex
on $(0,1/2)$. Moreover,
\[
V'(b)
=
-\frac{g'(b)}{K\omega g(b)^2}.
\]
Hence $V'(b)<0$ near $0$ and $V'(b)>0$ near $1/2$. A differentiable convex
function has a nondecreasing derivative. If $V'(b)$ vanished at two distinct
interior points, it would vanish throughout the interval between them.
Since $V(b)$ is a nonconstant rational, and therefore real-analytic, function
on $(0,1/2)$, this would force $V$ to be constant on the whole connected
interval, contradicting the strict derivative signs above. Therefore
$P_\eta(b)=0$ has exactly one root
\[
b^*\in(0,1/2),
\]
and this root is the unique global minimiser of $V(b)$.

\paragraph{Closed-form representation of the admissible root.}
For the cubic in \eqref{eq:supp_P_s4_R2}, define
\begin{equation}\label{eq:supp_delta0_s4_R2}
\Delta_{0,4}^{(2)}
=
B_2^2-3B_3B_1,
\end{equation}
\begin{equation}\label{eq:supp_delta1_s4_R2}
\Delta_{1,4}^{(2)}
=
2B_2^3-9B_3B_2B_1+27B_3^2B_0,
\end{equation}
and
\begin{equation}\label{eq:supp_disc_s4_R2}
\mathcal D_{4,2}
=
\left(\Delta_{1,4}^{(2)}\right)^2
-
4\left(\Delta_{0,4}^{(2)}\right)^3.
\end{equation}
The sign of $\mathcal D_{4,2}$ determines the number of real roots of the
cubic, but the preceding argument shows that, irrespective of this sign,
exactly one root belongs to $(0,1/2)$.

For completeness, the Cardano discriminant factorizes as
\begin{align}
\mathcal D_{4,2}
={}&
226492416\,\eta^4(\eta-1)
(2\eta^2-\eta+1)^2
(2\eta^3-5\eta^2+3\eta+2)^2
Q_8(\eta),
\label{eq:supp_disc_factor_s4_R2}
\end{align}
where
\[
Q_8(\eta)
=
2\eta^8+65\eta^7-281\eta^6+319\eta^5
+25\eta^4-127\eta^3-25\eta^2+5\eta+1.
\]
A Sturm-sequence calculation shows that $Q_8$ has exactly one zero in
$(-1/3,0)$,
\[
\eta_0\approx-0.1758181653.
\]
Since the remaining nonsquared factor in
\eqref{eq:supp_disc_factor_s4_R2} is $\eta-1<0$, it follows that
\[
\mathcal D_{4,2}>0
\quad\text{for}\quad
-\frac13<\eta<\eta_0,
\]
\[
\mathcal D_{4,2}=0
\quad\text{at}\quad
\eta=\eta_0,
\qquad\text{and}\qquad
\mathcal D_{4,2}<0
\quad\text{for}\quad
\eta_0<\eta<0.
\]

\medskip
\noindent
\emph{Case $\mathcal D_{4,2}>0$.}
The cubic has exactly one real root. Since
$P_\eta(0)<0<P_\eta(1/2)$, that root necessarily lies in $(0,1/2)$ and is
therefore $b^*$. To avoid the exceptional cancellation that can occur in
the usual fixed-sign Cardano expression, define the nonzero real quantity
\begin{equation}\label{eq:supp_C_s4_R2}
C_{4,2}
=
\sqrt[3]{
\frac{
\Delta_{1,4}^{(2)}
+
\operatorname{sgn}\!\left(\Delta_{1,4}^{(2)}\right)
\sqrt{\mathcal D_{4,2}}
}{2}
}.
\end{equation}
Here the real cube root is taken. Since
$\mathcal D_{4,2}>0$ implies $\Delta_{1,4}^{(2)}\neq0$, the definition in
\eqref{eq:supp_C_s4_R2} is nonzero. Cardano's formula then gives
\begin{equation}\label{eq:supp_opt4_R2_cardano}
b^*
=
-\frac{1}{3B_3}
\left(
B_2+C_{4,2}
+\frac{\Delta_{0,4}^{(2)}}{C_{4,2}}
\right).
\end{equation}

\medskip
\noindent
\emph{Case $\mathcal D_{4,2}\leq0$.}
In this case the cubic has three real roots, counting multiplicity when
$\mathcal D_{4,2}=0$. Moreover, $\Delta_{0,4}^{(2)}>0$, and we may define
\begin{equation}\label{eq:supp_theta_s4_R2}
\vartheta_{4,2}
=
\frac13
\arccos\left\{
\frac{\Delta_{1,4}^{(2)}}
{2\left(\Delta_{0,4}^{(2)}\right)^{3/2}}
\right\}.
\end{equation}
The three real roots are
\begin{equation}\label{eq:supp_trig_roots_s4_R2}
b_k
=
-\frac{1}{3B_3}
\left[
B_2+
2\sqrt{\Delta_{0,4}^{(2)}}
\cos\left(
\vartheta_{4,2}+\frac{2\pi k}{3}
\right)
\right],
\qquad k=0,1,2.
\end{equation}

It remains to identify the admissible member of
\eqref{eq:supp_trig_roots_s4_R2}. From
\eqref{eq:supp_disc_factor_s4_R2},
$\mathcal D_{4,2}\leq0$ implies $\eta_0\leq\eta<0$, and in particular
$-1/5<\eta<0$. Writing $x=-\eta\in(0,1/5)$,
\[
B_3
=
64\eta^2(2\eta^2-\eta+1)>0,
\]
and
\[
\frac{B_2}{8}
=
1-x+x^2-15x^3-22x^4
>
1-\frac15-\frac{15}{125}-\frac{22}{625}
=
\frac{403}{625}>0.
\]
Also $B_0<0$ throughout $-1/3<\eta<0$, as established above from
$P_\eta(0)=B_0<0$. Therefore the coefficient sequence
$(B_3,B_2,B_1,B_0)$ has exactly one sign change, irrespective of the sign of
$B_1$. By Descartes' rule of signs, $P_\eta$ has exactly one positive real
root. Since a root has already been shown to lie in $(0,1/2)$, this
admissible root is the unique positive root and hence the largest of the
three real roots.

For $0<\vartheta_{4,2}\leq\pi/3$, the quantity
\[
\cos\left(\vartheta_{4,2}+\frac{2\pi}{3}\right)
\]
is the smallest of the three cosine terms appearing in
\eqref{eq:supp_trig_roots_s4_R2}. Since $B_3>0$, the corresponding
Cardano root is the largest root. Consequently, the unique admissible root
is the $k=1$ member of \eqref{eq:supp_trig_roots_s4_R2}, namely
\begin{equation}\label{eq:supp_opt4_R2_trig}
b^*
=
-\frac{1}{3B_3}
\left[
B_2+
2\sqrt{\Delta_{0,4}^{(2)}}
\cos\left(
\vartheta_{4,2}+\frac{2\pi}{3}
\right)
\right].
\end{equation}

Combining the two cases, the exact optimal allocation is
\begin{equation}\label{eq:supp_opt4_R2}
\boxed{
\mathbf p^*
=
\left(
\frac12-b^*,\,
b^*,\,
b^*,\,
\frac12-b^*
\right)^\top,
}
\end{equation}
where $b^*$ is given by
\eqref{eq:supp_opt4_R2_cardano} when $\mathcal D_{4,2}>0$ and by
\eqref{eq:supp_opt4_R2_trig} when $\mathcal D_{4,2}\leq0$.

\section{Derivation of the exact optimal allocation for a four-sequence staircase design with $R\geq3$}
\label{sec:supp_s4_general}

\noindent
Consider a balanced closed-cohort staircase design with $S=4$ sequences and
$R_0=R_1=R\geq3$. Each sequence is observed over $2R$ consecutive periods,
and the total trial duration is $T=2R+3$. Throughout this derivation, all
clusters are assumed to have the same size $m$.

Under the block-exchangeable correlation structure, the covariance matrix of
the $2R$ cluster-period means from a single cluster has diagonal entry $v$ and
common off-diagonal entry $c$, and hence
\[
\mathbf V_*
=
(v-c)\mathbf I_{2R}
+
c\,\mathbf 1_{2R}\mathbf 1_{2R}^{\top}.
\]
Under the normalization $\sigma^2=1$,
\[
v=\frac{1+(m-1)\alpha_0}{m},
\qquad
c=\alpha_1+r(\alpha_0-\alpha_1)
+\frac{\alpha_2-\alpha_1}{m}.
\]
Define $\psi=c/v\in(0,1)$. The inverse covariance matrix can be written as
\[
\mathbf V_*^{-1}
=
w\mathbf I_{2R}
+
u\,\mathbf 1_{2R}\mathbf 1_{2R}^{\top},
\]
where
\[
w=\frac{1}{v-c},
\qquad
u=-\frac{c}{(v-c)\{v+(2R-1)c\}}.
\]
It is convenient to define
\begin{equation}\label{eq:supp_nu_s4_general}
\nu=\frac{u}{w}
=
-\frac{c}{v+(2R-1)c}
=
-\frac{\psi}{1+(2R-1)\psi},
\qquad
L=1+\nu.
\end{equation}
Since $0<\psi<1$ and $R\geq3$,
\[
-\frac{1}{2R}<\nu<0,
\qquad
1-\frac{1}{2R}<L<1,
\]
and therefore $5/6<L<1$.

\noindent By Lemma~\ref{lem:fixed_m_symmetry} we can write
\[
p_1=p_4=p,
\qquad
p_2=p_3=\frac12-p,
\qquad
0<p\leq \frac12.
\]
Let
\[
\mathbf P
=
\operatorname{diag}
\left(
p,\frac12-p,\frac12-p,p
\right).
\]

\noindent Let $\mathbf X=
\begin{pmatrix}
\mathbf 0_R\ \ 
\mathbf 1_R
\end{pmatrix}$ be the treatment indicator within a sequence-specific observation window, and let $\mathbf Z_s$ be the $2R\times T$ matrix selecting the calendar periods observed by sequence $s$. Define $\mathbf u^{(s)}
=
\mathbf Z_s^\top\mathbf 1_{2R}$, $\mathbf x^{(s)}
=
\mathbf Z_s^\top\mathbf X$ and $\mathbf U
=
\begin{bmatrix}
\mathbf u^{(1)}&
\mathbf u^{(2)}&
\mathbf u^{(3)}&
\mathbf u^{(4)}
\end{bmatrix}$.

\noindent Since
$\mathbf Z_s^\top\mathbf Z_s
=\operatorname{diag}(\mathbf u^{(s)})$, the weighted information matrix for
the categorical time effects is
\begin{align}
\mathbf A_{\mathbf p}
=
\sum_{s=1}^4
p_s\mathbf Z_s^\top\mathbf V_*^{-1}\mathbf Z_s 
=
w\mathbf D_{\mathbf p}
+
u\mathbf U\mathbf P\mathbf U^\top 
=
w\left(
\mathbf D_{\mathbf p}
+
\nu\mathbf U\mathbf P\mathbf U^\top
\right),
\label{eq:supp_A_s4_general}
\end{align}
where
\[
\mathbf D_{\mathbf p}
=
\operatorname{diag}
\left(
p,\,
\frac12,\,
1-p,\,
\underbrace{1,\ldots,1}_{2R-3\ {\rm entries}},\,
1-p,\,
\frac12,\,
p
\right).
\]
Thus $(\mathbf D_{\mathbf p})_{tt}$ is the total allocation proportion
observed in calendar period $t$.

\noindent Similarly,
\begin{align}
\mathbf b_{\mathbf p}
=
\sum_{s=1}^4
p_s\mathbf Z_s^\top\mathbf V_*^{-1}\mathbf X 
=
w\sum_{s=1}^4p_s\mathbf x^{(s)}
+
uR\sum_{s=1}^4p_s\mathbf u^{(s)} 
=
w\mathbf V_X
+
w\nu R\,\mathbf D_{\mathbf p}\mathbf 1_T,
\label{eq:supp_b_s4_general}
\end{align}
where
\[
\mathbf V_X
=
\left(
\underbrace{0,\ldots,0}_{R\ {\rm entries}},
p,\,
\frac12,\,
1-p,\,
\underbrace{1,\ldots,1}_{R-3\ {\rm entries}},
1-p,\,
\frac12,\,
p
\right)^\top.
\]
For $R=3$, the block containing $R-3$ entries is empty.

Because $\mathbf X^\top\mathbf X=R$ and
$\mathbf 1_{2R}^\top\mathbf X=R$, the treatment-information scalar is
\begin{equation}\label{eq:supp_cp_s4_general}
c_{\mathbf p}
=
\mathbf X^\top\mathbf V_*^{-1}\mathbf X
=
wR+uR^2
=
wR(1+R\nu),
\end{equation}
which is independent of $\mathbf p$.

\noindent Let
\[
W(\mathbf p)
=
\mathbf b_{\mathbf p}^{\top}
\mathbf A_{\mathbf p}^{-1}
\mathbf b_{\mathbf p}.
\]
Define
\[
\mathbf\Omega
=
\mathbf U^\top\mathbf D_{\mathbf p}^{-1}\mathbf U,
\qquad
\mathbf H
=
\frac{1}{\nu}\mathbf P^{-1}
+
\mathbf\Omega.
\]
Applying the Woodbury identity to \eqref{eq:supp_A_s4_general} yields
\begin{equation}\label{eq:supp_woodbury_s4_general}
\mathbf A_{\mathbf p}^{-1}
=
\frac{1}{w}
\left[
\mathbf D_{\mathbf p}^{-1}
-
\mathbf D_{\mathbf p}^{-1}\mathbf U
\mathbf H^{-1}
\mathbf U^\top\mathbf D_{\mathbf p}^{-1}
\right].
\end{equation}
Therefore,
\begin{equation}\label{eq:supp_Wsplit_s4_general}
W(\mathbf p)
=
\frac{1}{w}
\left[
\mathbf b_{\mathbf p}^{\top}
\mathbf D_{\mathbf p}^{-1}
\mathbf b_{\mathbf p}
-
\mathbf y^\top\mathbf H^{-1}\mathbf y
\right],
\qquad
\mathbf y
=
\mathbf U^\top\mathbf D_{\mathbf p}^{-1}\mathbf b_{\mathbf p}.
\end{equation}

From \eqref{eq:supp_b_s4_general},
\[
\mathbf b_{\mathbf p}^{\top}
\mathbf D_{\mathbf p}^{-1}
\mathbf b_{\mathbf p}
=
w^2
\left[
\mathbf V_X^\top
\mathbf D_{\mathbf p}^{-1}\mathbf V_X
+
2R\nu\,\mathbf V_X^\top\mathbf 1_T
+
R^2\nu^2
\mathbf 1_T^\top\mathbf D_{\mathbf p}\mathbf 1_T
\right].
\]
Direct calculation gives
\[
\mathbf V_X^\top
\mathbf D_{\mathbf p}^{-1}\mathbf V_X
=
2p^2-2p+R-\frac14,
\qquad
\mathbf V_X^\top\mathbf 1_T=R,
\qquad
\mathbf 1_T^\top\mathbf D_{\mathbf p}\mathbf 1_T=2R,
\]
so that
\begin{equation}\label{eq:supp_firstterm_s4_general}
\mathbf b_{\mathbf p}^{\top}
\mathbf D_{\mathbf p}^{-1}
\mathbf b_{\mathbf p}
=
w^2
\left[
2p^2-2p+R-\frac14
+
2R^2\nu(1+R\nu)
\right].
\end{equation}

Next define
\[
\mathbf S_E
=
\mathbf U^\top
\mathbf D_{\mathbf p}^{-1}\mathbf V_X.
\]
Summing over the four sequence windows gives
\[
\mathbf S_E
=
\left(
R-\frac32,\,
R-\frac12,\,
R+\frac12,\,
R+\frac32
\right)^\top,
\qquad
\mathbf U^\top\mathbf 1_T
=
2R\,\mathbf 1_4.
\]
Consequently,
\begin{equation}\label{eq:supp_y_s4_general}
\mathbf y
=
w\left(
\mathbf S_E
+
2R^2\nu\,\mathbf 1_4
\right).
\end{equation}

To simplify the second term in \eqref{eq:supp_Wsplit_s4_general}, define
\[
\mathbf z
=
\mathbf P\mathbf U^\top
\mathbf A_{\mathbf p}^{-1}
\mathbf b_{\mathbf p}.
\]
As in the three-sequence derivation, \eqref{eq:supp_woodbury_s4_general}
implies
\[
\mathbf z
=
\frac{1}{w\nu}
\mathbf H^{-1}\mathbf y,
\]
and therefore
\begin{equation}\label{eq:supp_zsystem_s4_general}
\left(
\mathbf I_4+\nu\mathbf P\mathbf\Omega
\right)\mathbf z
=
\mathbf P
\left(
\mathbf S_E+2R^2\nu\,\mathbf 1_4
\right).
\end{equation}

Premultiplying \eqref{eq:supp_zsystem_s4_general} by
$\mathbf 1_4^\top$ gives
$\mathbf 1_4^\top\mathbf z=R$. Indeed,
\[
\mathbf p^\top\mathbf\Omega
=
(\mathbf U\mathbf p)^\top
\mathbf D_{\mathbf p}^{-1}\mathbf U
=
\mathbf 1_T^\top\mathbf U
=
2R\,\mathbf 1_4^\top,
\]
while
\[
\mathbf p^\top\mathbf S_E=R,
\]
where
$\mathbf p=(p,\frac12-p,\frac12-p,p)^\top$. Hence
\[
(1+2R\nu)\mathbf 1_4^\top\mathbf z
=
R(1+2R\nu).
\]
Since $1+2R\nu>0$,
\begin{equation}\label{eq:supp_zsum_s4_general}
\mathbf 1_4^\top\mathbf z=R.
\end{equation}

It remains to determine the antisymmetric differences
\[
\Delta z_{14}=z_1-z_4,
\qquad
\Delta z_{23}=z_2-z_3.
\]
For $R\geq3$, direct evaluation of the sequence-window overlap sums gives
\[
\Omega_{11}-\Omega_{14}
=
\frac{1}{p(1-p)}+2,
\]
and
\[
\Omega_{12}-\Omega_{13}
=
\Omega_{21}-\Omega_{31}
=
\Omega_{22}-\Omega_{23}
=
2.
\]
Subtracting the fourth row of \eqref{eq:supp_zsystem_s4_general} from the
first, and the third row from the second, gives
\begin{align}
\left(
1+\frac{\nu}{1-p}+2\nu p
\right)\Delta z_{14}
+
2\nu p\,\Delta z_{23}
&=-3p,
\label{eq:supp_diff1_s4_general}\\
\nu(1-2p)\Delta z_{14}
+
\{1+\nu(1-2p)\}\Delta z_{23}
&=-\left(\frac12-p\right).
\label{eq:supp_diff2_s4_general}
\end{align}
The determinant of this $2\times2$ system is
\[
\Delta_{\rm sys}
=
\frac{
(1+\nu)^2-p(1+\nu+2\nu^2)
}{
1-p
}.
\]
Writing $L=1+\nu$, define
\begin{equation}\label{eq:supp_D_s4_general}
D(p)
=
L^2-p(2L^2-3L+2).
\end{equation}
For $0<p<1/2$,
\[
D(p)
>
L^2-\frac12(2L^2-3L+2)
=
\frac{3L-2}{2}>0,
\]
because $L>5/6$.

Solving \eqref{eq:supp_diff1_s4_general}--\eqref{eq:supp_diff2_s4_general}
for the particular linear combination required below gives
\begin{equation}\label{eq:supp_diffcomb_s4_general}
\frac32\Delta z_{14}
+
\frac12\Delta z_{23}
=
\frac{
-\frac14(1+\nu)
-p\left(\frac{15}{4}+\frac32\nu\right)
+p^2(4+6\nu)
-4\nu p^3
}{
(1+\nu)^2-p(1+\nu+2\nu^2)
}.
\end{equation}

From the definition of $\mathbf z$,
$\mathbf H^{-1}\mathbf y=w\nu\mathbf z$. Using
\eqref{eq:supp_y_s4_general} and \eqref{eq:supp_zsum_s4_general},
\begin{align}
\mathbf y^\top\mathbf H^{-1}\mathbf y
&=
w^2\nu
\left(
\mathbf S_E+2R^2\nu\mathbf 1_4
\right)^\top\mathbf z \nonumber\\
&=
w^2\nu
\left[
R^2
-\frac32\Delta z_{14}
-\frac12\Delta z_{23}
+
2R^3\nu
\right].
\label{eq:supp_secondterm_s4_general}
\end{align}
Substituting \eqref{eq:supp_firstterm_s4_general},
\eqref{eq:supp_diffcomb_s4_general}, and
\eqref{eq:supp_secondterm_s4_general} into
\eqref{eq:supp_Wsplit_s4_general} yields
\begin{align}
W(p)
=
w\Bigg[
&R-\frac14+\nu R^2+2p^2-2p \nonumber\\
&\quad
+\nu
\frac{
-\frac14(1+\nu)
-p\left(\frac{15}{4}+\frac32\nu\right)
+p^2(4+6\nu)
-4\nu p^3
}{
(1+\nu)^2-p(1+\nu+2\nu^2)
}
\Bigg].
\label{eq:supp_Wfinal_s4_general}
\end{align}

Combining \eqref{eq:supp_cp_s4_general} and
\eqref{eq:supp_Wfinal_s4_general}, define
\begin{equation}\label{eq:supp_H_s4_general}
H(p)
=
2p^2-2p
+
\nu
\frac{
-\frac14(1+\nu)
-p\left(\frac{15}{4}+\frac32\nu\right)
+p^2(4+6\nu)
-4\nu p^3
}{
(1+\nu)^2-p(1+\nu+2\nu^2)
}.
\end{equation}
Then
\begin{equation}\label{eq:supp_f_s4_general}
f(p)
=
c_{\mathbf p}-W(p)
=
w\left\{
\frac14-H(p)
\right\},
\end{equation}
and hence
\[
\operatorname{Var}(\hat\theta)
=
\frac{1}{
Kw\{\frac14-H(p)\}
}.
\]
Since $Kw>0$, minimizing the treatment-effect variance is equivalent to
minimizing $H(p)$ over $0<p<1/2$.

\noindent Differentiating \eqref{eq:supp_H_s4_general} and simplifying gives
\begin{equation}\label{eq:supp_Hprime_s4_general}
H'(p)
=
\frac{
Q(p)
}{
2D(p)^2
},
\end{equation}
where $D(p)$ is defined in \eqref{eq:supp_D_s4_general} and
\begin{equation}\label{eq:supp_cubic_s4_general}
Q(p)
=
A_3p^3+A_2p^2+A_1p+A_0,
\end{equation}
with
\begin{align*}
A_3&=64L^4-208L^3+296L^2-208L+64,\\
A_2&=-96L^4+212L^3-204L^2+92L-24,\\
A_1&=48L^4-56L^3+24L^2,\\
A_0&=-8L^4+L^3+2L^2+L.
\end{align*}
Thus the stationary points are precisely the roots of
$Q(p)=0$.

We next show that this cubic has exactly one admissible root. Define the
Cardano discriminant quantities
\[
\Delta_0=A_2^2-3A_3A_1,
\qquad
\Delta_1=
2A_2^3-9A_3A_2A_1+27A_3^2A_0.
\]
A direct factorization gives
\begin{align}
\Delta_1^2-4\Delta_0^3
={}&
110592\,L(L-1)^2(L+2)^2
(2L^2-3L+2)^2 
(4L^2-7L+4)^2
P(L),
\label{eq:supp_disc_s4_general}
\end{align}
where $P(L)=1439L^5-5646L^4+8819L^3
-6660L^2+2268L-216$.

\noindent For $5/6<L<1$, all factors outside $P(L)$ in
\eqref{eq:supp_disc_s4_general} are strictly positive. To establish the
positivity of $P(L)$, let $t=1-L$, so that $0<t<1/6$. Then
\[
P(1-t)
=
(4-16t)
+t^2(311-625t)
+t^4(1549-1439t),
\]
and each term on the right-hand side is strictly positive for
$0<t<1/6$. Therefore,
\[
\Delta_1^2-4\Delta_0^3>0,
\]
and $Q(p)$ has exactly one real root.

It remains to locate this root. At the endpoints of the admissible interval,
\[
Q(0)
=
-L(8L^3-L^2-2L-1)<0.
\]
Indeed,
$8L^3-L^2-2L-1$ is positive at $L=5/6$ and is strictly increasing for
$L>5/6$. Moreover,
\[
Q\!\left(\frac12\right)
=
2(1-L)>0.
\]
Thus, by continuity, the unique real root of $Q$ lies in $(0,1/2)$.
Since $D(p)^2>0$ on this interval, \eqref{eq:supp_Hprime_s4_general} implies
that $H'(p)<0$ before this root and $H'(p)>0$ after it. Hence the unique
admissible stationary point is the global minimiser of $H(p)$ and therefore
minimizes $\operatorname{Var}(\hat\theta)$.

\noindent Define
\[
\Delta_{0,4}=A_2^2-3A_3A_1,
\qquad
\Delta_{1,4}=
2A_2^3-9A_3A_2A_1+27A_3^2A_0,
\]
and
\[
C_4=
\sqrt[3]{
\frac{
\Delta_{1,4}+\sqrt{\Delta_{1,4}^2-4\Delta_{0,4}^3}
}{2}
},
\]
where the real cube root is taken. Cardano's formula therefore gives the
unique admissible solution
\begin{equation}\label{eq:supp_opt4general}
\boxed{
p^*
=
-\frac{1}{3A_3}
\left(
A_2+C_4+\frac{\Delta_{0,4}}{C_4}
\right)
}.
\end{equation}
Consequently, the exact optimal allocation vector is $\mathbf p^*
=
\left(
p^*,\,
0.5-p^*,\,
0.5-p^*,\,
p^*
\right)^\top.$